\documentclass[floatfix,
prab,
10pt,
amsmath,
amssymb,
aps,
twocolumn,
longbibliography,
]{revtex4-2}

\pdfoutput=1 % If your are submitting a pdflatex
\usepackage{textcomp,gensymb} % Symbols (including textcomp resolves a warning)
\usepackage{mathptmx} % Font
\usepackage{caption} % Figures
\usepackage[labelformat=simple]{subcaption}

\usepackage[number-unit-separator=~]{siunitx} % SI-units
\usepackage{graphicx} % Include figure files
\usepackage{dcolumn} % Align table columns on the decimal point
\usepackage{url} % For transforming links
\usepackage{comment} % For commenting larger sections out
\usepackage{xcolor}
\usepackage{ragged2e}
\usepackage{hyperref}

\newdimen\figrasterwd
\figrasterwd\textwidth

\definecolor{APSBlue}{RGB}{46, 48, 146}
\hypersetup{
    colorlinks,
    linkcolor = APSBlue,
    citecolor = APSBlue,
    urlcolor = APSBlue
}
\makeatletter\def\Hy@Warning#1{}\makeatother % Remove all hyperref warnings
\usepackage{microtype} % Helps to prevent text overflowing outside the columns.

\DeclareUrlCommand\url{\urlstyle{rm}}
\DeclareUrlCommand\nolinkurl{\urlstyle{rm}}
\renewcommand{\doi}[1]{\href{https://doi.org/#1}{\nolinkurl{#1}}}
\renewcommand{\url}[1]{\href{#1}{\nolinkurl{#1}}}

\DeclareCaptionFormat{figures}{\justifying #1#2\quad{#3}}
\DeclareCaptionFormat{tables}{\justifying #1#2\quad{#3}}
\DeclareCaptionFormat{subfigures}{\centering#1#2#3}  % Center subfigure captions
\begin{document}

\title{Decommissioning and post-irradiation examination of the LHC beam dumps}

\author{N.~Solieri} %STI-TCD
\email{nicola.solieri@cern.ch}
\author{A.~Lund} %STI-TCD
\email{asbjoern.lassen.lund@cern.ch}
\author{A.-P.~Bernardes} %STI-TCD
\author{L.~R. Buonocore} %BE-CEM
\author{A.~Cherif} %EN-MME
\author{S.~De Man} %STI-TCD
\author{M.~Di Castro} %BE-CEM-MRO
\author{S.~Di Giovannantonio} %BE-CEM
\author{G.~Dumont} %HSE-RP
\author{S.~El-Idrissi} %HSE-RP
\author{E.~Farina} %STI-BMI
\author{D.~Grenier} %STI-TCD
\author{E.~Grenier-Boley} %STI-TCD
\author{M.~Himmerlich} %TE-VSC
\author{A.~Infantino} %HSE-RP
\author{A.~Lechner} %STI-BMI
\author{R.~Mouret} %HSE-RP
\author{D.~Pazem} %TE-VSC
\author{A.~T.~Perez-Fontenla} %EN-MME
\author{E.~Romagnoli} %BE-CEM
\author{S.~Sgobba} %EN-MME
\author{C.~Tromel} %HSE-RP
\author{C.~Veiga Almagro} %BE-CEM
\author{M.~Calviani} %STI-TCD
\email{marco.calviani@cern.ch}

\affiliation{CERN, 1211 Geneva 23, Switzerland}

\begin{abstract}
The LHC beam dumps are responsible for the safe absorption of the Large Hadron Collider (LHC) particle beams. In 2018, the two 6.4-tonne beam dumps that had been in operation since the LHC's startup in 2008 were removed and replaced with upgraded versions. Endoscopic inspections of these beam dumps and experimental high-intensity proton-beam irradiation of material samples raised concerns about the structural integrity of the carbon-based materials in their cores. It was therefore decided to undertake an accelerated project of dismantling and post-irradiation examination of the removed dumps as part of a wider program of work to ensure the safe operation of the LHC beam dumps in the coming years. This paper describes the decommissioning process for the two beam dumps carried out at CERN between 2021 and 2023, covering the preparatory studies, practical challenges encountered, and solutions implemented. It details the establishment of an operational framework, including the preparation of the working environment, the development of a method for cutting the irradiated \SI[number-unit-separator=\text{-}]{12}{\milli\meter}-thick duplex stainless-steel vessel, and the cut sequencing. Additionally, the paper presents the findings derived from the post-irradiation examination of the different carbon-based core materials subjected to deposited energy densities up to \SI{1.5}{\kilo\joule\per\gram}. The extruded graphite plates within the vessel exhibited a cracking pattern, which was likely due to the dynamic response of the device upon beam impact, and their retaining rings were found to be displaced. Despite minor signs of surface deterioration, the expanded graphite sheets were intact, and the isostatic graphite blocks showed no evidence of material degradation. These findings were consistent across both beam dumps. Finally, this manuscript discusses the implications of these results for the safe operation of the beam dumps during LHC Run~3 (2022--2026) and for the design and construction of Run 3 operational spares.

\end{abstract}

%\keywords{Radioactive decommissioning, Beam intercepting devices, beam dumps, carbon-based materials} % PRAB does not use keywords

\maketitle

\section{Introduction} \label{introduction}
\subsection{Context} \label{context}
The beam dumps are key components of the Large Hadron Collider (LHC) Beam Dump System (LBDS); they are responsible for the safe extraction and absorption of the LHC particle beams in all possible scenarios, including instances of abnormal machine operation~\cite{Bracco2019LBDS2}.

From Run~1 (2009--2013) to the current Run~3 (2022--2026), the beam parameters of the LHC have placed increasing demands on the beam dumps, and this will continue with the initiation of High Luminosity LHC (HL-LHC) operations in Run~4 (2030--2041)~\cite{Aberle2020High-LuminosityReport}. The HL-LHC modifications will lead to an overall 4.5-fold increase in the total stored energy per beam, from an initial \SI{150}{\mega\joule} in Run~1 to \SI{680}{\mega\joule} in Run~4~\cite{Maestre2021DesignMj/Dump}. As such, an ongoing program of research and development (R\&D) is needed to ensure that the beam dumps can continue to comply with these increasingly challenging beam parameters~\cite{Karastathis2019LHCReport}. Over the years, this program has encompassed a wide range of activities, including mechanical testing~\cite{Solfiti2023FIB-SEMGraphite}, in-beam experiments examining the core materials~\cite{Maestre2021SigraflexPerspective, AndreuMunoz2022IrradiationExperiment}, and numerical simulations, with a view to designing the next generation of beam dumps.

To the authors' best knowledge, post-operation examinations of beam-intercepting devices subjected to levels of peak and integrated energy deposition comparable to those found in the LHC have never been performed. Consequently, dismantling the beam dumps was an essential step in evaluating the materials within their cores. While the International Atomic Energy Agency (IAEA) has compiled substantial information on the decommissioning of particle accelerators \cite{2020DecommissioningAccelerators}, none of the documented cases, nor those found in the broader literature, is directly comparable to the LHC beam dumps. Additionally, the options for executing the decommissioning of the beam dumps were severely constrained by the levels of induced radioactivity and the complex nature of the dismantling activity, which required the careful extraction of activated carbon-based materials from a large duplex stainless-steel vessel. Finally, the potential impact of the results of the assessment of the beam dumps on their operation during Run~3 imposed a strict timeline. This timeline could only be met by the development of specialized tooling and techniques, as well as the construction of a temporary facility at CERN to carry out both the decommissioning operations and the post-irradiation examination.

\begin{figure*}[thbp]
\centering
\includegraphics[width=0.90\textwidth]{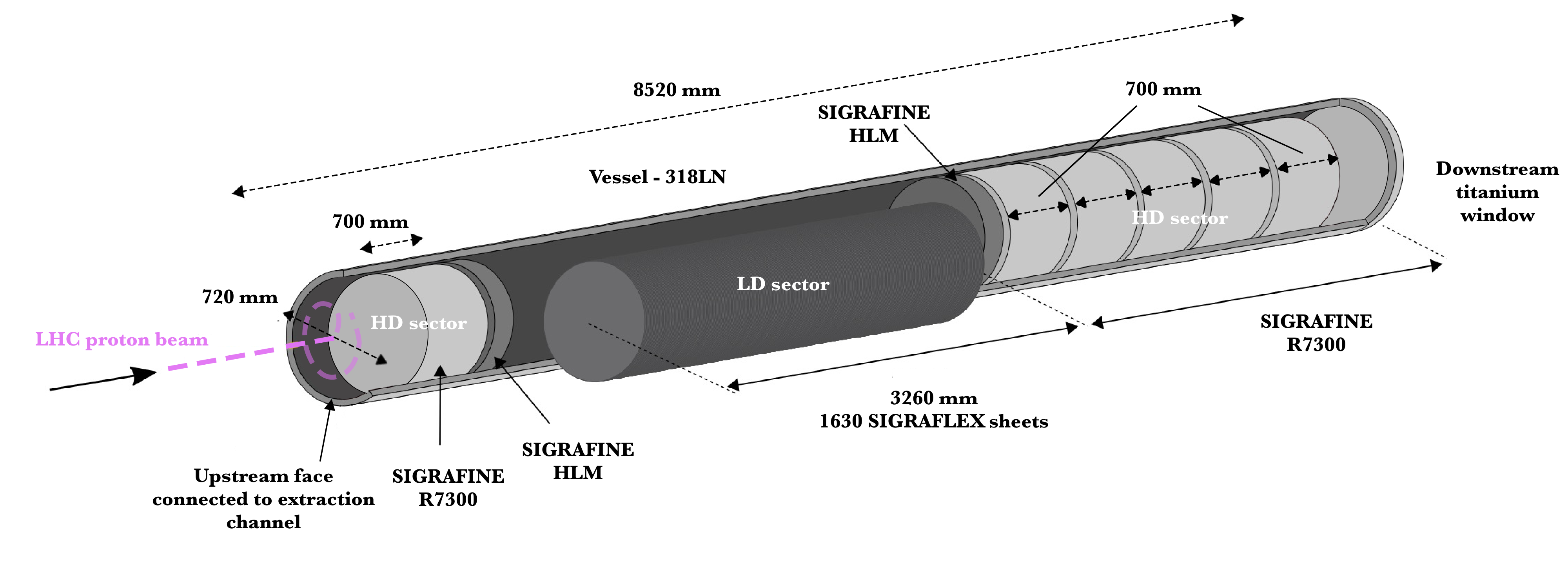}
\caption{Schematic illustration of a beam dump core in the configuration used for Run~1 and Run~2. In the direction of the beam, this includes: the upstream face, which is directly connected via a bellows to the extraction beam line; an HD isostatic graphite block; an LD sector comprising \SI{3.26}{\meter} of expanded graphite sheets, compressed at both ends by extruded graphite discs; five blocks of HD isostatic graphite; and a downstream titanium Grade~2 window. The core is encased in a \SI[number-unit-separator=\text{-}]{12}{\milli\meter}-thick duplex stainless-steel vessel.}
\label{fig:dump-cross-section}
\end{figure*}

\subsection{Structure and materials of the LHC beam dumps} \label{structure-and-materials}
The original dump-block design dates back to the late 1990s~\cite{Zazula1996LHCSystem, Zazula1996LHCSystemb}. Figure~\ref{fig:dump-cross-section} shows a schematic cross-sectional view of the core of the operational beam dumps used from Run~1 to the current Run~3. The dump block has a diameter of \SI{720}{\milli\meter}, a length of \SI{8.5}{\meter}, and a weight of \SI{6.4}{\tonne}~\cite{Maestre2021DesignMj/Dump}. Duplex stainless-steel seamless blanks (318LN, AISI~1.4462, EN~10088-2, also known by its commercial name URANUS\textsuperscript{\tiny\textregistered}~45), each approximately \SI{0.7}{\meter} in length, are circumferentially joined using partial-penetration butt welds to form the up- and downstream high-density (HD) sectors. The \SI[number-unit-separator=\text{-}]{3.5}{\meter}-long low-density (LD) sector is constructed from a plate of the same material, which is bent, longitudinally welded, and then joined to the high-density sectors at both ends using circumferential welds.

Different grades of carbon-based materials with varying densities were selected to progressively absorb the proton beam and the generated secondary particles~\cite{Zazula1996LHCSystem, Zazula1996LHCSystemb}. Six HD blocks of \SI[number-unit-separator=\text{-}]{0.7}{\meter} length, made of \SI[number-unit-separator=\text{-}]{1.73}{\gram\per\cubic\centi\meter} isostatic graphite (SIGRAFINE\textsuperscript{\tiny\textregistered}, SGL grade R7300~\cite{SGL_R7300}), are shrink-fitted inside the vessel to ensure efficient thermal contact and facilitate heat extraction. A stack of approximately 1630 expanded graphite sheets forms the central LD sector, with a total length of \SI{3.26}{\meter}; these \SI[number-unit-separator=\text{-}]{2}{\milli\meter}-thick expanded graphite sheets (commercial name SGL SIGRAFLEX\textsuperscript{\tiny\textregistered} L20012C~\cite{SGL_SIGRAFLEX}) have a density of \SI{1.2}{\gram\per\cubic\centi\meter}. Due to their low rigidity, the sheets cannot be shrink-fitted into the vessel. The LD sector is held under slight longitudinal compression by two extruded graphite discs (SGL SIGRAFINE HLM~\cite{SGL_SIGRAFINE_HLM_US}). These are retained by rings inserted into grooves within the vessel and spot-welded around the circumference. Figure~\ref{fig:TDE-installation} shows the configuration of the beam dumps during Run~1 and Run~2, where the vessel was connected to an extension of the extraction line with a beam pipe at its upstream end and sealed at the downstream end with a titanium Grade~2 window. The beam dump is filled with nitrogen (N$_2$) at 1.2~bar to create an inert atmosphere to protect the carbon-based materials from oxidation~\cite{Maestre2021DesignMj/Dump}.

\subsection{Brief history of the beam dumps and their integration} \label{history}
\begin{figure}[tbp]
    \centering
        \begin{subfigure}{\columnwidth}
        \centering
        \includegraphics[width=\textwidth]{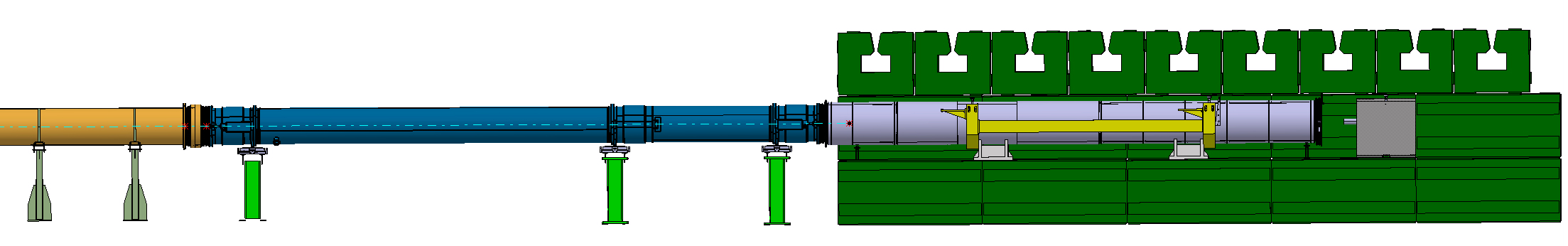}
        \caption{Schematic cross-sectional view of the installation of the beam dump in its shielding before LS2. The extraction line (orange and blue) was connected to the beam dump (grey) via a bellows. The beam pipe (shown in blue) and the bellows were removed during the upgrade works in 2020~\cite{Maestre2021DesignMj/Dump}. This resulted in the decoupling of the beam dump from the LHC extraction line.}
        \label{fig:TDE-carven-illustration}
    \end{subfigure}
    \begin{subfigure}{\columnwidth}
        \centering
        \includegraphics[width=\textwidth]{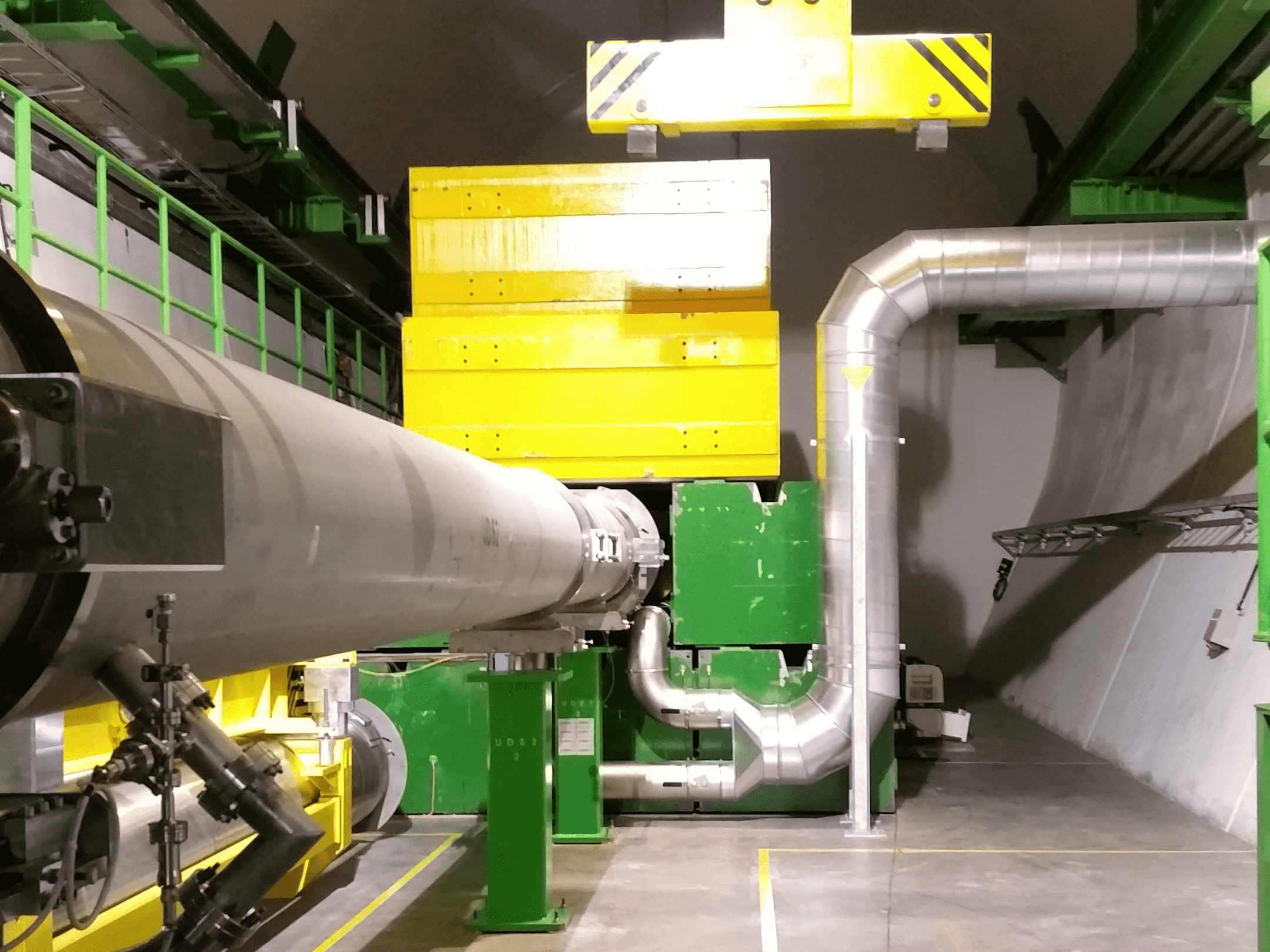}
        \caption{Photograph showing the beam dump installed inside the \SI[number-unit-separator=\text{-}]{900}{\tonne} shielding (green/yellow) prior to the upgrade program in 2020.}
        \label{fig:TDE-cavern}
    \end{subfigure}
    \caption{Beam dump installation before LS2.}
    \label{fig:TDE-installation}
\end{figure}

The two LHC beam dumps are housed within purpose-built underground caverns (UD62 and UD68). These are located at the end of the \SI[number-unit-separator=\text{-}]{630}{\meter}-long extraction channels that are tangential to the LHC ring at point~6~\cite{Goddard2003LHCAcceptance}. Figure~\ref{fig:TDE-cavern} shows the beam dump installed in the UD62 cavern inside \SI{900}{\tonne} of radiation-shielding blocks. Until the end of Run 2, the dumps were physically connected to the beam-transfer line arriving from the LHC ring via a beam pipe and bellows, as shown in Fig.~\ref{fig:TDE-carven-illustration}. They rested on steel supports within the shielding, and only frictional forces resisted their movement when impacted by the beam~\cite{Maestre2021DesignMj/Dump}.

\begin{figure}[tbp]
\centering
\includegraphics[width=\columnwidth]{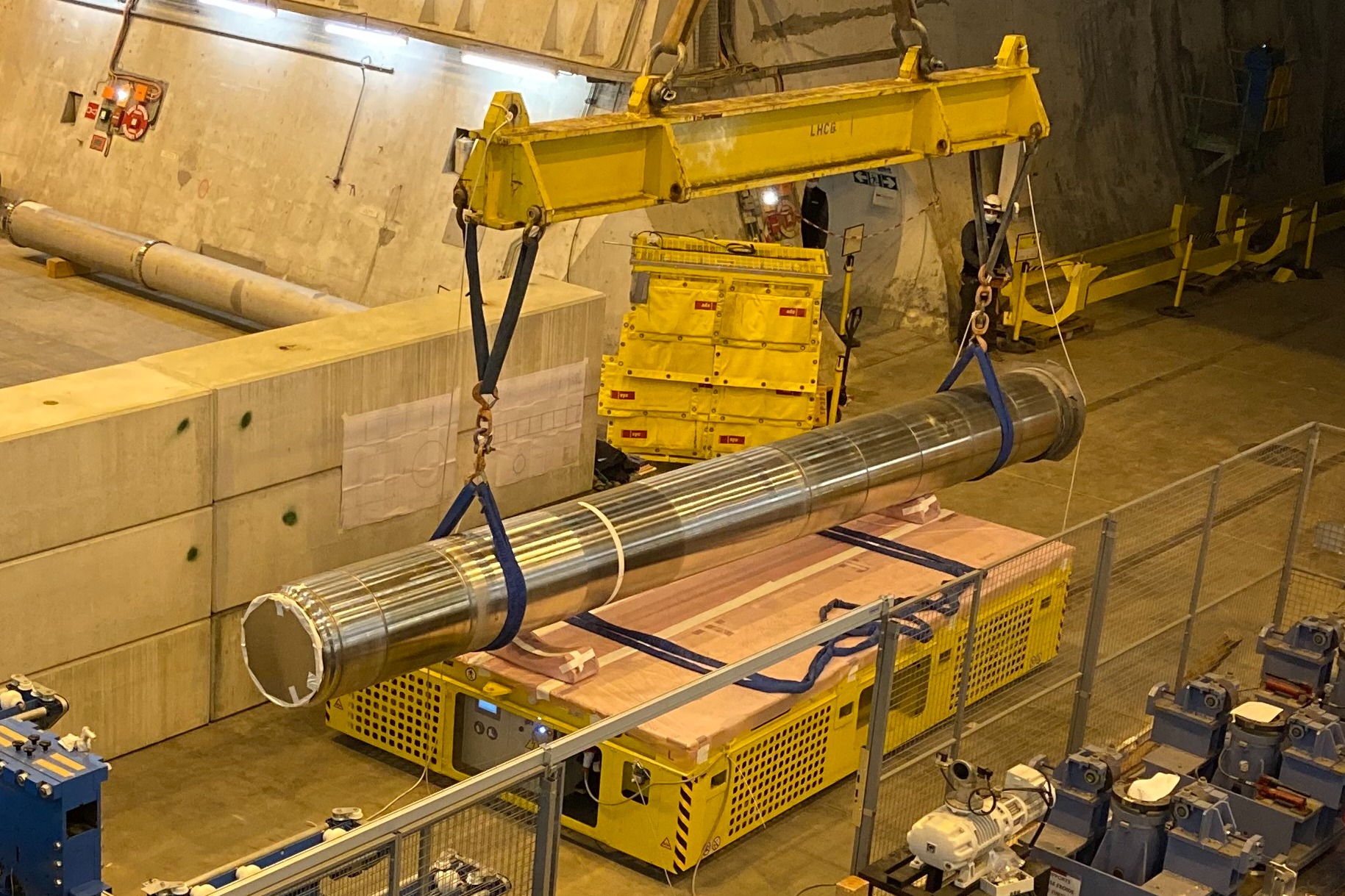}
\caption{Transporting one of the beam dump cores that was irradiated during Run~1 and Run~2 after its removal from the UD68 cavern.}
\label{fig:TDE-removal}
\end{figure}

In 2015, after 6~years of operation, nitrogen leaks were observed concurrently with the intensity ramp-up of the LHC~\cite{Ruiz2021PracticalUpgrades}. Rapid interventions identified the source of the leaks as a damaged Helicoflex gasket located in the connection between the upstream flange of the dump block and the extraction line as well between the downstream flange of the dump block and the downstream window. The cause of the damage was attributed to the dynamic response of the dump blocks upon beam impact. The existing configuration did not allow for any mechanical vibration or displacement of the dump block since it was rigidly supported and connected to the extraction line. The interventions ensured the continued operation of the beam dumps until the beginning of Long Shutdown~2 (LS2) at the end of 2018. During LS2, the dumps were replaced with former spare dumps that had been upgraded to mitigate nitrogen leaks and ensure safe operation with increased beam energy in subsequent years~\cite{Ruiz2021PracticalUpgrades}. This upgrade program is described in detail in Ref.~\cite{Maestre2021DesignMj/Dump}. As a result of these operations, the two irradiated dumps that had been uninstalled required decommissioning. Figure~\ref{fig:TDE-removal} shows the beam dump that had previously been installed in the UD68 cavern after its removal.

\section{Beam-induced energy deposition and response of the dump to beam impact} \label{beam-induced-energy-deposition-on-dump-block-assembly}
Upon beam impact, a secondary particle shower develops along the length of the dump due to nuclear interactions between the particle beam and the carbon-based materials. 

A crucial component of the LBDS is the set of four horizontal and six vertical kicker magnets, known as dilution kickers, located immediately downstream of the LHC extraction kicker magnets. These magnets apply a time-dependent transverse kick to the beam bunches, creating a sweeping pattern across the front face of the beam dump. This pattern evolves laterally along the extraction tunnel, increasing the transverse separation between bunches and thereby reducing the peak energy density in the dump core~\cite{Maestre2021DesignMj/Dump}.

\begin{figure*}[btp]
    \centering
    \begin{subfigure}{1\textwidth}
        \includegraphics[width=0.9\textwidth]{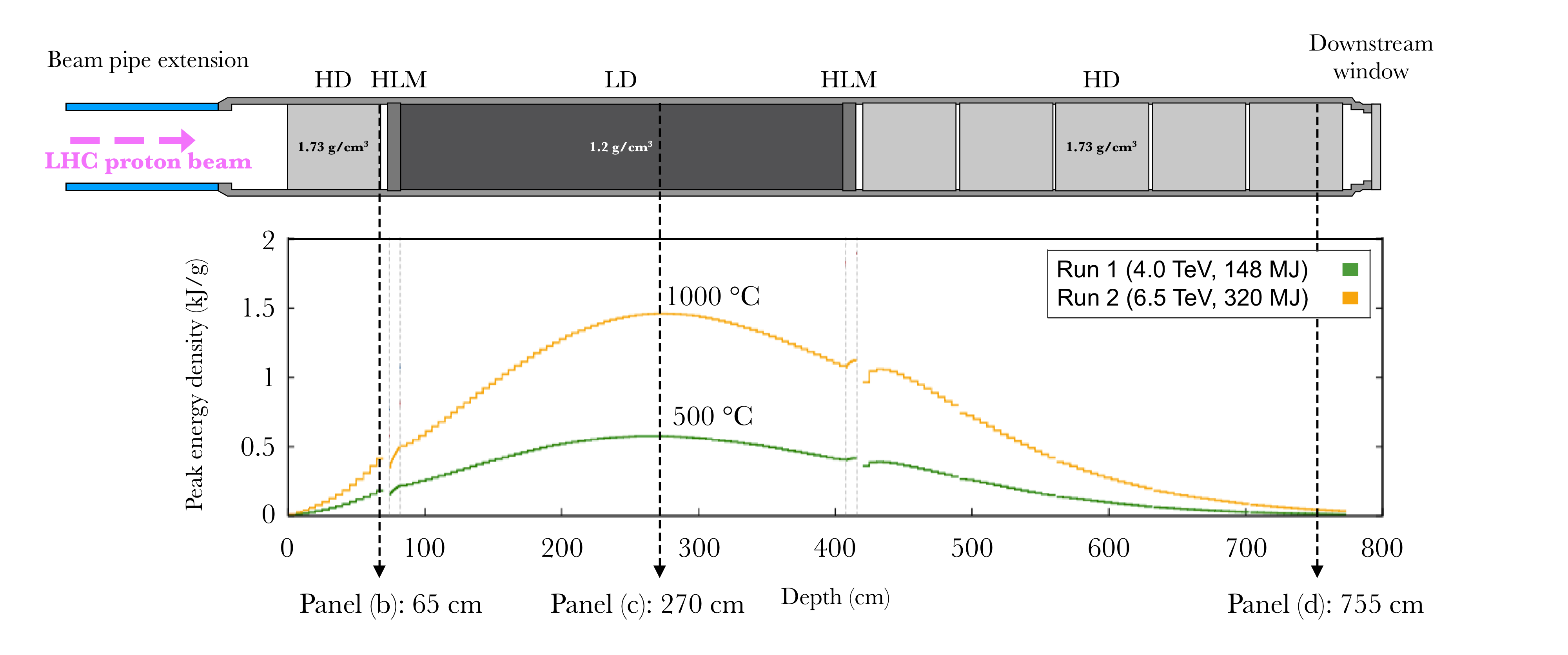}
        \caption{}
        \label{fig:energy-density-along-beamline}
    \end{subfigure}
    \begin{subfigure}{0.32\textwidth}
        \centering
        \includegraphics[width=\textwidth]{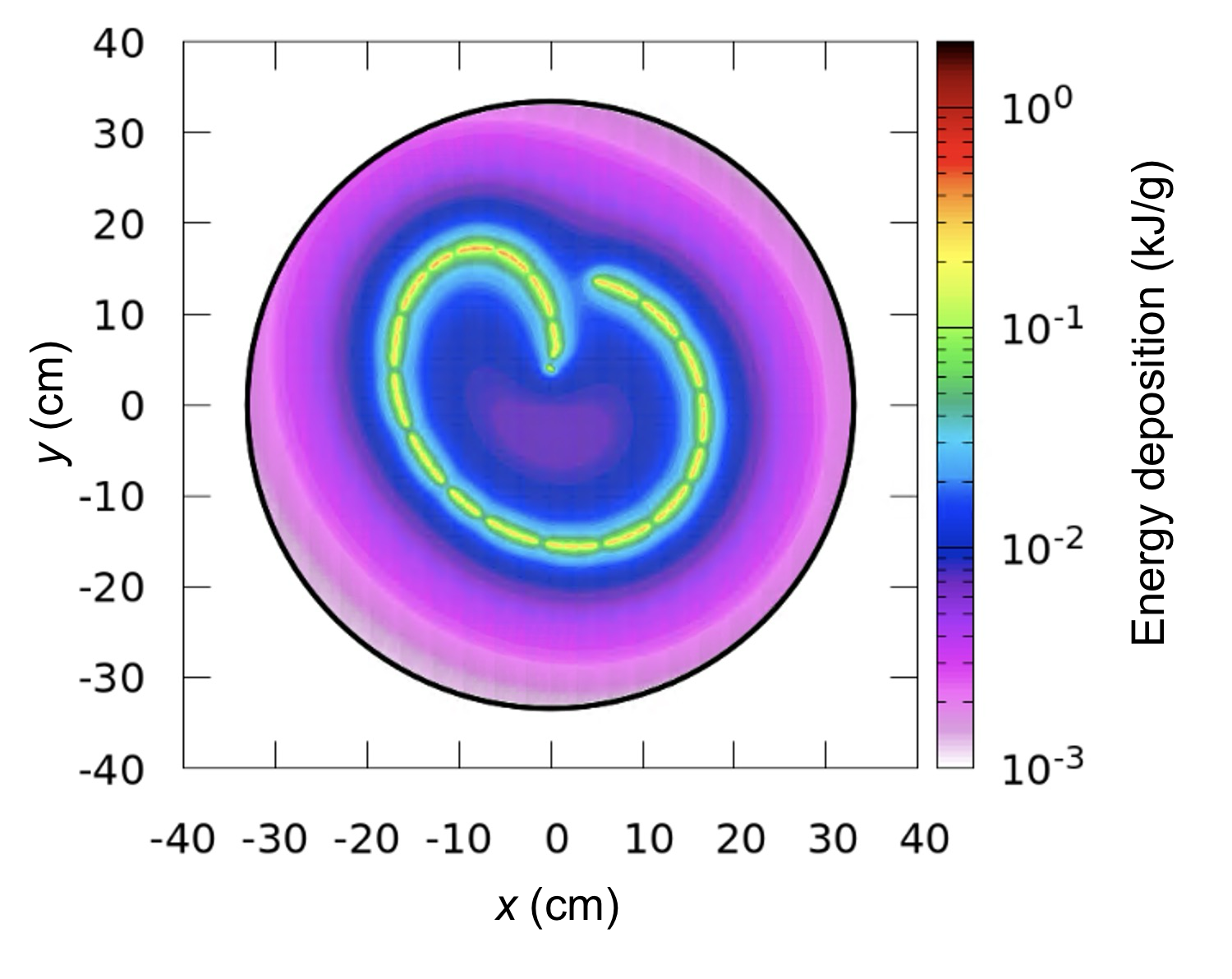}
        \caption{}
        \label{fig:Peak65}
    \end{subfigure}
    \begin{subfigure}{0.32\textwidth}
        \centering
        \includegraphics[width=\textwidth]{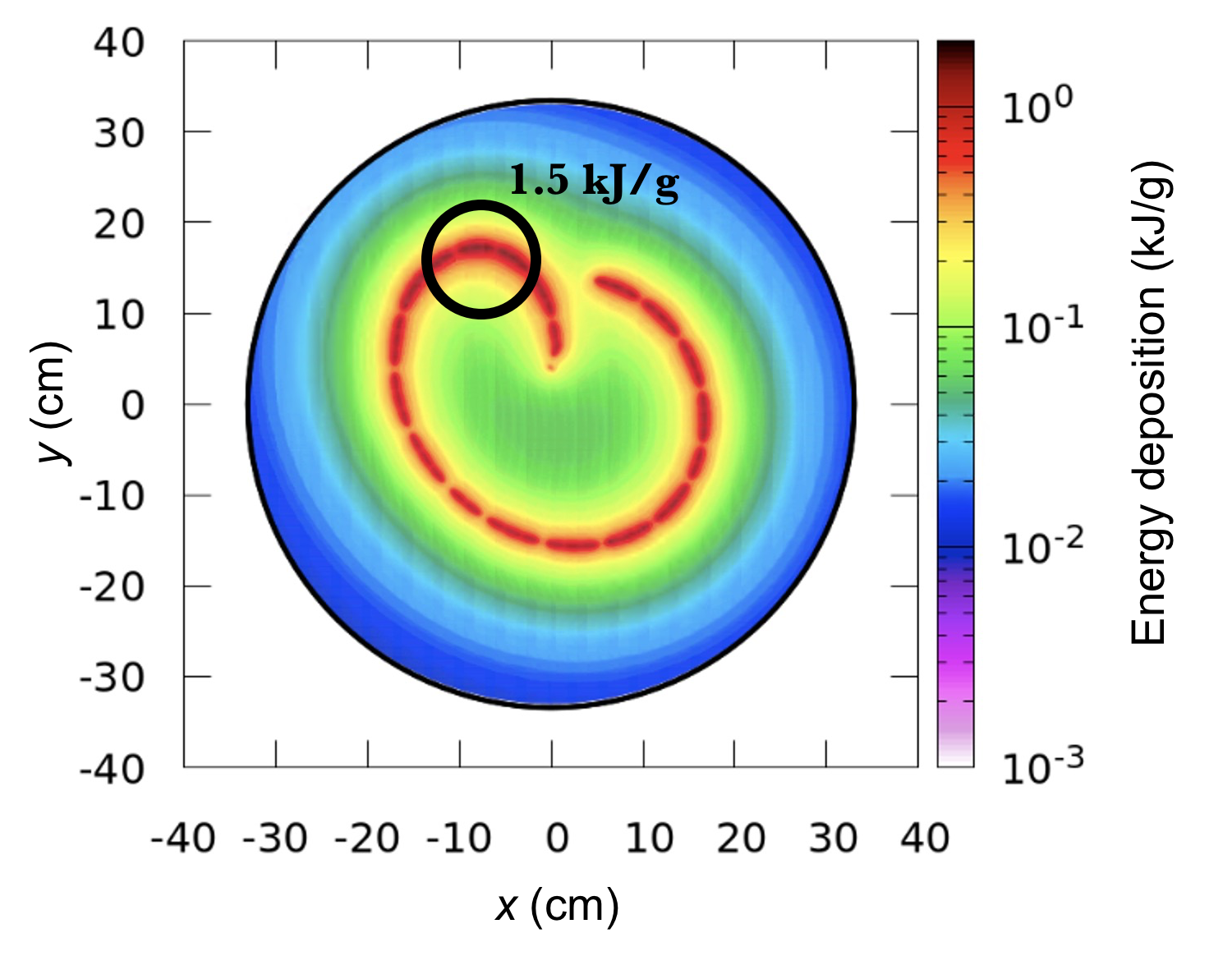}
        \caption{}
        \label{fig:Peak270}
    \end{subfigure}
    \begin{subfigure}{0.32\textwidth}
        \centering
        \includegraphics[width=\textwidth]{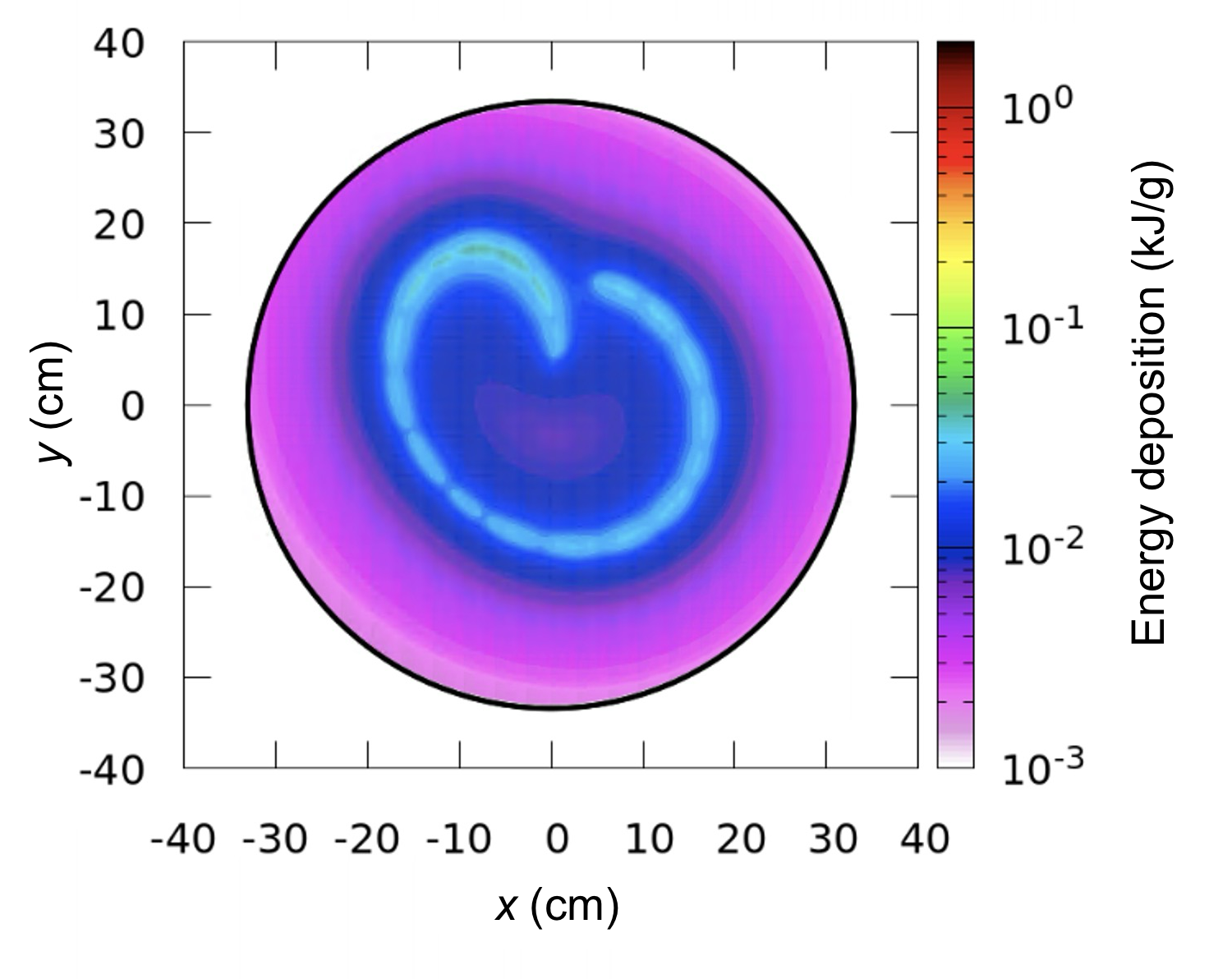}
        \caption{}
        \label{fig:Peak755}
    \end{subfigure}
    \caption{Overview of energy deposition as a function of longitudinal distance and transverse coordinates. (a)~Peak energy density profiles achieved at different positions along the length of the dump core during LHC Run~1 and Run~2. During Run~2, the maximum energy density of \SI{1.5}{\kilo\joule\per\gram} occurred in the LD sector at a distance of approximately \SI{270}{\centi\meter} from the upstream window, resulting in an adiabatic temperature rise of up to \SI[number-unit-separator=\,]{1000}{\celsius}. Run~2 transverse energy density maps at depths of (b)~65~cm, (c)~270~cm (location of peak energy density), and (d)~755~cm. The swept patterns in panels (b)--(d) are a consequence of the horizontal and vertical dilution kickers in the LBDS system, which dilute the beam to reduce the peak energy density in the beam dump~\cite{Bracco2016LBDSSystem}. \label{fig:energy-deposition-overview}}
\end{figure*}

The energy deposited in the dump core materials can be computed using the Monte Carlo-based FLUKA simulation package~\cite{FLUKA.CERNWebsite, Ahdida2022NewCode, Battistoni2015OverviewCode}, as shown in Figs.~\ref{fig:Peak65}--\ref{fig:Peak755}. Figure~\ref{fig:energy-density-along-beamline} shows the maximum transverse energy density as a function of the longitudinal position within the dump across Run~1 and Run~2. The sweeping pattern, visible in these figures, originates at the center of the transverse plane and progresses in a counter-clockwise direction.

Although the LBDS was designed and verified to meet a high level of dependability (rated SIL4 according to IEC 61508~\cite{Filippini2006DependabilityCERN}), partial dilution failures, where some of the dilution kicker magnets fail to operate, cannot be entirely excluded~\cite{Carlier2003DesignSystems}. Such failures can lead to reduced lateral bunch separation, resulting in higher peak energy densities in the beam dump.

Three dilution failures with dumped energy exceeding \SI{150}{\mega\joule} occurred during the 9-year operation of the dumps~\cite{Lechner2021Beam-InducedAssembly}. The most critical of these, with a dumped energy of \SI{196}{\mega\joule}, resulted in energy densities in the core comparable to those produced during a full-intensity dump (\SI{320}{\mega\joule}) in nominal operation during Run~2. A full summary of the beam parameters for different LHC runs is presented in Table~\ref{table:BeamParameters}.

\begin{table*}[btp]

\caption{\centering{Overview of the beam parameters during different LHC runs~\cite{Bailey2003StandardModes}.}\label{table:BeamParameters}}
\begin{tabular}{lcccc}
\hline \hline
\multicolumn{1}{c}{\textbf{}} & \begin{tabular}[c]{@{}c@{}}Run~1\\ (2009--2013)\end{tabular} & \begin{tabular}[c]{@{}c@{}}Run~2\\ (2015--2018)\end{tabular} & \begin{tabular}[c]{@{}c@{}}Run~3\\ (2022--2026)\end{tabular} & \begin{tabular}[c]{@{}c@{}}Run~4/HL-LHC\\ (2030--)\end{tabular} \\ \hline
Beam dumps/year & \multicolumn{4}{c}{$220~\text{days} \times 2 = 440$} \\ \hline
Energy/proton (TeV) & 4.0 & 6.5 & 6.8 & 7.0 \\
Nominal bunch intensity (protons/bunch) & $1.7\times 10^{11}$ & $1.2\times 10^{11}$ & $1.8\times 10^{11}$ & $2.2\times 10^{11}$ \\
Maximum number of bunches/beam & 1380 & 2556 & 2748 & 2760 \\
Maximum stored energy (MJ) & 150 & 320 & 539 & 680 \\ \hline \hline
\end{tabular}
\end{table*}

The response of the beam dumps to beam impact occurs in two phases: a highly dynamic phase during the first milliseconds after impact, followed by a quasistatic response lasting several hours. The sudden deposition of energy in the vessel excites several vibrational modes in the beam dumps~\cite{Maestre2021DesignMj/Dump}. These vibrations, which cause accelerations of up to 2000~g at the longitudinal extremities, dissipate over the course of a few milliseconds. After this, the dump cores expand quasistatically due to diffusion of heat from the core to the vessel~\cite{Maestre2021DesignMj/Dump}. An initial longitudinal expansion of up to \SI{5}{\milli\meter} occurs during the first \SI{45}{\minute} after beam impact, followed by contraction back to the undeformed state as heat is extracted by the cooling system. After \SI{12}{\hour}, about 10\% of the deposited energy remains within the core.

A more detailed description of the response of the beam dumps to beam impact can be found in Ref.~\cite{Maestre2021DesignMj/Dump}.

\section{Motivations for dump autopsy} \label{motivation-behind-dump-autopsy}
In 2018, after 9~years of operation and a substantial increase in the intensity and energy of the particle beams, no information was available that would allow the assessment of the condition of the core materials and guide the HL design. This was particularly true for the expanded graphite sheets in the central section of the beam dumps, as these are the only known beam-intercepting devices to employ this material in their core. To address this uncertainty, the HiRadMat-43~\cite{Maestre2021SigraflexPerspective} experiment in 2018 included four \SI[number-unit-separator=\text{-}]{0.5}{\milli\meter}-thick expanded graphite sheets with a density of \SI{1.0}{\gram\per\cubic\centi\meter} (SIGRAFLEX grade F05010TH~\cite{SGL_SIGRAFLEX_Foil}).

The four samples were exposed in sets of two to high-intensity proton beams depositing energy densities of 1.85 and \SI{2.50}{\kilo\joule\per\gram} (respectively 20\% and 70\% higher than the peak values achieved during Run~2). Post-irradiation examination revealed damage in all samples, with greater severity observed in those subjected to the higher energy density. Contactless profilometry analysis revealed out-of-plane deformations of up to 200--\SI{250}{\micro\meter}, and micro-computed tomography scans revealed the development of large interlaminar voids in the material~\cite{Maestre2021SigraflexPerspective}.

If the same response were exhibited by the SIGRAFLEX sheets in the beam dump core, this would lead to a reduction in the density of the material in the beam-impacted area, and this would have critical consequences for the absorption efficacy of the devices and their overall reliability.

\begin{figure*}[tbp]
  \centering
  \parbox{\figrasterwd}{
    \parbox{.65\figrasterwd}{%
      \subcaptionbox{\label{fig:endoscopy-overview}}{\includegraphics[width=\hsize]{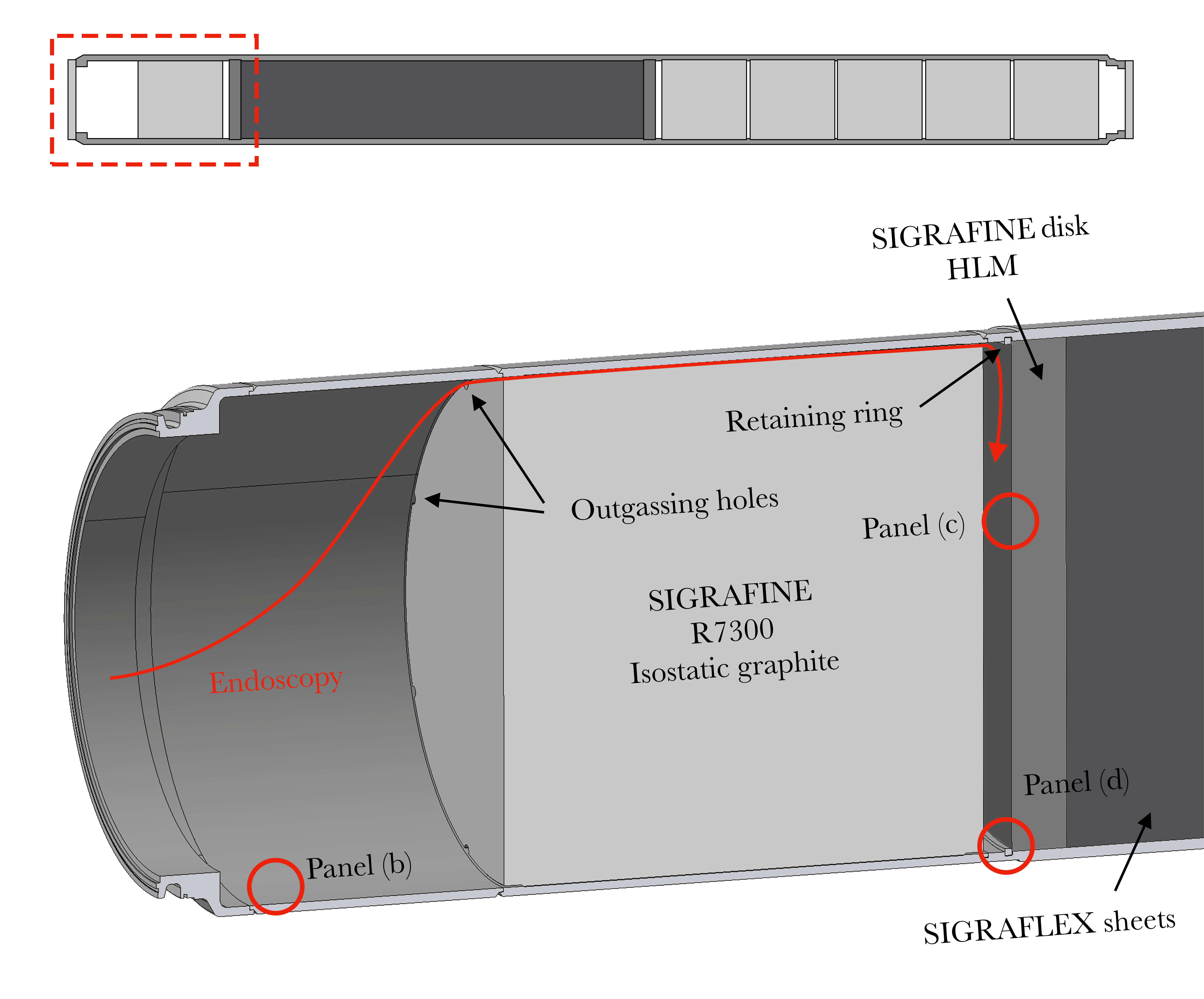}}

    }
    \hskip1em
    \parbox{.30\figrasterwd}{%
      \subcaptionbox{\label{fig:GraphitePowder}}{\includegraphics[width=\hsize]{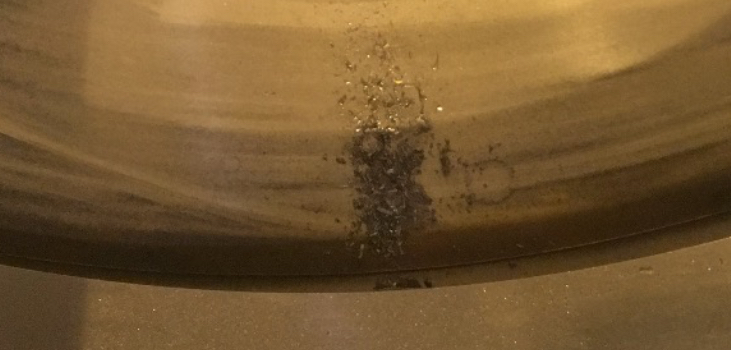}}
      \subcaptionbox{\label{fig:HLMCrack}}{\includegraphics[width=\hsize]{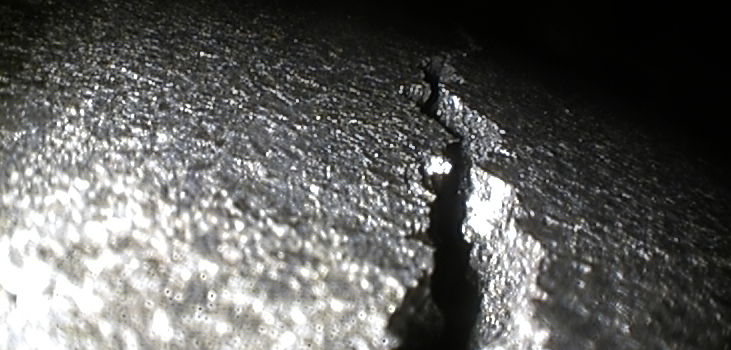}}
      \subcaptionbox{\label{fig:RetaingingRing}}{\includegraphics[width=\hsize]{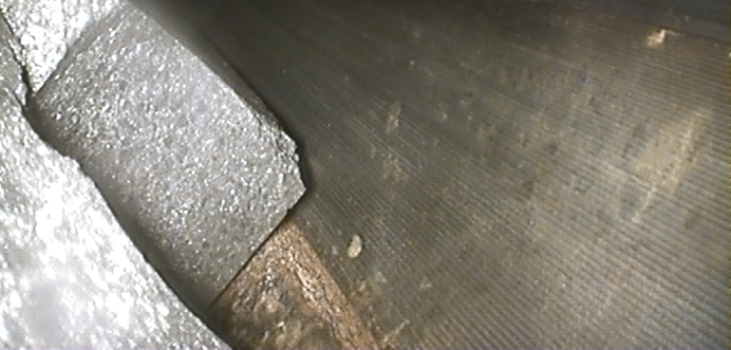}}
    }
  }
\caption{Overview of the endoscopic assessment performed from the upstream end of the beam dump through outgassing holes to reveal the HLM disc. (a)~Schematic views. (b)~Graphite powder found inside of vessel after removal of window. (c)~Crack in HLM disc of approximately 40--50~cm. (d)~Dislocation of the HLM disc with respect to the retaining ring.}
\end{figure*}

The concerning results from the HiRadMat-43 experiment prompted an endoscopic examination of the Run~2 operational beam dumps once they had been removed from the caverns in the spring of 2020. Figure~\ref{fig:GraphitePowder} illustrates the presence of graphite powder, which was found inside the vessel after accessing its upstream end, suggesting potential damage to the core materials. The endoscopic procedure offered limited access for assessment, providing only a partial view of the core's exterior through an outgassing hole, as shown in Fig.~\ref{fig:endoscopy-overview}. Figure~\ref{fig:HLMCrack} shows a crack of approximately 40--\SI{50}{\centi\meter} that was found on the upstream SIGRAFINE HLM extruded graphite disc~\cite{Maestre2021SigraflexPerspective}. Moreover, the retaining ring was dislodged from its groove, preventing adequate compression and containment of the LD sector, as shown in Fig.~\ref{fig:RetaingingRing}. Due to the inaccessibility of the LD sector, the condition of the expanded graphite sheets could not be assessed.

The potential impact of these results on operations during Run~3 highlighted the need for an extensive assessment of the condition of the core. This ultimately led to the decision, made in the summer of 2020, to dismantle the devices to allow full access to all parts of the core.

In late 2020, an initial attempt to cut the duplex stainless-steel vessel using a handheld angle grinder quickly revealed that the challenges of cutting the material had been underestimated. The chosen cutting method proved inadequate and slow, possibly leading to an unacceptably prolonged exposure of the operator to the high dose rates of the beam dump.

%An evaluation of the industrial landscape was conducted with the objective of selecting an external contractor to manage the dismantling and post-irradiation activities; however, the responses received exceeded the established budget and timeframe, leading to a decision to internalize the project.

In light of the significant interest in evaluating the state of the cores within the dumps, along with the inherent difficulties associated with decommissioning such devices, a comprehensive protocol was devised for the internal execution of the autopsy and decommissioning processes. This also served as a case study to expand the capabilities for the dismantling and waste packaging of large radioactive devices at CERN. The evaluation of the dump core, combined with the HiRadMat-56 experiment~\cite{AndreuMunoz2022IrradiationExperiment}, sought to establish potential operational limits for Run~3, inform the decision on the use of LD graphite for the Run~3 operational beam dump spares, and provide input for the development of the next generation of beam dumps for Run~4.

\section{Radiation protection} \label{radioprotection challenges}
\subsection{Legal framework and challenges}
The legal framework within which CERN operates is defined by a tripartite agreement between the laboratory and the authorities of its host states: the Autorité de Sûreté Nucléaire (ASN) for France and the Office Fédéral de la Santé Publique (OFSP) for Switzerland. The set of radiation-protection rules is outlined in ``Safety Code~F''~\cite{SafetyCodeF}, which defines the rules for the protection of personnel, the population, and the environment from ionizing radiation produced at the CERN accelerator complex~\cite{BOZZATO2024111573}.

CERN introduced a formalized approach to the ``As Low As Reasonably Achievable'' (ALARA) principle at the end of 2006, and this is applied to all CERN facilities. The ALARA approach aims to optimize work coordination, work procedures, handling tools, and even the design of entire facilities. Consequently, all work in Radiation Areas must be optimized. In particular, work in Controlled Radiation Areas~\cite{BOZZATO2024111573} must be carefully planned and optimized, including making estimates of the collective and individual effective doses to workers participating in the completion of a task. Five criteria have been established to determine the ``ALARA Level''~\cite{BOZZATO2024111573}, which provides a radiological risk assessment for a given intervention.

During the operation of high-energy accelerators, the interaction of radiation with matter can lead to the activation of machine components and of the surrounding infrastructure. Radioactive waste (RW) that is produced as a result of maintenance operations and during the decommissioning of an installation must be managed and disposed according to the regulatory framework defined in the tripartite agreement. RW produced at CERN is then disposed to either in France or Switzerland, depending on its classification (see Ref.~\cite{BOZZATO2024111573}).

In this framework, the decommissioning of the beam dumps and their preconditioning as RW posed significant radiation protection (RP) challenges due to the residual activation of the equipment and the need to operate in a highly radioactive environment. To ensure high safety standards and to respect the ALARA principle, these challenges were addressed through the use of state-of-the-art simulation techniques to predict the residual ambient dose equivalent rate and the radionuclide inventory at various stages of the interventions. The CERN Occupational Health \& Safety and Environmental Protection Unit, Radiation Protection Group (HSE-RP) makes extensive use of FLUKA~\cite{FLUKA.CERNWebsite, Ahdida2022NewCode}, along with other tools that complement or extend FLUKA's internal capabilities, such as ActiWiz~\cite{Vincke:2018} and SESAME~\cite{BOZZATO2024111573}. These simulations have been covered in depth in previous reports~\cite{rp_LHC_TDE_autopsy, rp_LHC_TDE_characterization}. For completeness, the main relevant points will be reported in the following, together with additional data not previously reported in other publications.

\subsection{FLUKA simulations for residual studies}
\begin{figure}
\centering
\includegraphics[width=\columnwidth]{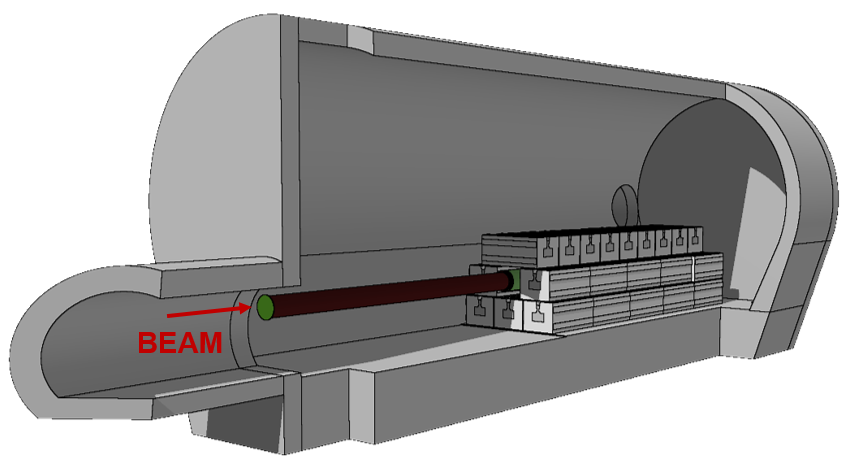}
\caption{FLUKA model of the UD cavern hosting the beam dump and its shielding. This shows the pre-LS2 configuration, including the connecting line.} \label{fig:FLUKA_model_UD}
\end{figure}

\begin{figure*}[tbp]
\centering
\includegraphics[width=0.8\textwidth]{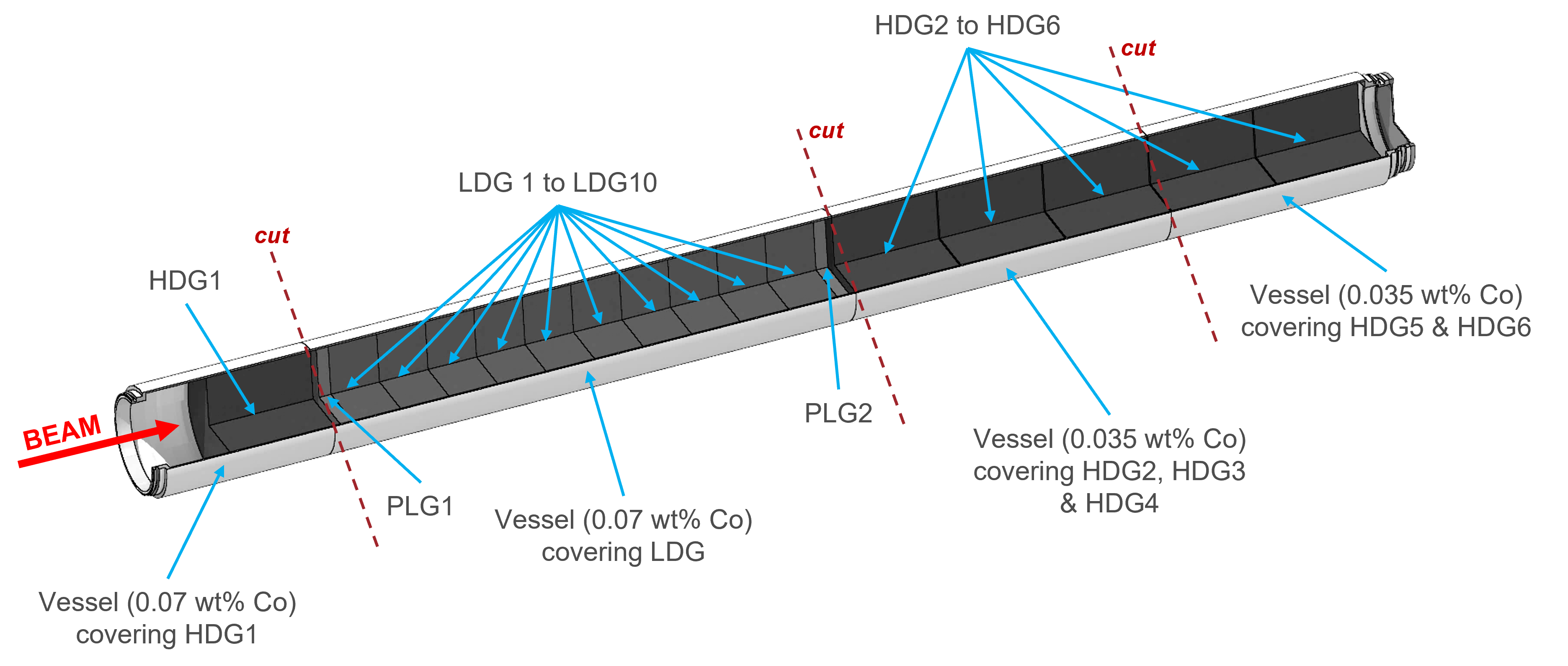}
\caption{FLUKA model of the LHC beam dump. The different regions represent the high- and low-density graphite (HDG and LDG, respectively) sectors and the stainless-steel vessel. The three dashed lines represent the main transverse cuts performed during the decommissioning.}
\label{fig:TDE_FLUKA}
\end{figure*}

A detailed FLUKA model of the beam dump cavern and the dump core, including the stainless-steel vessel, the carbon-based materials of the core, and the surrounding shielding, was used throughout the project. A detailed description of this FLUKA model is provided in Ref.~\cite{rp_LHC_TDE_autopsy}, and it is illustrated in Figs.~\ref{fig:FLUKA_model_UD} and~\ref{fig:TDE_FLUKA}. The model of the beam dump core (Fig.~\ref{fig:TDE_FLUKA}) is divided into several subregions, according to the autopsy cutting sequence and the RW characterization. The chemical compositions of the graphite and the steel vessel were modeled based on the best available information, including material certificates provided by the manufacturers. Traces of impurities were included in both the graphite and the vessel. Concerning the latter, a variable cobalt (Co) content between 0.035 and 0.070~wt\,\% was used based on \textit{in situ} measurements, discussion with material experts at CERN, and assumptions drawn from comparisons between simulations and experimental RP surveys.

A key aspect of the operation of the LHC is the repeated cycle of run and shutdown periods over the years. While the operational phases define the prompt source term for any LHC-related FLUKA RP assessment, the shutdown phases present a major challenge for predicting residual dose rates. During scheduled stops, both the accelerator and detectors may change configurations to allow for maintenance, replacement, and subsystem upgrades. As a result, the positions in which structures and devices might become activated during a run can be different from those in which they are stored, or simply moved to, during a shutdown. This is also the case for the de-installation of the beam dumps.

\begin{figure}[tbh]
\includegraphics[width=\columnwidth]{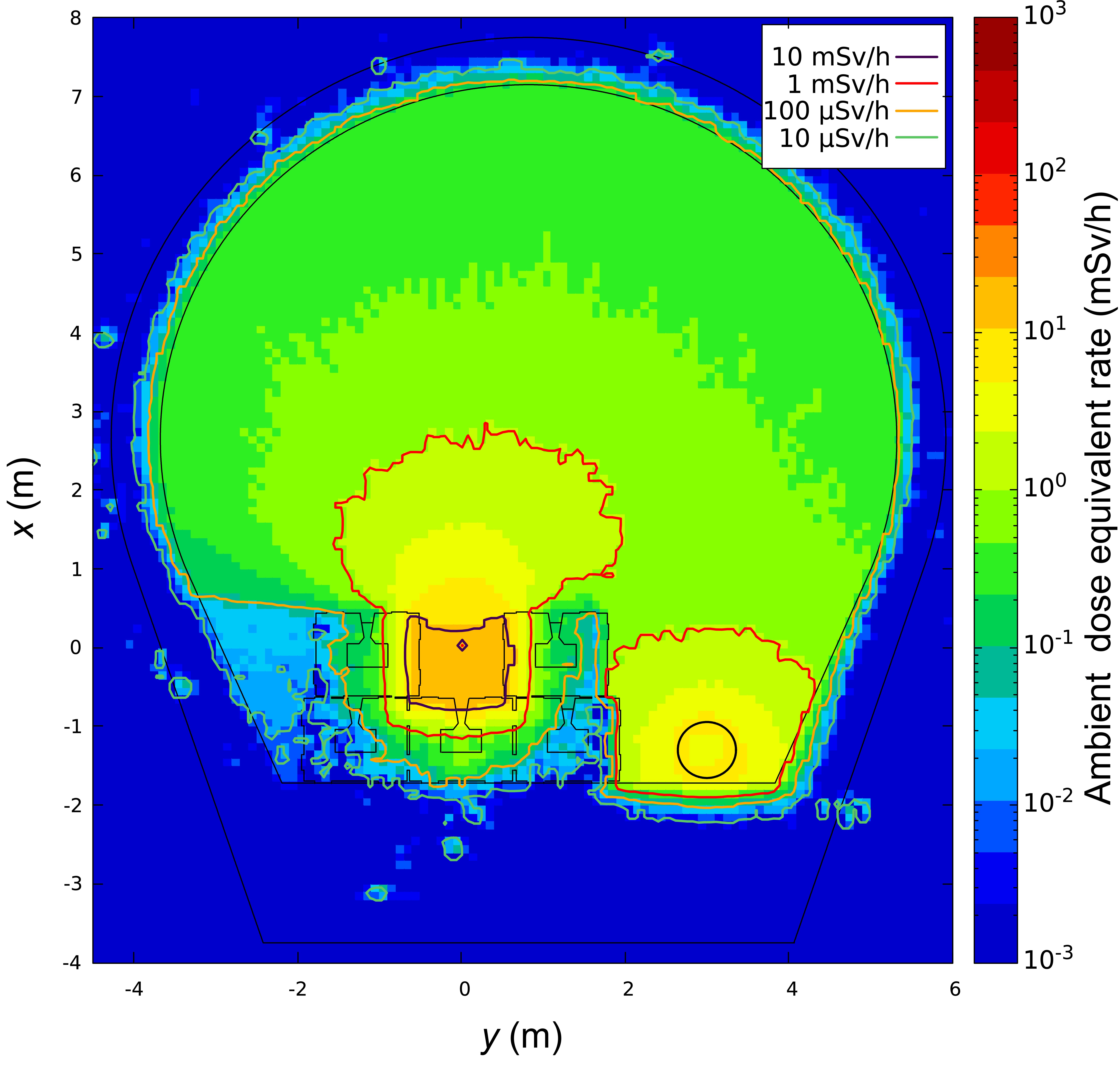}
\caption{FLUKA-simulated residual radiation field (after a 1-year cool down from the end of Run~2) in the UD cavern when the top shielding is removed and the beam dump is extracted from its operational position~\cite{rp_LHC_TDE_autopsy}.}\label{fig:FLUKA_TDE_removal}
\end{figure}

\begin{figure}[tbh]
\includegraphics[width=\columnwidth]{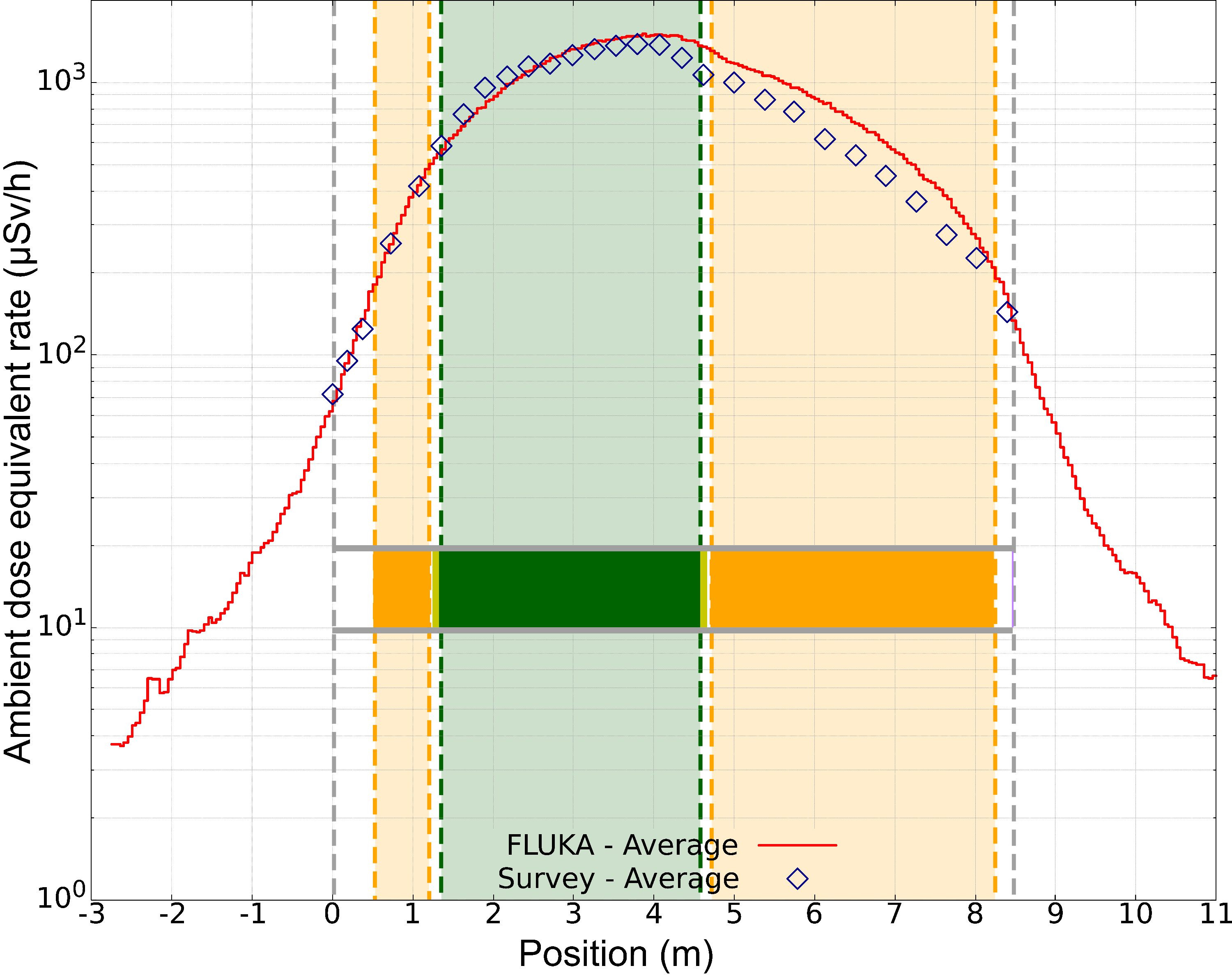}
\caption{Comparison between the results of FLUKA-SESAME simulations and an RP survey conducted in June~2021~\cite{rp_LHC_TDE_autopsy}. Both the measurements and the simulations consider positions $\pm$30~cm from the surface of the steel vessel. The results from the FLUKA simulations have a statistical uncertainty of $<$1\% while a conservative 10\%--20\% uncertainty can be attributed to the measurements. It is important to note that the measurements were performed for operational RP purposes and were not originally intended for benchmark validation.}\label{fig:FLUKA-CERNBot_comparison}
\end{figure}

For the prompt phase, FLUKA was used to ``activate'' the dump according to LHC operations over several years; for the residual phase, FLUKA-SESAME was used to account for geometry changes during decay and to model the opening of the beam dump shielding. An example of this process is shown in Fig.~\ref{fig:FLUKA_TDE_removal}. The plot illustrates the configuration after the top part of the beam dump shielding has been removed, with the beam dump core placed on the floor of the UD cavern. This represents an intermediate step in the beam dump removal process, during which the CERN transport team must manipulate the shielding and temporarily store the dump in the cavern before transporting it using appropriate vehicles. The results of these FLUKA simulations were compared with RP surveys~\cite{rp_LHC_TDE_autopsy} conducted during the removal phase, and they showed an agreement better than 20\% (see Fig.~\ref{fig:FLUKA-CERNBot_comparison}).

The results of the FLUKA simulations were compared with surveys conducted at various cool-down times following the end of Run~2 operation. This comparison was crucial to validate the irradiation profile used to compute both the residual dose field and the radionuclide inventory. Moreover, it provided highly accurate estimates, which formed the foundation for planning the dismantling operations.

Remote RP surveys were performed using a dedicated robot (CERNBot~\cite{DiCastro2017}), a rover equipped with a movable arm and a Geiger Müller probe, to minimize the exposure of personnel to radiation. The resulting measurements were also used to retrospectively fine-tune the chemical composition of the steel vessel, particularly regarding the Co content in the 318LN steel jacket. Despite its low mass content, traces of cobalt may cause radiological risks through the production of the radionuclide cobalt-60 ($^{60}$Co), which has a rather long half-life of 5.72~years. Naturally occurring cobalt consists of the stable isotope $^{59}$Co, which transforms into $^{60}$Co via low-energy neutron capture [$^{59}$Co(n,$\gamma$)$^{60}$Co]. Over long decay periods ($>$1~year), $^{60}$Co can dominate the radiological environment, becoming the primary contributor to the residual dose field.

FLUKA-SESAME simulations of the different steps of the autopsy of the beam dumps were performed to plan the activity and to produce an initial radiological risk assessment of the intervention (i.e., the ALARA classification). The detailed results of these simulations are reported in Ref.~\cite{rp_LHC_TDE_autopsy}, and the ALARA classification of the intervention will be discussed in the next section.

\subsection{Operative radiation protection}
\begin{figure*}[b!t]
    \centering
    \includegraphics[width=0.8\linewidth]{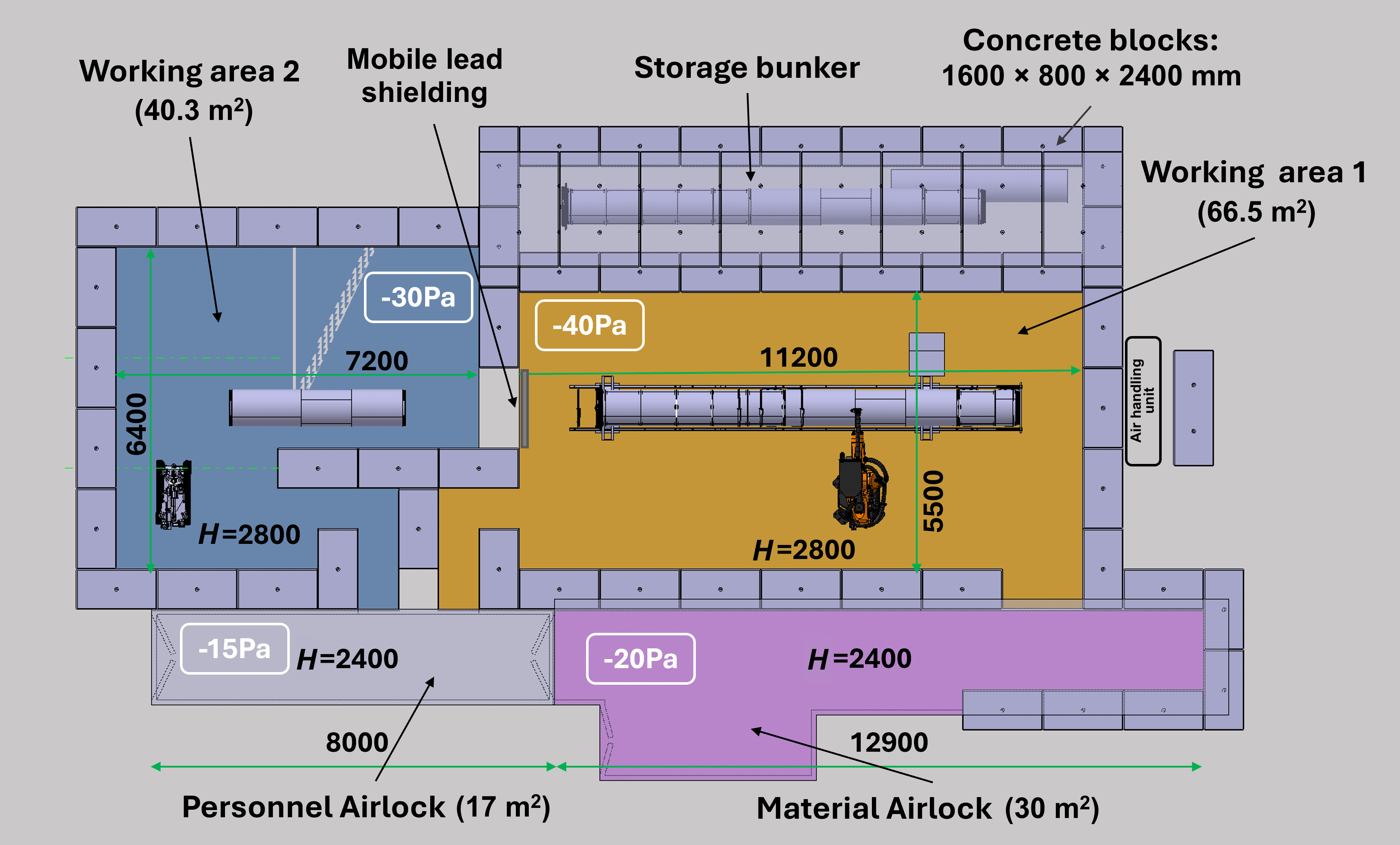}
    \caption{Schematic of the beam dump autopsy worksite showing the different working areas established along with their sizes and differential pressure.}
    \label{fig:worksite_aut}
\end{figure*}

Figure~\ref{fig:worksite_aut} shows a schematic of the beam dump dismantling worksite. This was enclosed by concrete blocks measuring \SI{1600}{\milli\meter} in length, \SI{800}{\milli\meter} in thickness, and \SI{2400}{\milli\meter} in height. These blocks served to reduce the radiation background in the surrounding area and to comply with the building's area classification requirements, i.e., Supervised Radiation Area [$\dot{H}^{*}(10)<15$~\SI{}{\micro\sievert\per\hour}]~\cite{BOZZATO2024111573}. Here, $\dot{H}^{*}(10)$ denotes the dose at a depth of \SI{10}{\milli\meter} in tissue, and is used to assess radiation exposure in controlled areas.

The worksite was divided into different subareas: the main working area (working area~1, covering \SI{66.5}{\meter\squared}), where the main cutting activities were performed; an area dedicated to the RW preconditioning (working area~2, covering \SI{40.3}{\meter\squared}); a storage bunker for RW, completely shielded to reduce the ambient dose equivalent rate in the two main working areas; and two airlocks, one dedicated to personnel and one for materials. The worksite area delimited by the shielding blocks was classified as a Limited Stay Radiation Area, with a maximum allowed dose rate of $\dot{H}^{*}(10)<2000$~\SI{}{\micro\sievert\per\hour}~\cite{BOZZATO2024111573}.

An-air handling unit equipped with HEPA filters was used to establish a cascade of pressure differentials among the different compartments, maintaining differences in the range \SI{-15}{} to \SI{-40}{\pascal}. Air contamination was continuously monitored using an $\alpha$/$\beta$ counting system (ABPM 203M, Mirion Technologies) with a control unit and visual alarm. In addition, the air-monitoring unit included an Automess AD6 (6150AD-k probe) and was remotely connected to CERN's Radiation and Environment Monitoring Unified Supervision (REMUS)~\cite{REMUS_web} system. Real-time tritium ($^{3}$H) activity monitoring in the ambient air was carried out using a $\beta$ ionix\ (Mirion Technologies) portable $^{3}$H monitor. 

\begin{figure}[tbp]
    \centering
    \includegraphics[width=1.0\linewidth]{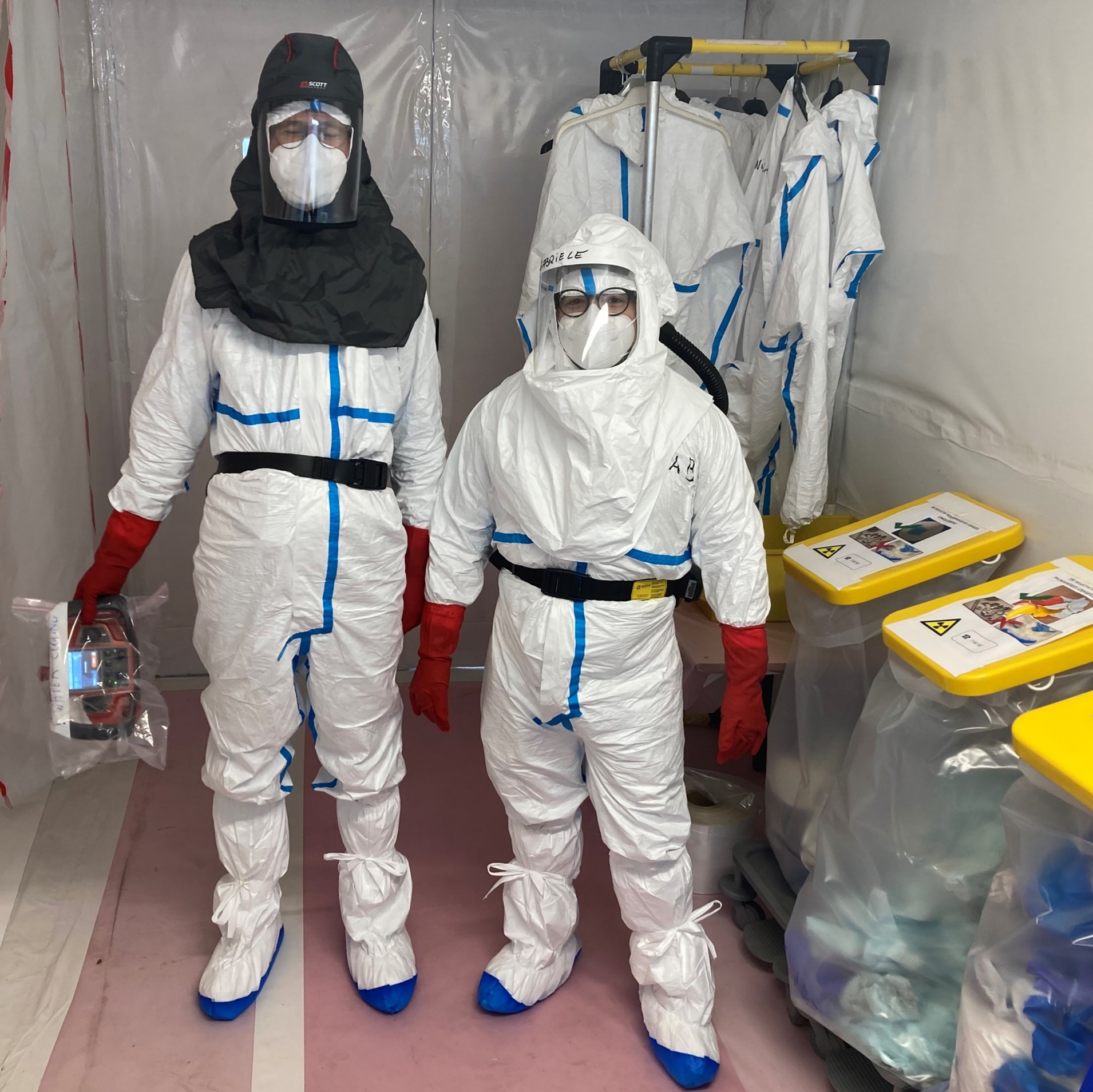}
    \caption{Personal protective equipment (PPE) used during the LHC beam dump autopsy.}
    \label{fig:PPE}
\end{figure}

Different personal protective equipment (PPE; Fig.~\ref{fig:PPE}) was worn by personnel working in the main areas, including: a disposable hooded coverall and boot covers, both made of Tyvek; gloves; overshoes; FFP2 masks; 3M\ respiratory protection masks with filtering cartridges; and personal passive dosimeters (DIS-1, Mirion Technologies) and active dosimeters (DMC~3000k, Mirion Technologies) for $\gamma/\beta$ monitoring. The personnel airlock, used as changing room, was equipped with a hand-foot monitor (LB~147, Berthold Technologies), to check for potential contamination when personnel exited the worksite.

Radiological control, including radiation surveys and contamination measurements using smear tests, was performed on equipment leaving the material airlock, as well as on tools and PPE exiting the working area. After each intervention involving a risk of contamination, the area was cleaned using dedicated RP vacuum cleaners for classified areas. Smear tests and surface contamination measurements were then conducted. Prior to any roof opening, a waiting period was observed to allow for a complete air exchange in the working area.

\begin{figure*}[thbp]
\centering
\includegraphics[width=0.9\linewidth]{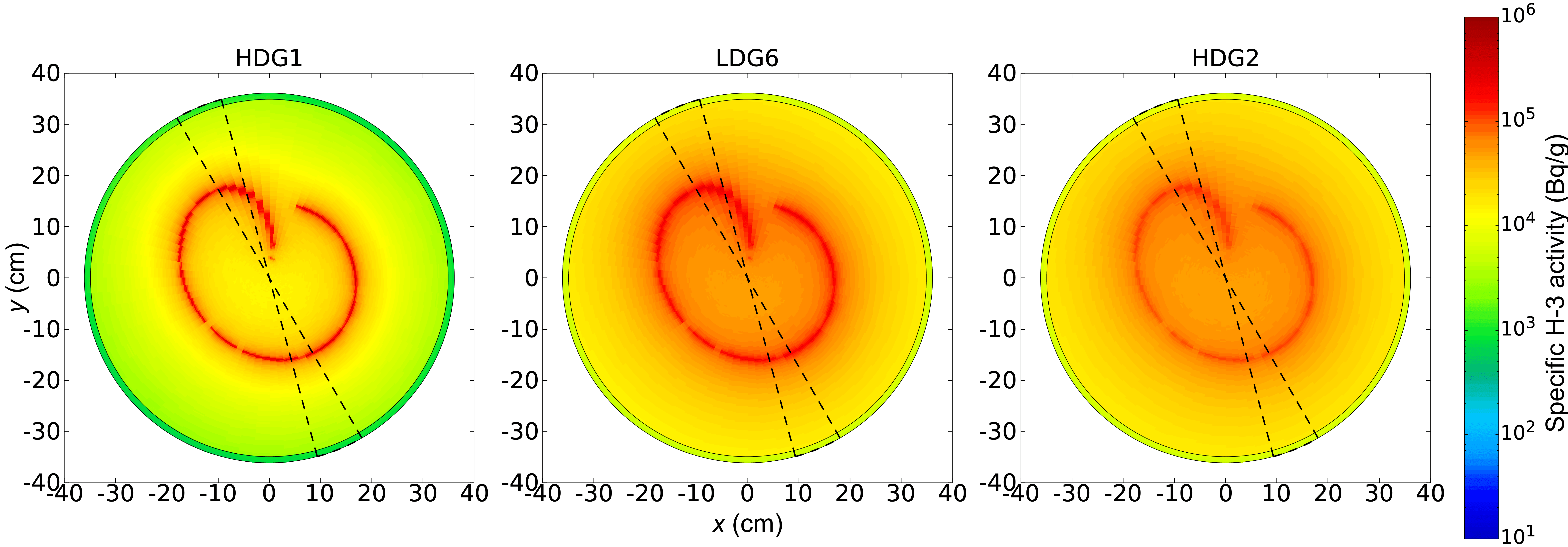}
\caption{Specific $^{3}$H activity over different sections of high- and low-density graphite (HDG and LDG), after 1724~days of irradiation, {$\sim$}$3\times10^{17}$ cumulative protons, and 1245~days of cool down. The dashed lines show the sectors that were used to produce the longitudinal plot in Fig.~\ref{fig:SA_H-3_TDE}.}
\label{fig:SA_H-3_HDG_LDG}
\end{figure*}

\begin{figure*}[thbp]
\centering
\includegraphics[width=0.9\linewidth]{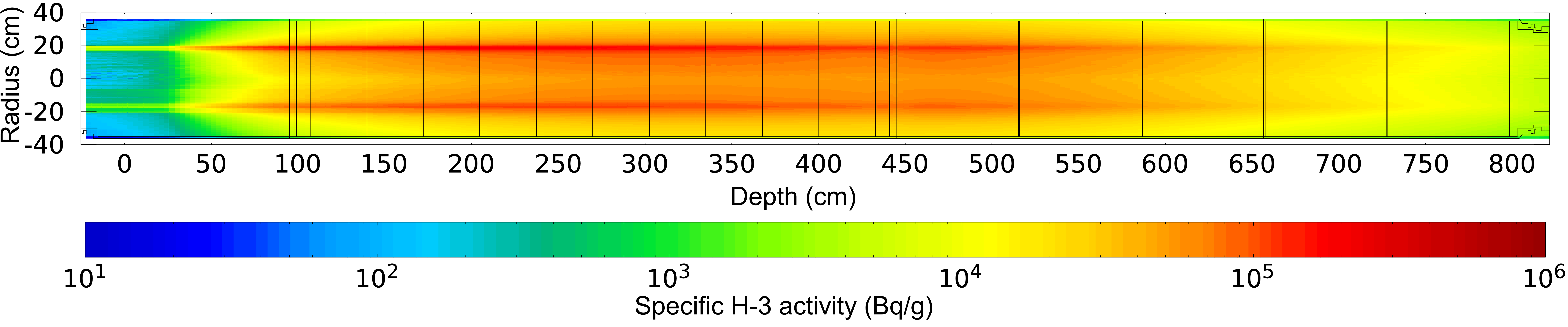}
\caption{Specific $^{3}$H activity over the longitudinal axis of the beam dump after 1724~days of irradiation, {$\sim$}$3\times10^{17}$ cumulative protons, and 1245~days of cool down, showing the average over the sectors $105^{\circ} \le \phi \le 120^{\circ}$ and $285^{\circ} \le \phi \le 300^{\circ}$ from the positive $x$ axis.}
\label{fig:SA_H-3_TDE}
\end{figure*}

Regarding the ALARA classification of the intervention, the estimated integrated doses were \SI{4800}{person\text{-}\micro\sievert} for the collective dose and \SI{558}{\micro\sievert} for the maximum individual dose, resulting in an ALARA Level~II intervention. At the end of the intervention, the integrated doses were \SI{2900}{person\text{-}\micro\sievert} or the collective dose and \SI{316}{\micro\sievert} for the maximum individual dose. This final optimization was achieved through careful preparation of each step, including the analysis of associated radiological risks and possible optimization strategies (e.g., cutting methods, remote handling and manipulation, time per step, number of workers, PPE, etc.). Dry runs on nonactivated spares facilitated strategic decisions, such as selecting the most suitable tools for cutting the beam dump core and training the personnel involved on site.

The second beam dump (UD62) was cut and preconditioned about 1~year after the first one. The estimated integrated doses for this intervention were \SI{3536}{person\text{-}\micro\sievert} for the collective dose and \SI{346}{\micro\sievert} for the individual dose, resulting once more in an ALARA Level~II intervention. At the end of the second intervention, the integrated doses were \SI{1243}{person\text{-}\micro\sievert} for the collective dose and \SI{201}{\micro\sievert} for the individual dose. The reduction in dose during the preconditioning of the second dump was attributed to the additional cool-down time and the experience gained from the first dismantling operations.

\subsection{Pre-characterization of radioactive waste}

The need for a post-mortem analysis of the beam dump core provided an opportunity to combine the dismantling with the procedure for RW packaging. This had a direct impact in terms of optimization of the interventions to be performed on the beam dump, namely the dismantling, post-irradiation examination, characterization, and disposal. This led to significant cost reductions for materials, services, and resources allocated for the dismantling and packaging. Examples include the use of a combined worksite, an optimized cutting sequence and tools designed with packaging in mind, and the implementation of remote handling and robotic tools. Furthermore, this integration minimized radiological risks (i.e., through optimization according to the ALARA principle) by reducing the exposure of personnel through combining two interventions into one.

Although the activation of the graphite in graphite-moderated nuclear reactors has been extensively discussed in scientific literature (for example Refs.~\cite{international2006iaea,international2010iaea,international2016iaea}), until the detailed analyses conducted by CERN's Radiation Protection group~\cite{rp_LHC_TDE_characterization}, no previous studies had examined the radiological characteristics of a beam dump exposed to a \SI[number-unit-separator=\text{-}]{6.5}{\tera\electronvolt\per\textit{c}} proton beam. Indeed, the LHC beam dump is subjected to a unique radiation environment due to the unique characteristics of the CERN accelerator complex. This environment arises from the interaction of \SI[number-unit-separator=\text{-}]{6.5}{\tera\electronvolt} primary protons with the graphite core, as well as from high-energy secondary particles (neutrons, charged pions, and more) generated by the hadronic and electromagnetic particle showers within the dump. This radiation environment is fundamentally different from that found in graphite-moderated nuclear reactors and therefore cannot be directly compared. A comprehensive summary of the RP studies performed to characterize the activation of the LHC beam dump can be found in Ref.~\cite{rp_LHC_TDE_characterization}.

FLUKA was employed to calculate the energy-dependent fluence distributions of neutrons ($n$), protons ($p$), photons ($\gamma$), positive pions ($\pi^+$), and negative pions ($\pi^-$) in each of the main components of the beam dump core, as illustrated in Fig.~\ref{fig:TDE_FLUKA}. These fluences were then incorporated into the analytical code ActiWiz Creator~\cite{Vincke:2018} to estimate the radionuclide inventory of the different parts of the beam dump. This estimation was based on the irradiation profile (i.e., the number of primary protons dumped during Run~1 and Run~2 operations) and the material composition.

In addition, the production yields of key radionuclides such as $^{3}$H, $^{60}$Co, $^{54}$Mn, and $^{55}$Fe were estimated by means of the fluence conversion coefficient~\cite{Froeschl:2018} method. This allowed the use of \textit{ad hoc} fluence-to-specific-activity conversion coefficients, which were tailored to account for the material composition of the beam dump components, the irradiation profile, and a suitable cool-down time. These coefficients were applied in real time to the FLUKA simulations via a dedicated user routine, enabling the generation of 3D spatial distribution maps for the aforementioned radionuclides. Specifically, the coefficients were implemented on cylindrical $r$--$\phi$--$z$ meshes that covered the entire radius and length of the beam dump core, with a \SI{5}{\degree} step size in $\phi$ (the azimuthal angle around the longitudinal $z$ axis).

Figures~\ref{fig:SA_H-3_HDG_LDG} and \ref{fig:SA_H-3_TDE} show the spatial distribution of the specific activity for $^{3}$H within the beam dump core after a cool-down period of 1245~days from the end of Run~2 operations. This cool-down time is referenced to the day when dedicated samples were extracted from the beam dump core for analysis. The cross sections displayed in the plots (HDG1, LGD6, and HDG2) align with specific sampling locations. The figures demonstrate that the $^{3}$H distribution is predominantly driven by high-energy spallation reactions occurring in the graphite matrix, with activation patterns closely following the trajectory of the primary proton beam.

To determine the waste classification of the LHC beam dump, the activities estimated using ActiWiz were compared to the limits specified for waste classes, as described in Ref.~\cite{rp_LHC_TDE_characterization}. Calculations were performed for each region of the beam dump shown in Fig.~\ref{fig:TDE_FLUKA} (four sections of the steel vessel, six HD graphite sections, and ten LD graphite sections) using reported specific activities plus 2$\sigma$ uncertainties. These uncertainties, which are well below $1\%$ for most nuclides of interest, account for statistical errors only (arising from both the nuclear data library and fluence spectra).

The activity estimates were representative of February~2022, the period during which the beam dump autopsy was conducted. The primary limiting radionuclide, in terms of compliance with waste repository limits, was identified as $^{3}$H within the graphite core. Based on these findings and aided by radiochemical analysis of samples of graphite, according to CERN RW classification~\cite{rp_LHC_TDE_characterization} it was determined that the beam dump should be classed as medium-level-activity waste (Faible et Moyenne Activité - Vie Courte or FMA-VC), with the additional requirement that the waste should be encapsulated in a concrete matrix.

According to CERN's radioactive waste management strategy, the elimination pathway for this class of waste is the Centre de Stockage de l'Aube (CSA) in France. Only standard storage containers are allowed at the CSA. These containers serve as the first safety barrier, being designed to meet the general safety objectives of containment and environmental protection. Considering the dimensions of the beam dump, only two standard conditioning options are available for its storage at CSA while meeting the specific requirements of the Agence Nationale pour la Gestion des Déchets Radioactifs (Andra): an injectable container of \SI{5}{\meter\cubed} and an injectable container of \SI{10}{\meter\cubed}. It is important to note that CERN is only responsible for preconditioning of the dump (characterization, cutting of the dumps into appropriate pieces, and preparation of the packages following Andra specifications); the final conditioning (e.g., filling the container with concrete) and storage of the RW package will be performed by Andra.

While the present paper provides the methodology for the radiological characterization of the LHC beam dump as RW, the details of the final conditioning at the Andra CSA center are outside its scope. Other criteria must be adhered to in accordance with Andra's specifications, as described in Ref.~\cite{rp_LHC_TDE_characterization}, including a maximum limit of \SI{50}{\giga\becquerel} $^{3}$H per container.

\begin{figure}
    \centering
    \includegraphics[width=0.4\textwidth]{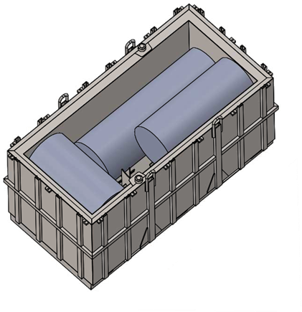}
    \caption{Schematic of the final container with three HD graphite blocks inside.}
    \label{fig:RIBA_FULL}
\end{figure}

The three blocks of HDG were placed inside a ``lost mold'' equipped with shimming supports sized to perfectly fit the dimensions of the dump pieces, , allowing for safe stowing. This lost mold is contained within the injectable \SI{10}{\meter\cubed} container, with a \SI[number-unit-separator=\text{-}]{150}{\milli\meter} pre-concrete lining. Based on ActiWiz estimates, the total tritium activity inside this container is \SI{19.7}{\giga\becquerel}. This configuration allows for the addition of LDG disks in batches of ten units, until the \SI{50}{\giga\becquerel\per container} limit is reached. The filled container, as shown in Fig.~\ref{fig:RIBA_FULL}, will then be shipped to the CSA repository (see also Sec.~\ref{Sec:Cutting_of_the_HD_sector}).

\section{Cutting methods}
Given the unsuccessful preliminary cutting operation described in Sec.~\ref{motivation-behind-dump-autopsy}, a feasibility study was undertaken to assess potential methods for cutting the stainless-steel vessel. It was necessary for the optimal cutting method to be reliable and repeatable to minimize potential hazards and the exposure of personnel to radiation throughout the procedure; furthermore, the ability to undertake the cutting by remote control was required, to reduce the need for human intervention during the cutting process. Three 318LN stainless-steel blanks, each measuring \SI{1}{\meter} in length and with the same diameter and thickness as the dump vessel, were procured to conduct these tests.

The following options were analyzed and ultimately rejected during the study:
\begin{itemize}
\itemsep0em
  \item Plasma-arc cutting was discarded due to an unsuccessful cutting trial on a vessel mockup. During this trial, SIGRAFLEX sheets were placed behind a steel plate to reproduce the configuration of a longitudinal cut in the vessel, and the plasma torch was placed on the opposite side. The SIGRAFLEX sheets ignited at the start of the trial due to the large amount of heat deposited and the reduced chemical stability of the material in air at high temperatures.
  \item A cryomagnet dismantling bench, which was already installed at CERN, had cutting dimensions that were compatible with the dump core. Nevertheless, this option was abandoned prior to trials due to the difficulty of decontaminating the numerous movable components of the bench after the cutting operations.
  \item It was found that the use of an industrial pipe cutter mounted around the circumference of the blank was ineffective due to its fixed rotational speed, which resulted in frequent tooling breakage due to the hardness of the material. Additionally, the process of installing the pipe cutter was found to be excessively time-consuming, and it exhibited an excessive level of sensitivity to small diameter variations in the blanks.
  \item Finally, a milling tool attached to a rail was found to experience significant vibrations, leading to accelerated wear of the tool. Moreover, the gearbox of the milling tool began to leak following a cutting distance of only \SI{500}{\milli\meter}.
\end{itemize}

Two cutting methods were, however, deemed compatible with the requirements on reliability, repeatability and remote operation: an industrial robot with a milling tool and a remotely actuated wall saw mounted on a rail. Each method exhibited distinct characteristics, offering specific advantages as well as associated limitations.

\subsection{Industrial robot with milling tool}
The KUKA KR~120 R2700~HA~\cite{KUKA_KR120_R2700} robot is a six-axis robotic arm with a maximum reach of \SI{2700}{\milli\meter} capable of milling operations. The milling method was specifically investigated due to its low operational temperature, which minimizes the risk of core ignition during cutting, and its tendency to produce large chips rather than fine dust particles, thereby simplifying post-process cleaning for radiation protection purposes.

The robot was equipped with a Keyence GT2-P12~\cite{Keyence_GT2_P12} high-accuracy (\SI{0.01}{\milli\meter}) digital contact sensor, enabling detection of the relative position of the workpiece with respect to the robot base reference frame. The accuracy of the milling process was successfully validated in preliminary trials by reproducing the milling trajectory on the cylindrical surface of the test blanks.A wide range of tests were performed to determine the optimal tooling and milling parameters, with the solid carbide end mill HPC TiAlN~\cite{Hoffmann_Group_22709} being selected as the most suitable option.

\begin{figure}[tbp]
\centering
\includegraphics[width=\columnwidth]{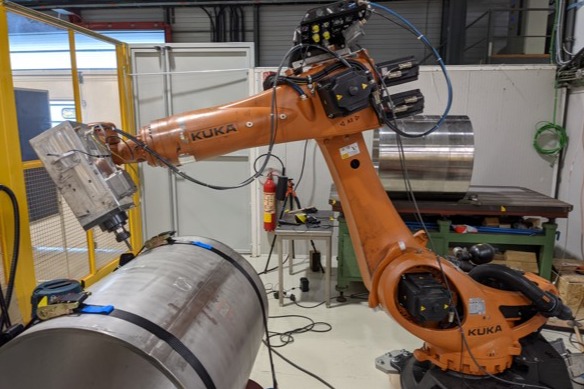}
\caption{KUKA KR~120 R2700~HA robotic arm performing a milling test on one of the vessel mockups.}
\label{fig:KUKA-robot}
\end{figure}

Figure~\ref{fig:KUKA-robot} shows one of the circumferential milling tests performed to the full depth of the vessel mockup. While the test was successful, the tool exhibited a limited lifespan, completing approximately one-third of the vessel’s circumference (around \SI{800}{\milli\meter}) before requiring replacement. Additionally, the milling tool was required to operate at a low cutting speed due to the inherent characteristics of the milling process prolonging the overall cutting time significantly.

%Although tool exchange could be completed in less than 2 minutes and at a distance of \SI{3.4}{\meter} from the vessel, multiple interventions and their necessary depressurization and repressurization of the airlock would significantly prolong the time required to perform the milling operation.

\begin{figure*}[thbp]
\centering
\includegraphics[width=\textwidth]{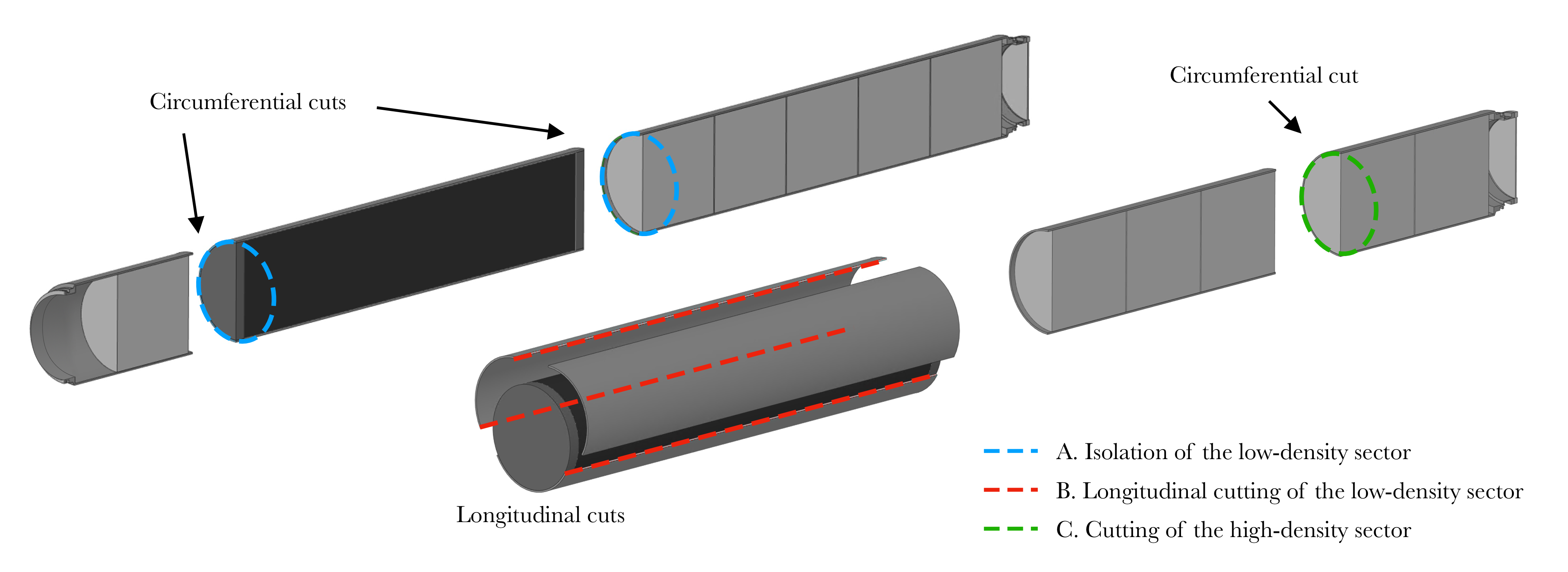}
\caption{Cross-sectional schematic illustration of the cutting sequence of the beam dumps, indicating the locations of the circumferential and longitudinal cuts.}
\label{fig:cutting-sequence}
\end{figure*}

\subsection{Remotely actuated wall saw mounted on guiding frame}
\begin{figure}[tbp]
\centering
\includegraphics[width=\columnwidth]{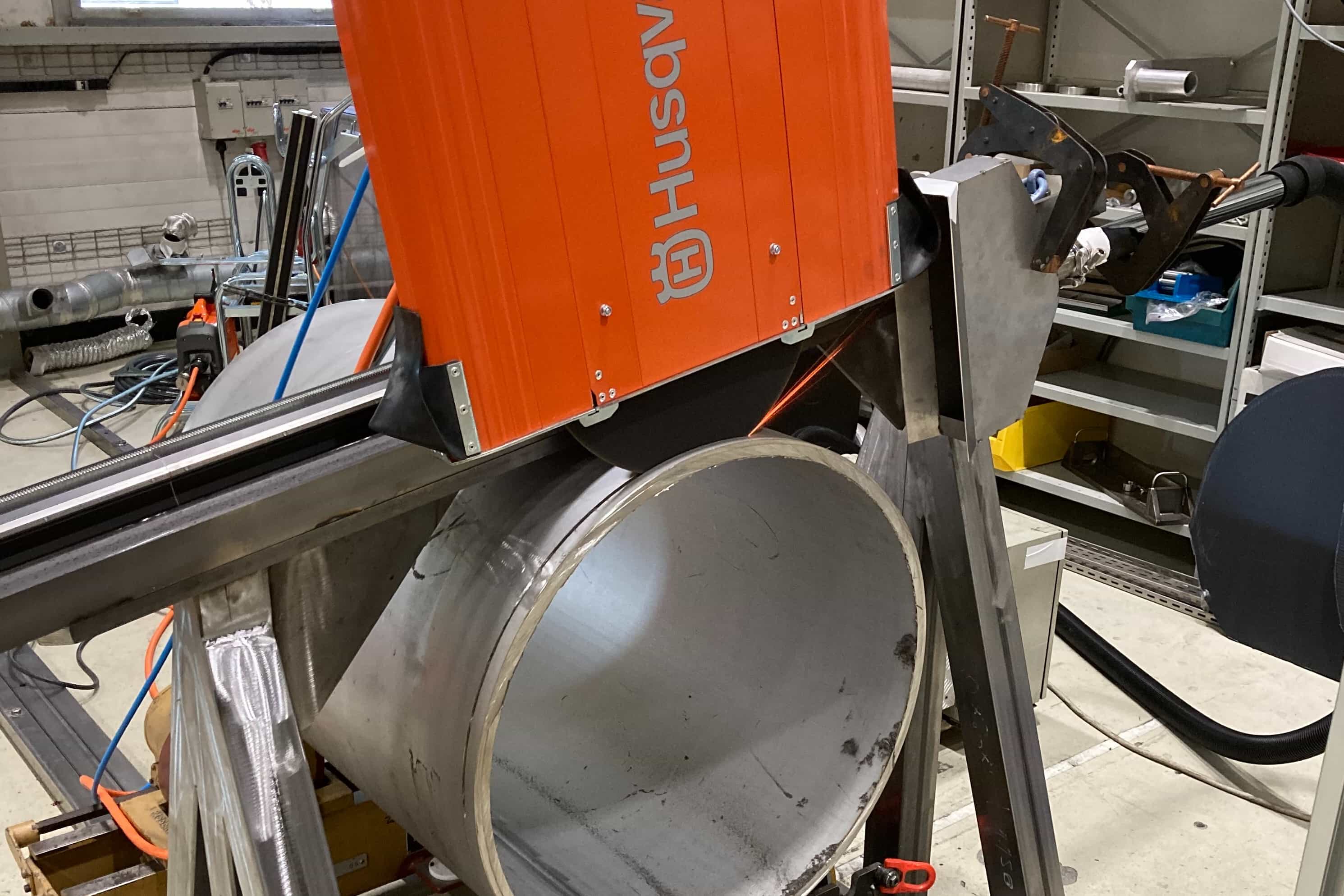}
\caption{Husqvarna WS~220 saw~\cite{Husqvarna2018IPLSAWS} installed on the inclined guiding frame while performing a circumferential cutting trial on one of the vessel mockups.}
\label{fig:Husqvarna-saw}
\end{figure}

An alternative approach to cutting was presented by the Husqvarna WS~220 saw~\cite{Husqvarna2018IPLSAWS}, which is typically used for cutting reinforced concrete. Figure~\ref{fig:Husqvarna-saw} shows the saw mounted on a frame with inclined guiding rails during one of the numerous tests performed on the vessel mockups to determine the optimal cutting setup. The guiding rails allowed remote control of the cutting position and adjustment of its depth by moving the saw along the rail system. Six rollers (five passive and one motorized with remote control) were used for the circumferential cutting to rotate the beam dump and provide the cutting feed.

Various types of saw blades were tested. The pre-mounted saw blade equipped with diamond segments experienced clogging due to molten steel, while a carbide blade underwent excessive wear. A grinding disc was able to successfully cut \SI{3}{\meter} of a 318LN stainless-steel plate. This test was successfully replicated on the blanks using three passes, resulting in a total cutting time of \SI{70}{\minute} per circumferential cut. As indicated by the sparks in Figure~\ref{fig:Husqvarna-saw}, the wall saw generated higher temperatures compared to the milling method. Multiple tests involving the placement of expanded graphite directly beneath the cutting area confirmed that no ignition occurred.

The saw was connected to a radiation protection vacuum cleaner, and its protective cover was extended to enhance vacuum efficiency and reduce the dispersion of contaminated metallic particles during cutting. This upgrade was found to be effective, with particle spread remaining minimal and posing no significant challenges for radiation protection when entering the airlock following the cutting process.

Ultimately, the wall-saw method was selected over the use of the industrial robotic arm due to its stability, simplicity, and more time-efficient cutting performance observed across multiple trials. While it is acknowledged that the milling method, or the use of a grinding disc with the robotic system, could have been further optimized for the task, the wall-saw solution was at a more advanced stage of development.

\section{Cutting, sequencing, and extraction of specimens}
Based on the RP assessment in Sec.~\ref{radioprotection challenges}, a strategy was adopted that involved isolating the LD sector, allowing for prolonged human intervention during the extraction of samples for post-irradiation examination. A cutting sequence plan was formulated, as shown in Fig.~\ref{fig:cutting-sequence}. This plan comprised three radial cuts and three longitudinal cuts:
\begin{itemize}
    \item[A.] Isolation of the LD sector.
    \item[B.] Longitudinal cutting of the LD sector.
    \item[C.] Cutting of the HD sector.
\end{itemize}
These steps are explained in detail in the next three subsections.

%As described in~\ref{radioprotection challenges}, Monte Carlo FLUKA code simulations based on beam dump operation in Run 1 and Run 2 were performed to characterize the radionuclide inventory produced in the dumps. The simulations demonstrated that the implementation of a strategy involving the isolation of the low-density-sector and the removal of the 318LN stainless steel vessel resulted in a substantial reduction in the residual dose rate of approximately 90\% from \SI{1000}{\micro\sievert\per\hour} to \SI{100}{\micro\sievert\per\hour} [X]. This woulow-density allow human intervention during the extraction process of samples for post-irradiation examination.

\subsection{Isolation of the LD sector}
\begin{figure}[tbp]
\centering
\includegraphics[width=\columnwidth]{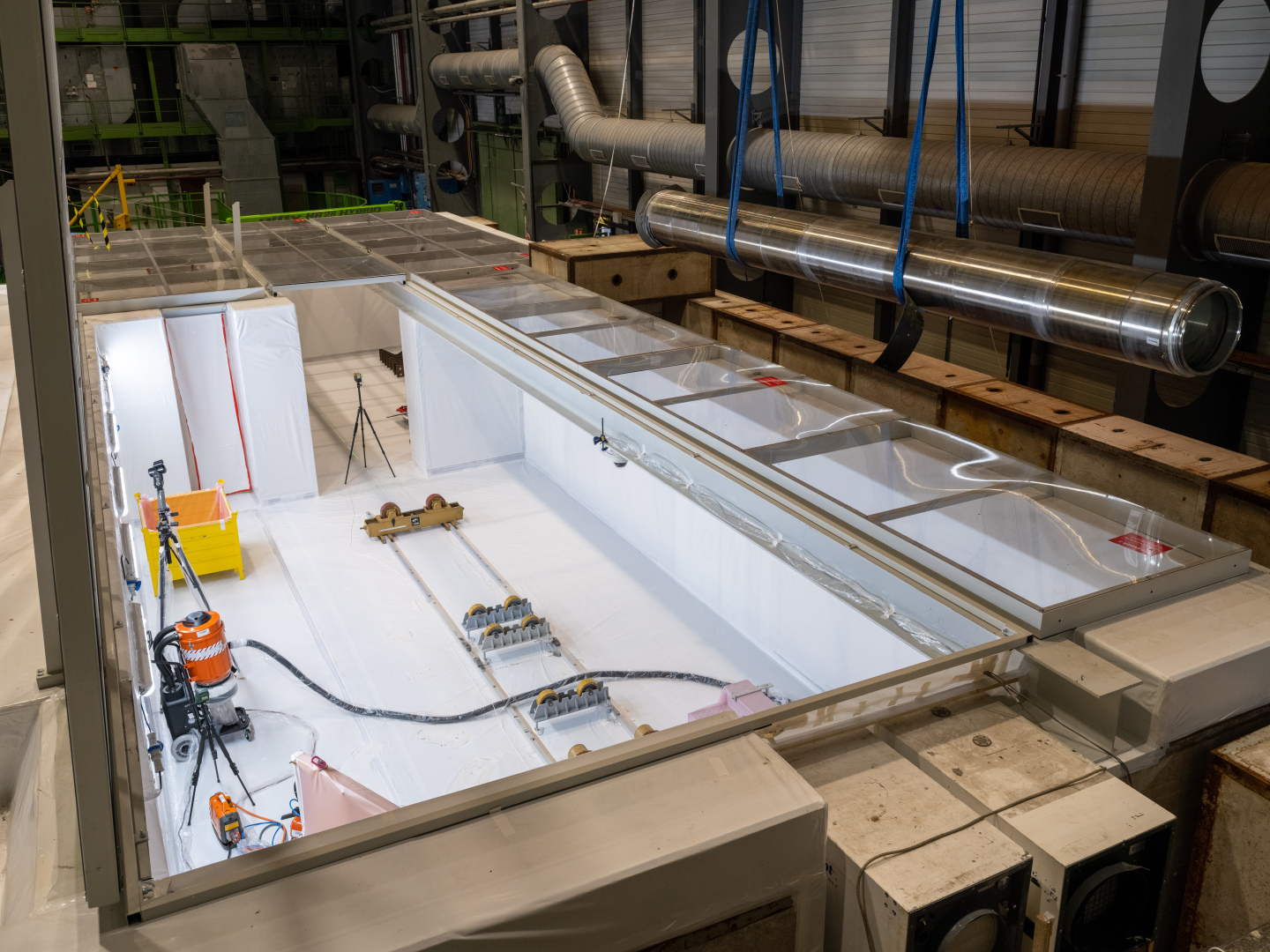}
\caption{View of the beam dump being transferred from the shielded storage bunker to the main working area for cutting to isolate the LD sector. Rollers are installed on premounted rails bolted to the floor.}
\label{fig:cutting-installation}
\end{figure}

\begin{figure}[tbp]
\centering
\includegraphics[width=\columnwidth]{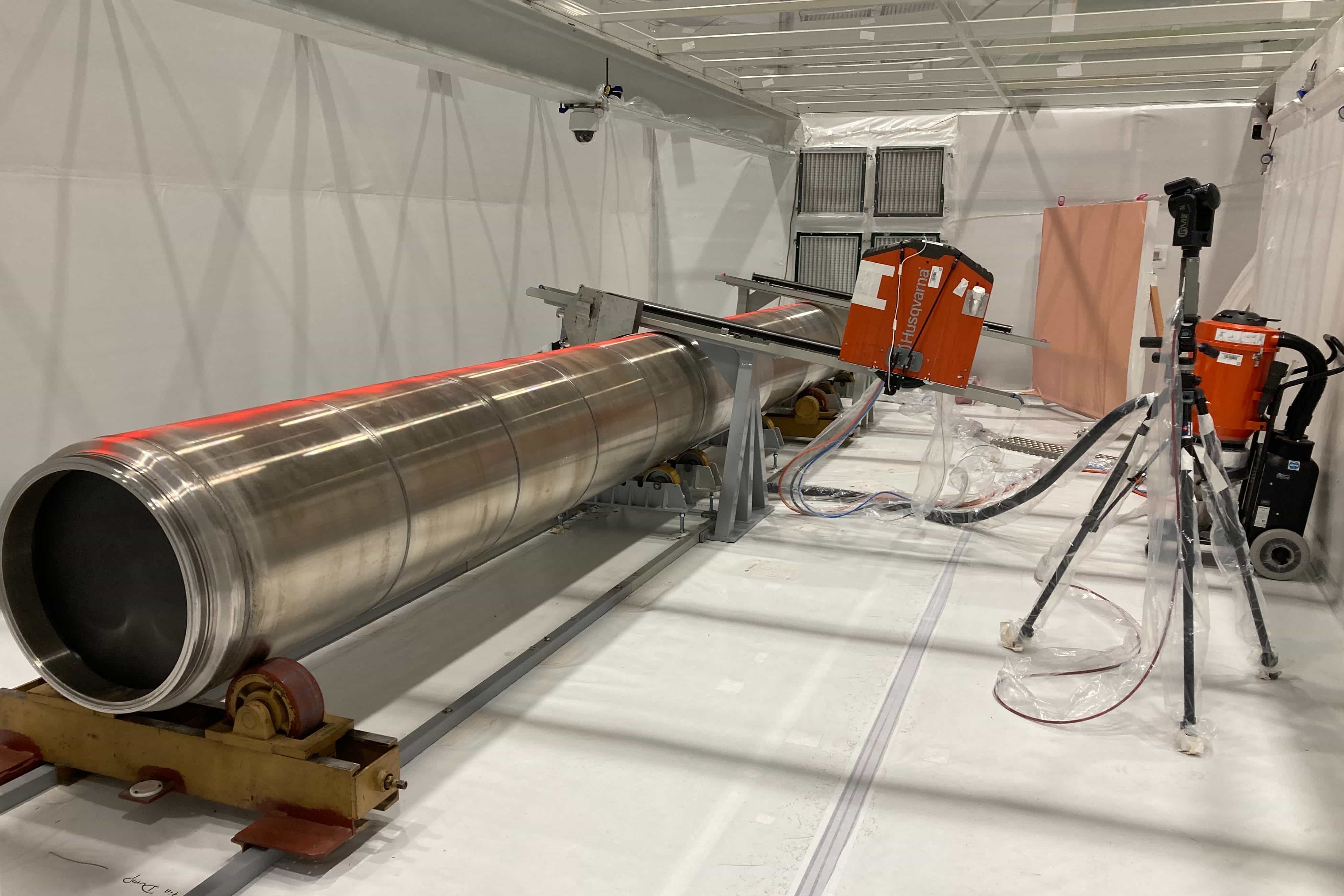}
\caption{View of the cutting setup inside the bunker prior to the first two circumferential cuts. The guiding rails are installed in their correct axial positions on either side of the LD sector.}
\label{fig:cirfumferential-cuts}
\end{figure}

Figure~\ref{fig:cutting-installation} shows the beam dump being transferred from the storage bunker to the primary working area. It is worth noting the open roof in the airlock that allows the crane to operate. Figure~\ref{fig:cirfumferential-cuts} shows the beam dump positioned on the rollers, ready for the initial circumferential cuts to isolate the LD sector. To ensure the traceability of the alignment of the transverse beam sweep relative to the vessel during the dismantling activities, the upper surface of the dump was marked along its entire length as a reference.

\begin{figure}[tbp]
    \centering
    \begin{subfigure}{0.26\textwidth}
        \centering
        \includegraphics[width=\textwidth]{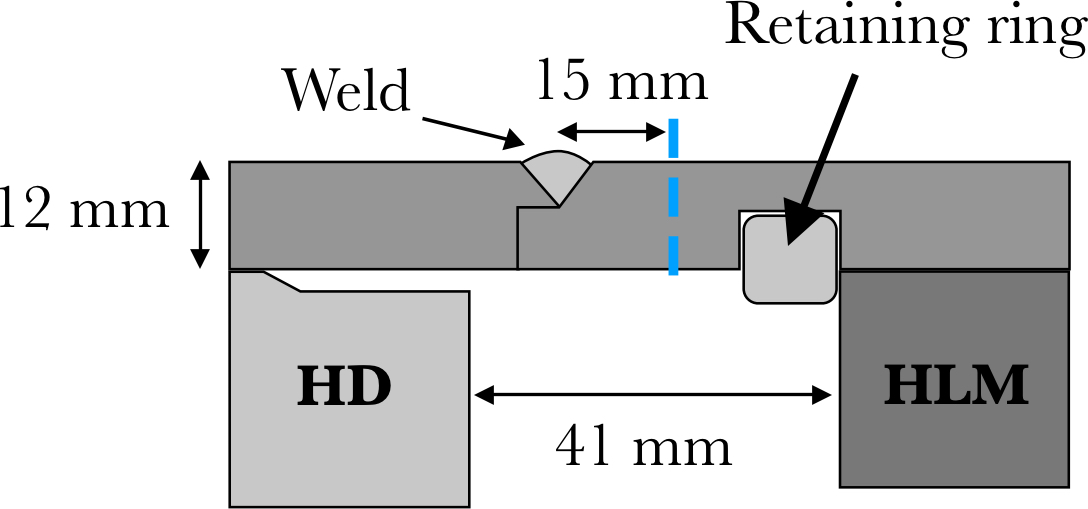}
        \caption{}
        \label{fig:cuttingcross}
    \end{subfigure}
    \begin{subfigure}{0.20\textwidth}
        \centering
        \includegraphics[width=\textwidth]{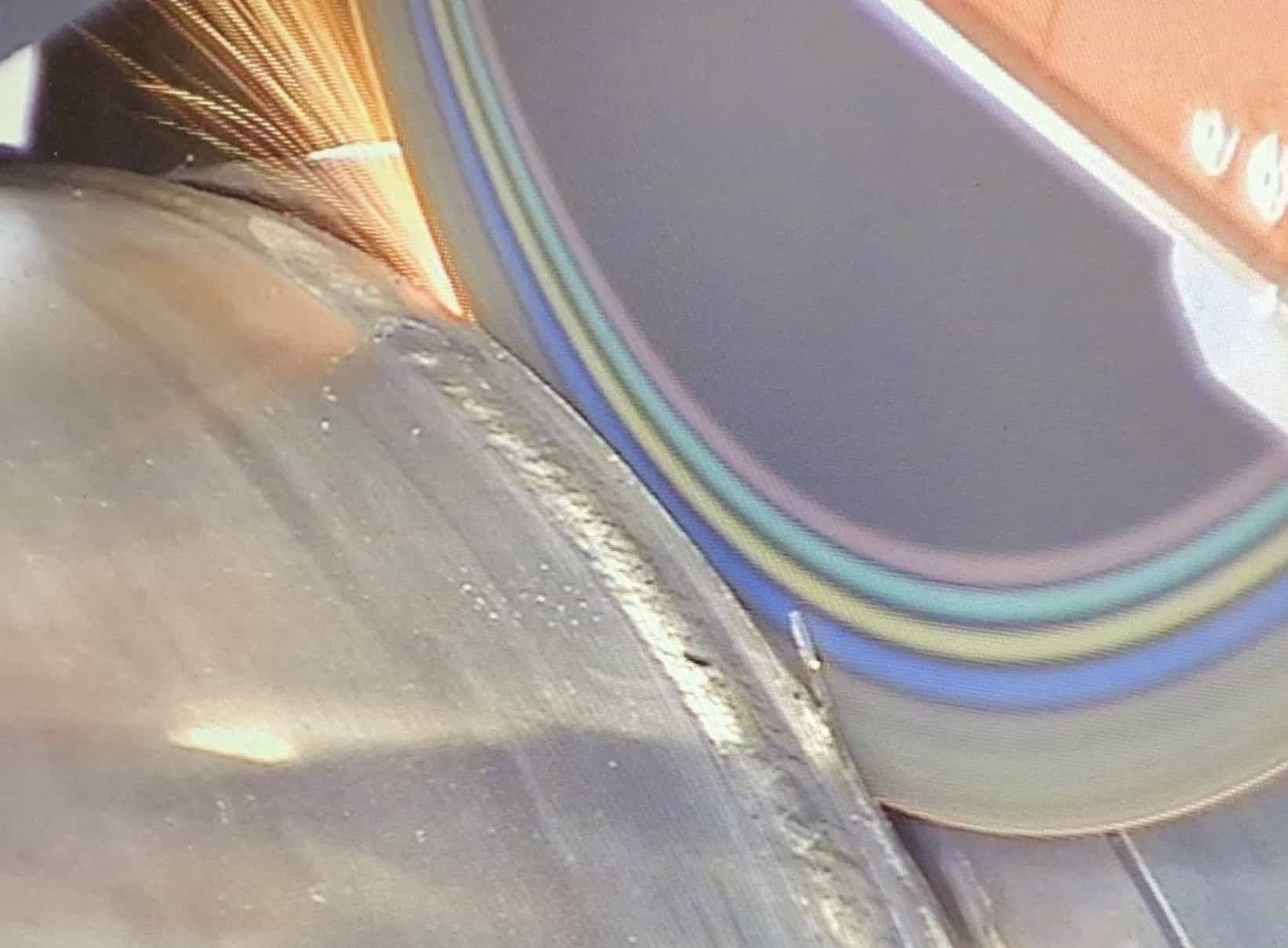}
        \caption{}
        \label{fig:cutting-monitoring}
    \end{subfigure}
    \caption{(a)~Cross-sectional view of beam dump with an indication of the position of the circumferential cut (blue dashed line) with respect to the circumferential weld and the retaining ring for the LD sector. (b)~Remote-monitoring view of the longitudinal cutting of the LD sector, showing the colored marks on the grinding disc to indicate the cutting depth of the saw.}
    \label{fig:wall-saw-longitudinal}
\end{figure}

Figure~\ref{fig:cuttingcross} shows the presence of a gap of \SI{41}{\milli\meter} between the extruded graphite plate and the HD block. Performing the cut at this location therefore prevented the saw from coming into contact with the core. The axial placement of the guiding frames was determined by referencing the position of the welds and the accompanying assembly drawings of the dumps. Figure~\ref{fig:cutting-monitoring} shows the wall saw correctly positioned relative to the circumferential weld during cutting.

To ensure precise control of the cutting depth during each pass, the grinding disc was marked with colors, as depicted in Fig.~\ref{fig:cutting-monitoring}. This coloring facilitated the determination of the cutting depth through remote monitoring.

\begin{figure}[tbp]
\centering
\includegraphics[width=\columnwidth]{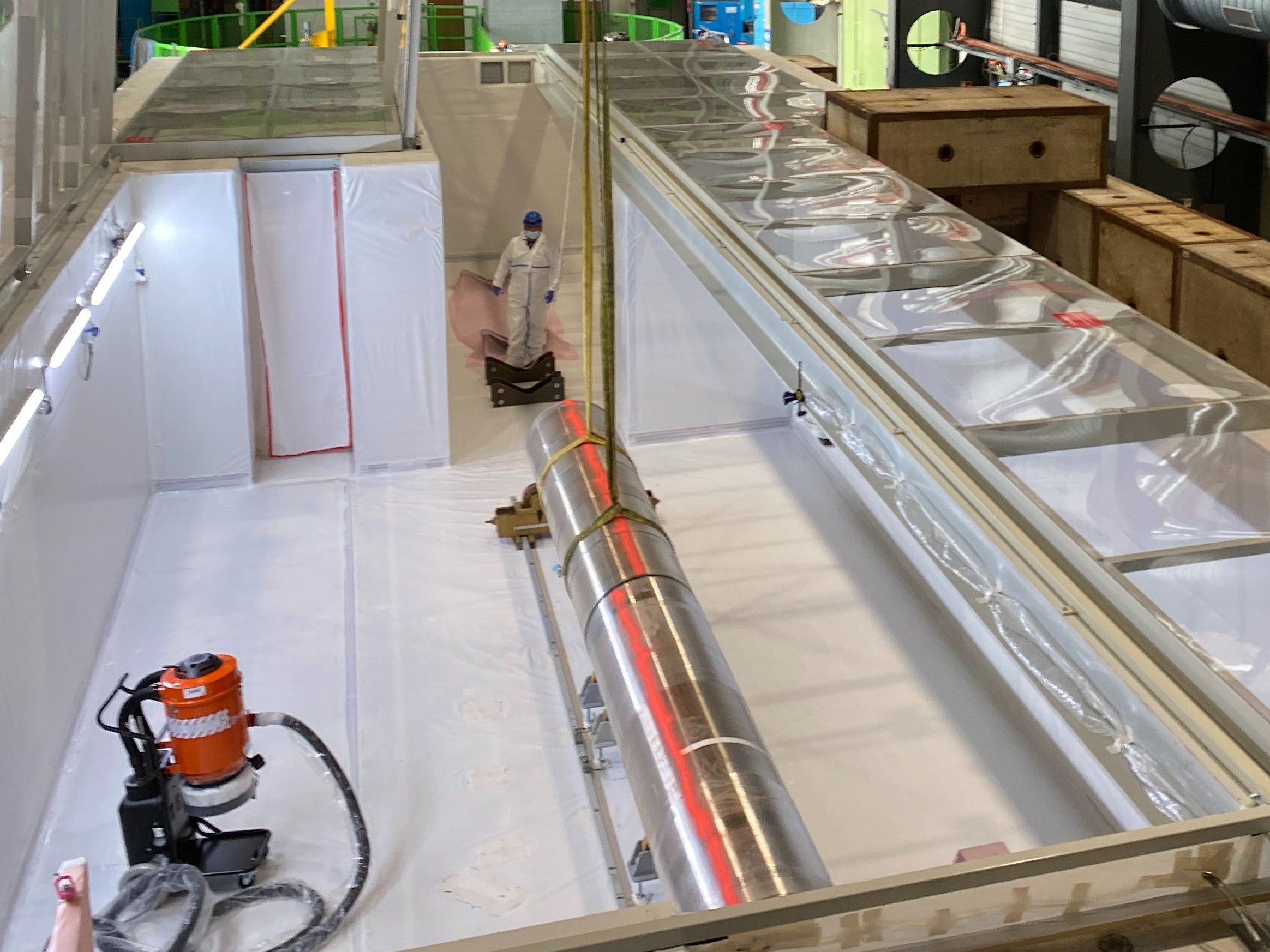}
\caption{View of separated LD sector. The downstream HD sector is being transferred to the storage bunker prior to the longitudinal cutting of the LD sector.}
\label{fig:Misaligned-sectors}
\end{figure}

Figure~\ref{fig:Misaligned-sectors} shows the beam dump after the two first circumferential cuts. The misalignment of the longitudinal red line indicates the complete separation of the LD sector. The upstream and downstream HD sectors were relocated to the storage bunker. Prior to initiating the longitudinal cutting of the LD sector, scans of the extruded graphite plates located at either end of the LD sector were conducted using an optical coordinate measuring machine (CMM; MetraSCAN~3D, Creaform~\cite{Creaform_MetraSCAN}). This assessment sought to evaluate any pre-existing failures before the possibility of damage resulting from the longitudinal cutting process. A more detailed description of this process can be found in Sec.~\ref{extruded graphite plate}.

\begin{figure}[tbp]
\centering
\includegraphics[width=\columnwidth]{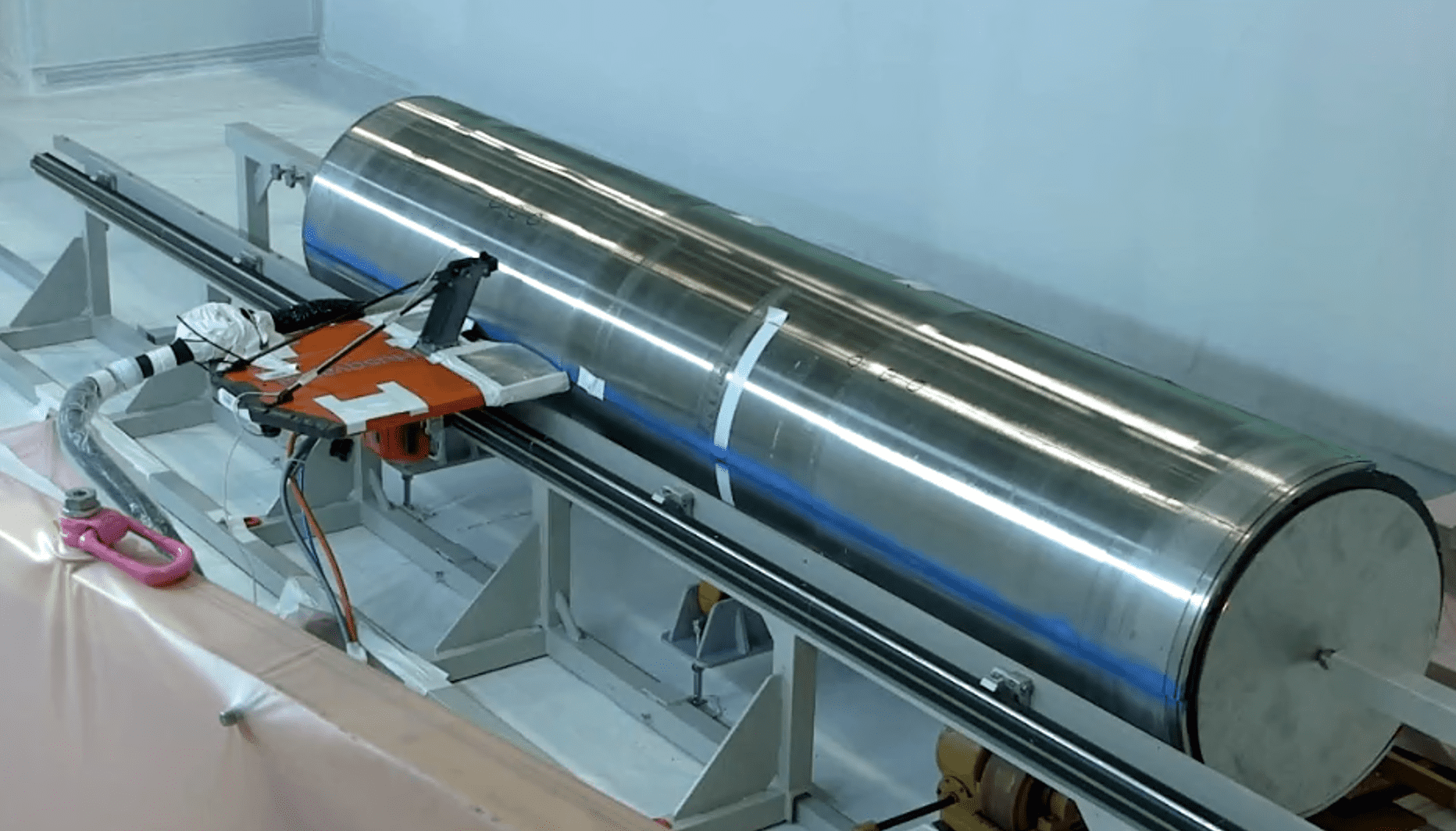}
\caption{Longitudinal cutting setup with tailstocks and aluminum discs installed to stabilize the LD sector during cutting, along with the longitudinal guide frame and saw.}
\label{fig:cutting-longitudinal}
\end{figure}

\subsection{Longitudinal cutting of the LD sector}
The configuration used for the longitudinal cutting process is illustrated in Fig.~\ref{fig:cutting-longitudinal}. Two tailstocks were positioned on either side of the LD sector, with aluminum (Al) discs placed between the tailstocks and the extruded graphite discs. This arrangement was designed to maintain the stability of the core within the vessel during the cutting. A longitudinal guide frame was also used for the wall saw, enabling its movement along the vessel's length. The mass of the LD sector, combined with the compression from the tailstocks, was sufficient to ensure the stability of its angular orientation during the operation.

Two additional marks were made at angles of \SI{120}{\degree} relative to the original longitudinal mark, serving as reference points for the longitudinal cuts. The longitudinal position of the saw, as well as the blade's depth, could be regulated by remote control. Rotation of the LD sector between cuts was achieved via the motorized roller supporting the sector.

Following the extraction of the first steel shell, two lifting studs were welded to its surface to facilitate its transportation to the storage bunker. Once the shell had been removed, the SIGRAFLEX was marked longitudinally to preserve information about the angular alignment of the sheets within the core.

\begin{figure}[tbp]
\centering
\includegraphics[width=\columnwidth]{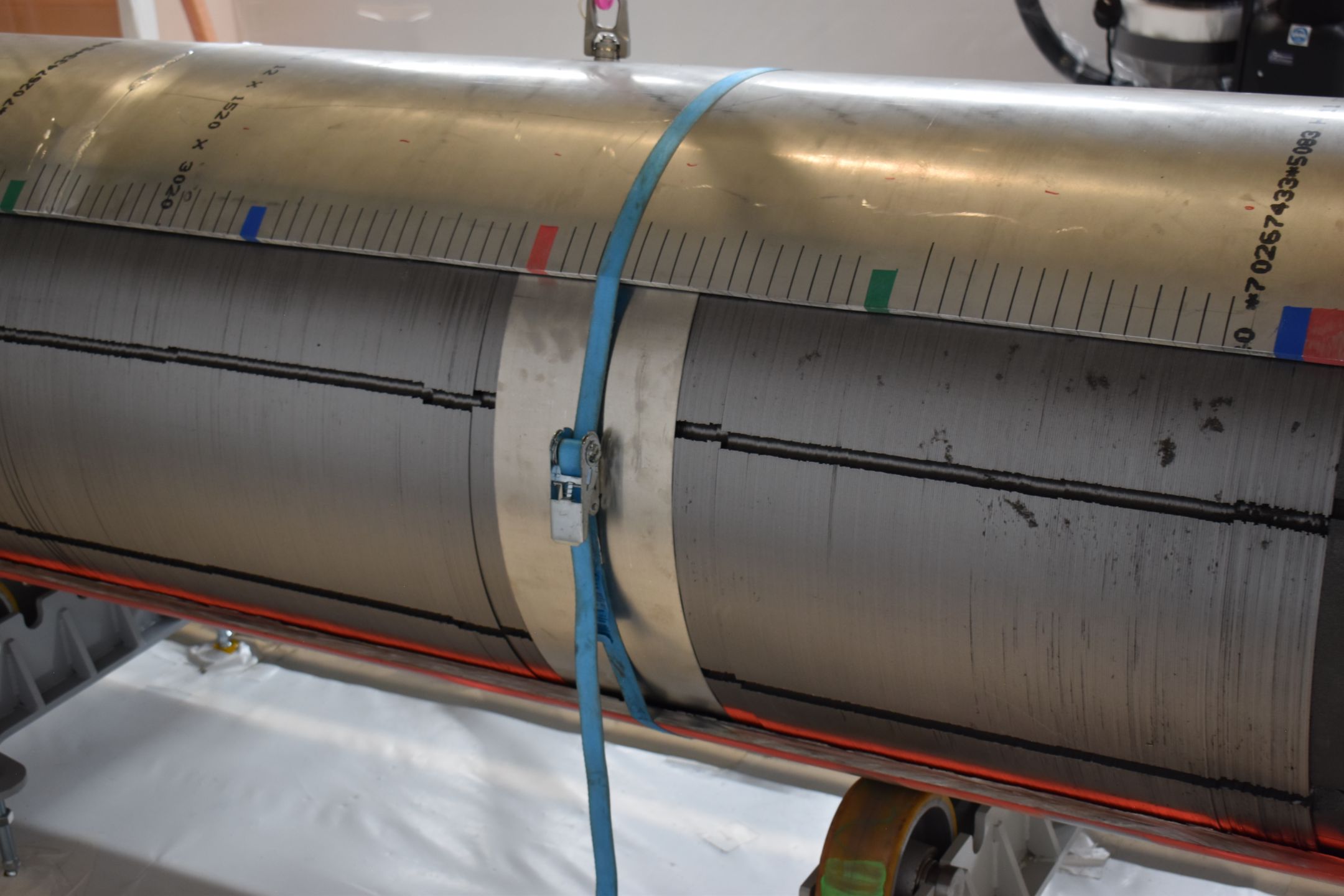}
\caption{View of the LD sector resting on the steel shell after two longitudinal cuts, prior to its replacement with an aluminum shell. The aluminum shell is marked with tape to indicate the extraction positions for the expanded graphite sheets for further post-irradiation examination.}
\label{fig:longitudinal-cuts}
\end{figure}

\begin{figure}[tbp]
    \centering
    \begin{subfigure}{0.23\textwidth}
        \centering
        \includegraphics[width=\textwidth]{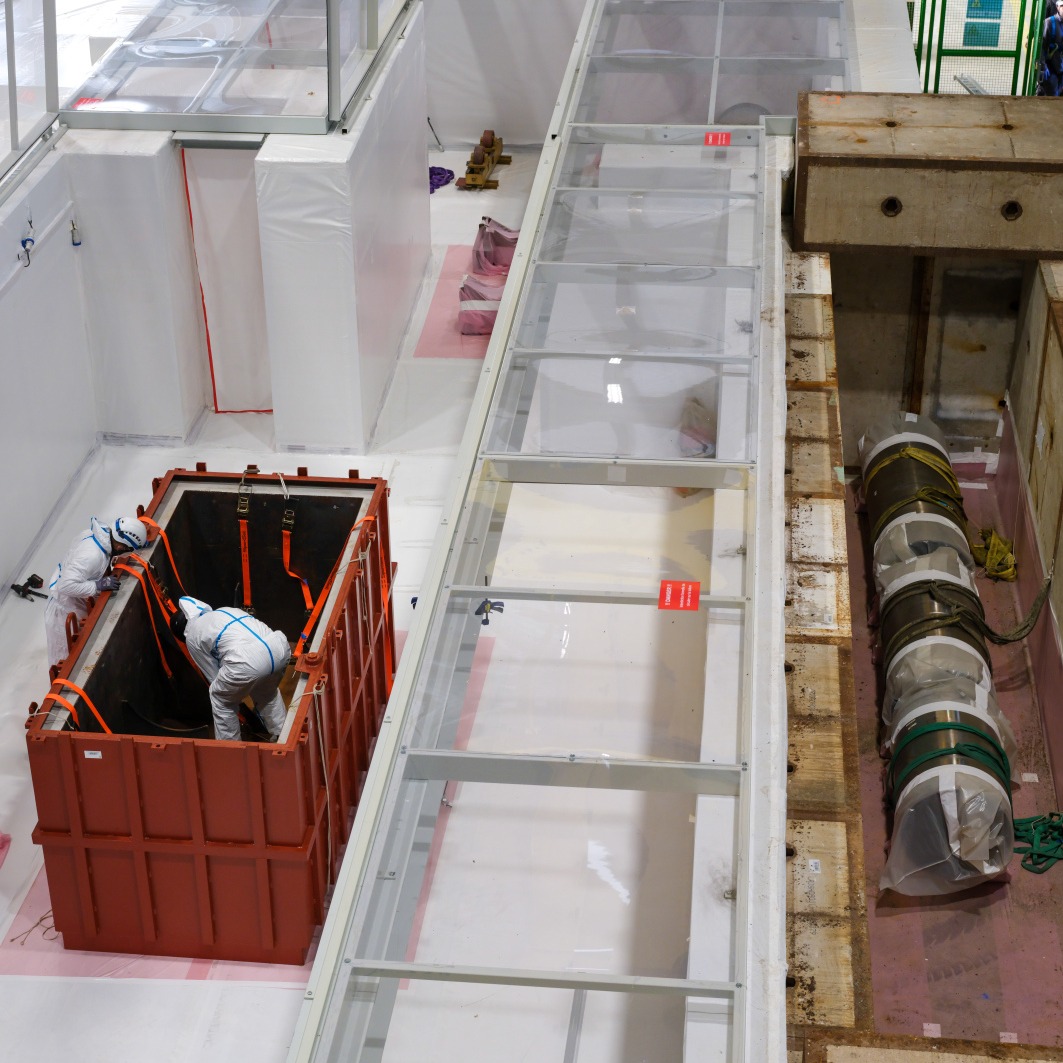}
        \caption{}
        \label{fig:waste-container:a}
    \end{subfigure}
    \begin{subfigure}{0.23\textwidth}
        \centering
        \includegraphics[width=\textwidth]{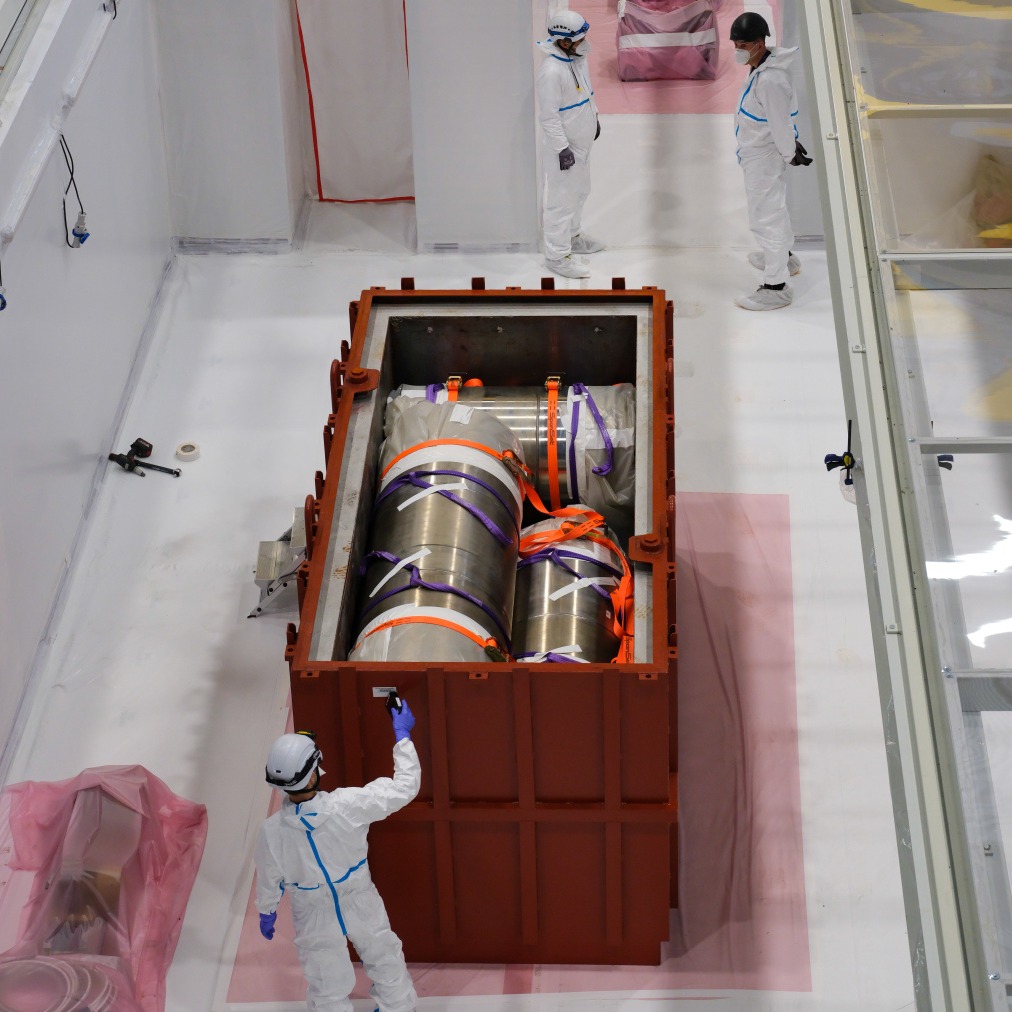}
        \caption{}
        \label{fig:waste-container:b}
    \end{subfigure}
    \caption{Photographs showing the process of transferring the cut segments of the beam dump to the waste container. (a)~Preparation of the waste container before the segments are transferred from the storage bunker. (b)~The waste container after the segments had been loaded inside.}
    \label{fig:waste-container}
\end{figure}

Following the completion of the third longitudinal incision, two-thirds of the LD vessel was successfully removed; however, the remaining third of the highly activated vessel continued to produce significant residual dose rates, rendering extended human inspection during the post-irradiation assessment of the LD sector unfeasible. Consequently, a decision was made to substitute the final steel shell with an aluminum alternative. As shown in Fig.~\ref{fig:longitudinal-cuts}, two slings were used to compress the shells against the core. The assembly was then gradually rotated to rest the LD sector on the new aluminum shell, allowing the removal of the last steel shell section.

\subsection{Cutting of the HD sector}\label{Sec:Cutting_of_the_HD_sector}
After performing the specimen extraction and post-irradiation examination (as detailed in Sec.~\ref{Sec:Extraction-and-PIE}), the downstream HD sector was reinstalled in the primary working area. This sector, comprising five isostatic graphite blocks with a total length of \SI{3.5}{\meter}, needed to be cut in two to allow it to be placed in a \SI[number-unit-separator=\text{-}]{10}{\cubic\meter} waste container, as shown in Fig.~\ref{fig:waste-container}.

\section{Specimen extraction and post-irradiation examination} \label{Sec:Extraction-and-PIE}
\subsection{Expanded graphite sheets}
\begin{figure}[tbp]
\centering
\includegraphics[width=\columnwidth]{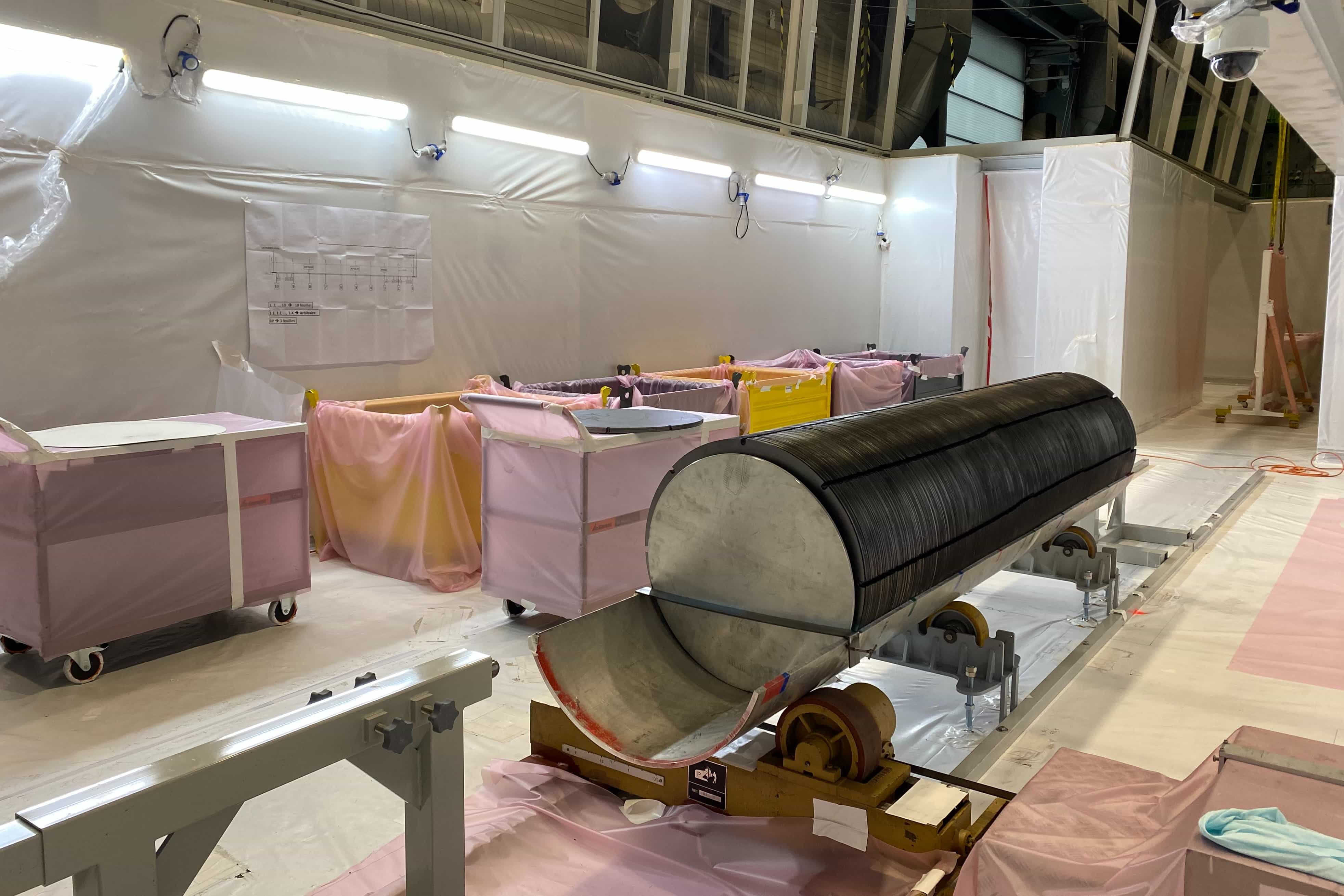}
\caption{View of the working conditions implemented during the sheet extraction. The LD sector is supported by the Al sleeve, with Al discs placed on either side to keep the sheets from tipping over. A fastening device attached to the aluminum sleeve was used to secure the discs. Behind the LD sector is a table used for dose measurements, with waste-disposal containers visible in the background.}
\label{fig:LDsectordisassembly}
\end{figure}

Figure~\ref{fig:LDsectordisassembly} shows the working conditions implemented during the process of sheet extraction. A threefold variation in peak energy density is observed across different longitudinal positions along the LD sector (Figure~\ref{fig:energy-density-along-beamline}), highlighting the necessity of sampling specimens from multiple locations for comprehensive analysis. This approach enables the study of material properties under varying energy densities, and it could help to identify a potential energy density threshold for maintaining the structural integrity of the material.

Based on the longitudinal and transverse energy deposition gradient shown in Fig.~\ref{fig:energy-deposition-overview}, three sheets were selected and catalogued every tenth of the length of the LD sector for the extraction of specimens for post-irradiation examination. The longitudinal positions for selecting the sheets were marked using blue, red, and green masking tape along the aluminum sleeve, as shown in Fig.~\ref{fig:longitudinal-cuts}. The remaining sheets were extracted in stacks of ten, systematically measured for their dose rates, and subsequently transferred to containers designated for waste disposal.

\begin{figure}[tbp]
    \centering
    \begin{subfigure}{0.23\textwidth}
        \centering
        \includegraphics[width=\textwidth]{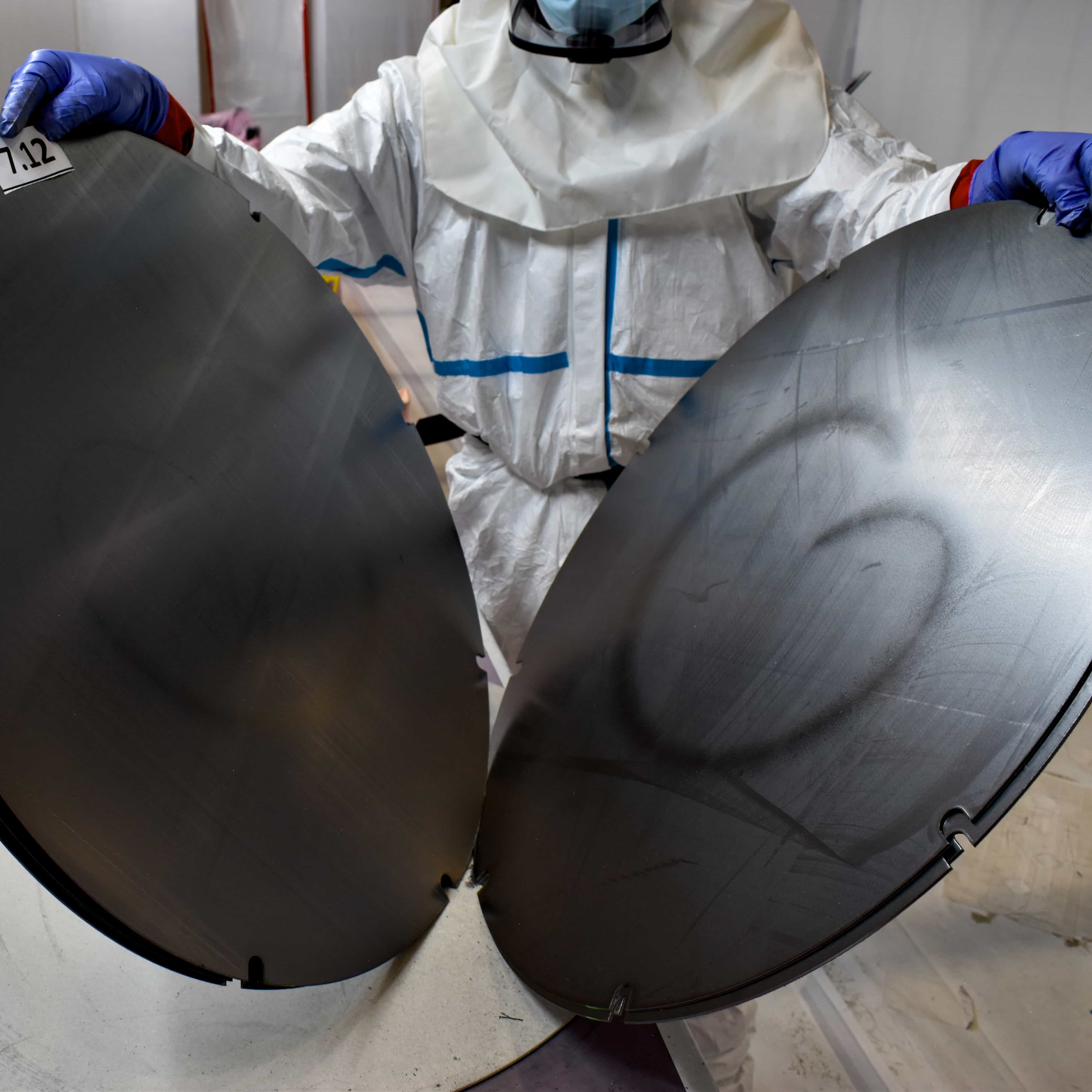}
        \caption{}
        \label{fig:sigraflex-extraction:a}
    \end{subfigure}
    \begin{subfigure}{0.23\textwidth}
        \centering
        \includegraphics[width=\textwidth]{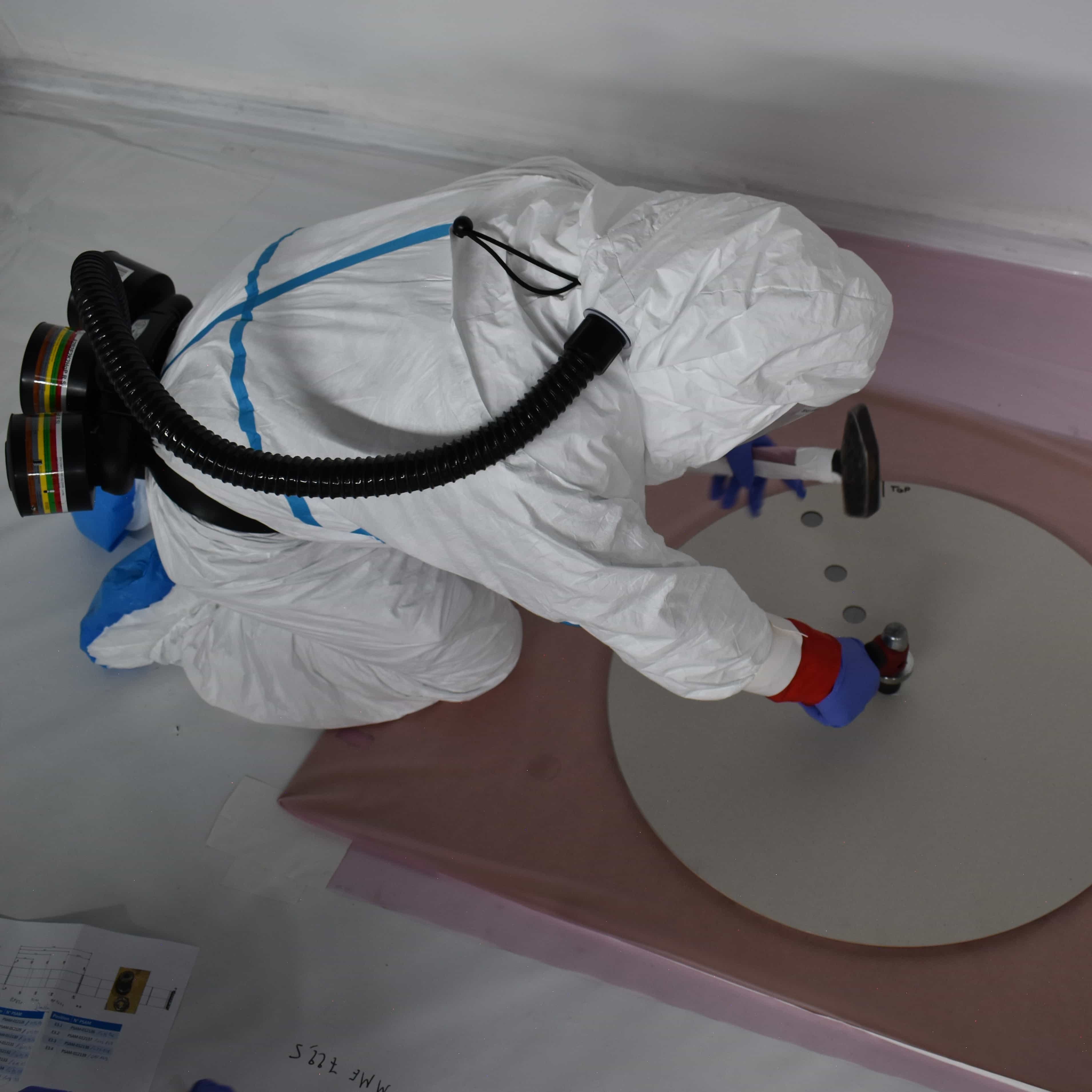}
        \caption{}
        \label{fig:sigraflex-extraction:b}
    \end{subfigure}
    \caption{Photographs showing the SIGRAFLEX extraction. (a)~Extraction of a stack of sheets clearly showing the beam sweep. Dose measurements were taken before the sheets were transferred to waste-disposal containers. (b)~Extraction of specimens for post-irradiation examination using a stencil and punching tool.}
    \label{fig:sigraflex-extraction}
\end{figure}

The post-irradiation specimens were extracted using a premade stencil that matched the diameter of the sheets. This stencil was carefully aligned with the red marks on the tops of the sheets to ensure accurate extraction locations relative to the beam sweep. A punching tool and hammer were used to punch out the specimens for post-irradiation examination. Figure~\ref{fig:sigraflex-extraction} shows photographs depicting the sheet-extraction process and the punching of specimens for post-irradiation examination.

As introduced in Sec.~\ref{motivation-behind-dump-autopsy}, the delamination of the expanded graphite sheets that were irradiated in the HiRadMat-43 experiment raised concerns about their structural integrity. A similar phenomenon was expected in the beam dump, potentially reducing the density of the expanded graphite sheets along the sweep. This could, in turn, impair the optimal energy deposition and performance of the beam dumps, and may render the material unsuitable for use in Run~3 and Run~4. However, various factors, as summarized in Table~\ref{tab:sigraflex-differences}, meant that the experiment was not directly comparable to the operational beam dumps.

\begin{table*}[tbp]
        \caption{Differences between the configuration of the expanded graphite in the HiRadMat-43 experiment and the Run~2 beam dumps.}
        \label{tab:sigraflex-differences}
    \begin{tabular}{l c c} 
        \hline \hline
        & HiRadMat-43 & Run~2 beam dumps \\
        \hline
        SIGRAFLEX grade & F05010TH~\cite{SGL_SIGRAFLEX_Foil} & L20012C~\cite{SGL_SIGRAFLEX} \\
        Sheet thickness (mm) & 0.5 & 2.0 \\
        Energy density (kJ/g) & 1.85--2.50 & Up to 1.50 \\
        Number of beam impacts & Single & Thousands \\
        Surface constraints & Free to deform & Stacked under compression \\
        Atmosphere & Air & Nitrogen \\ 
\hline \hline
        \end{tabular}
\end{table*}

\begin{figure}[tbp]
    \centering
    \begin{subfigure}{0.23\textwidth}
        \centering
        \includegraphics[width=\textwidth]{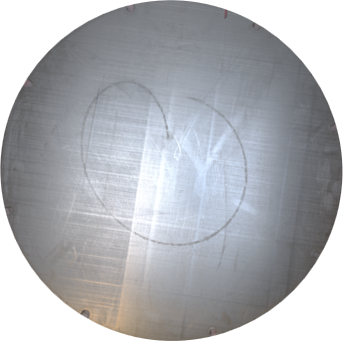}
        \caption{}
        \label{fig:darkening-profile:a}
    \end{subfigure}
    \begin{subfigure}{0.23\textwidth}
        \centering
        \includegraphics[width=\textwidth]{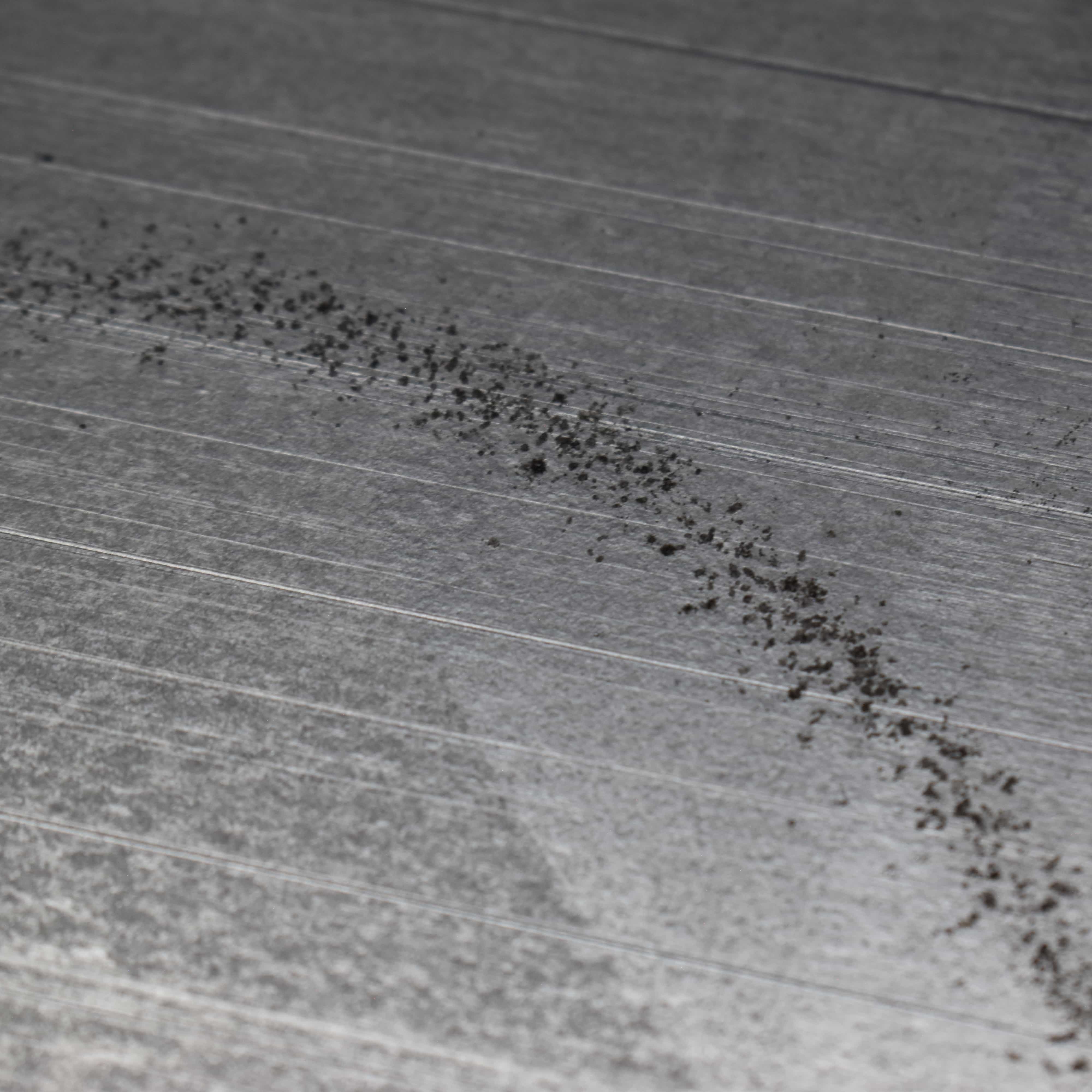}
        \caption{}
        \label{fig:darkening-profile:b}
    \end{subfigure}
    \caption{Transverse view of an expanded graphite sheet showing traces from the beam sweep: (a)~overall beam sweep; (b)~enlarged view of darkened area. The darkening appears as small dots on the surface.}
    \label{fig:darkening-profile}
\end{figure}

Initially, it was anticipated that the wave-dampening properties of expanded graphite would confine any deterioration to the regions directly affected by the beam sweep; however, upon extraction and visual inspection of the expanded graphite sheets, no signs of material delamination or deterioration were found. Figure~\ref{fig:darkening-profile} shows a darkening pattern consistent with the beam sweep on the surface of one of the extracted sheets. The degree of darkening varied across different longitudinal positions, yet there was no apparent correlation with energy density.

\begin{figure}[tbp]
    \centering
    \begin{subfigure}{\columnwidth}
        \centering
        \includegraphics[width=\textwidth]{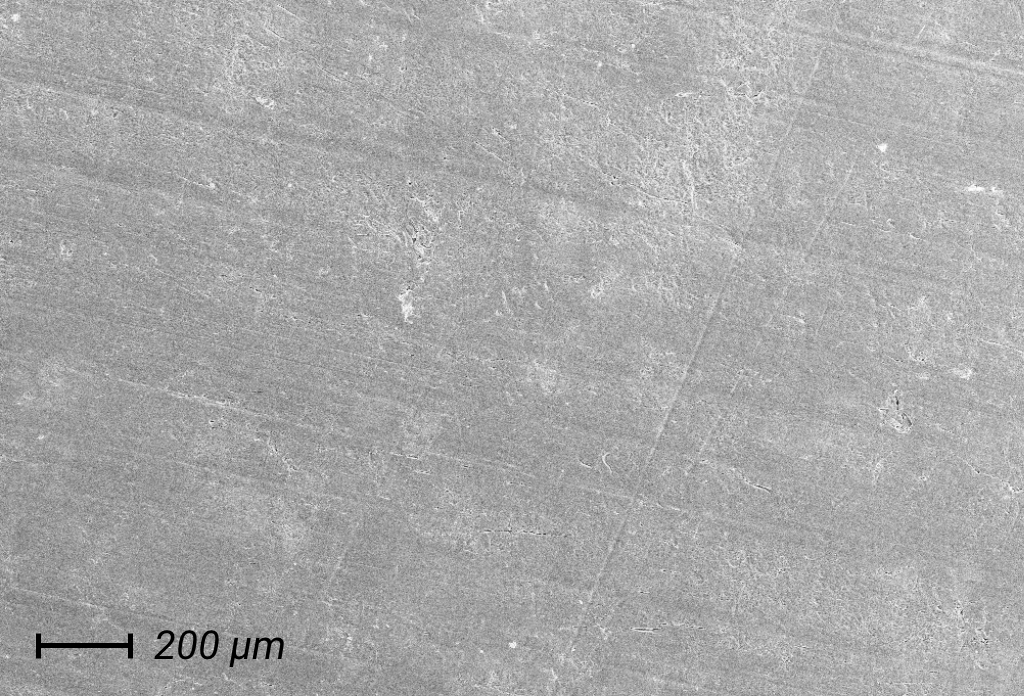}
        \caption{}
        \label{fig:SEM-top:a}
    \end{subfigure}
    \begin{subfigure}{\columnwidth}
        \centering
        \includegraphics[width=\textwidth]{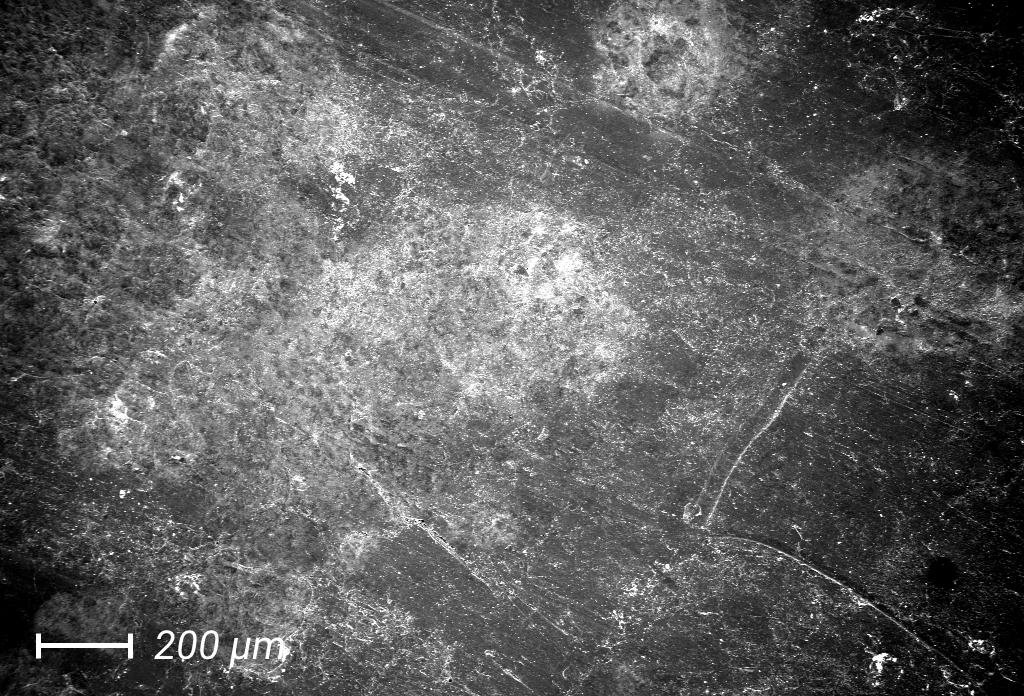}
        \caption{}
        \label{fig:SEM-top:b}
    \end{subfigure}
    \caption{Representative SEM images of the surface of expanded graphite sheets: (a)~Reference zone located outside the primary beam-sweep area; (b)~Beam-spot region at the location of peak energy density.}
    \label{fig:SEM-top}
\end{figure}

Optical and scanning electron microscope (SEM) images were used to examine the characteristics of the darkened profile~\cite{P.Stefan2022Post-mortemDump}. Figure~\ref{fig:SEM-top} shows the initial examination of the surface that revealed circular spots of different sizes and an increase in local roughness in the darkened region where the beam had impacted. To enhance the understanding of the material composition beneath the surface, focused ion beam (FIB) cross-section milling was used to analyze and compare the structural characteristics in different regions along the beam sweep, as well as in the unaffected reference areas.

\begin{figure}[tbp]
    \centering
        \begin{subfigure}{\columnwidth}
        \centering
        \includegraphics[width=\textwidth]{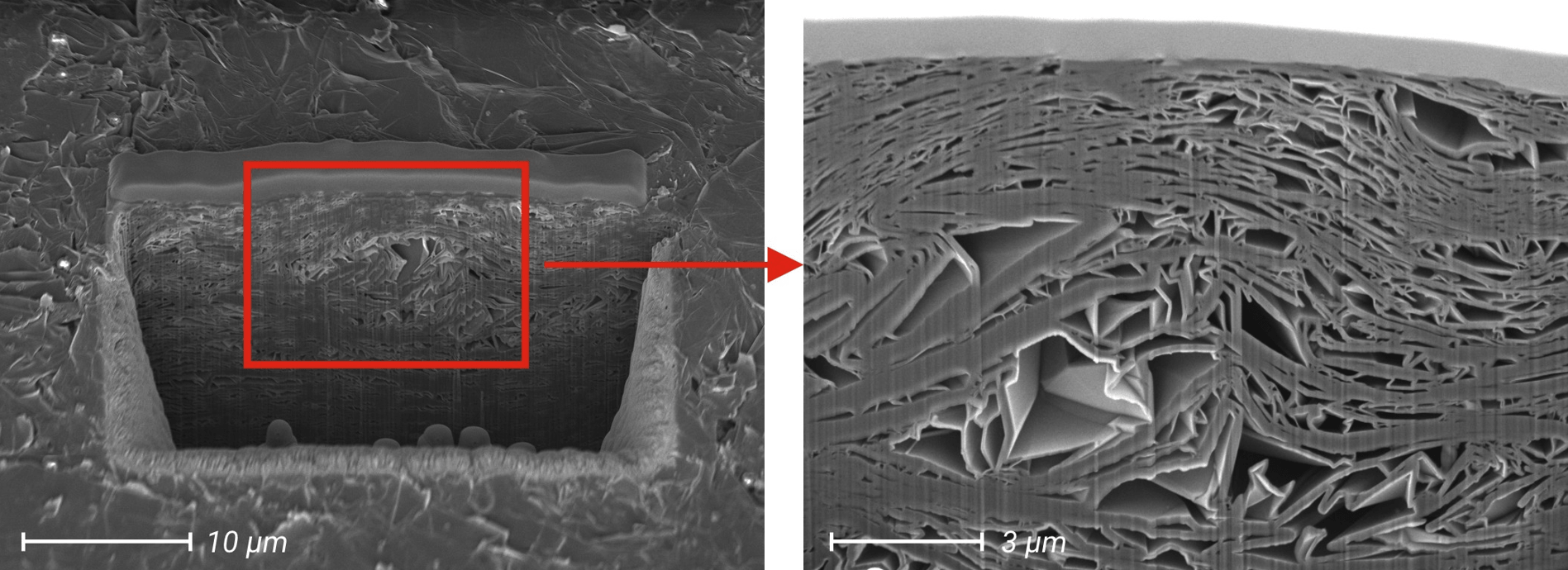}
         \caption{}
        \label{fig:SEMReference}
    \end{subfigure}
    \begin{subfigure}{\columnwidth}
        \centering
        \includegraphics[width=\textwidth]{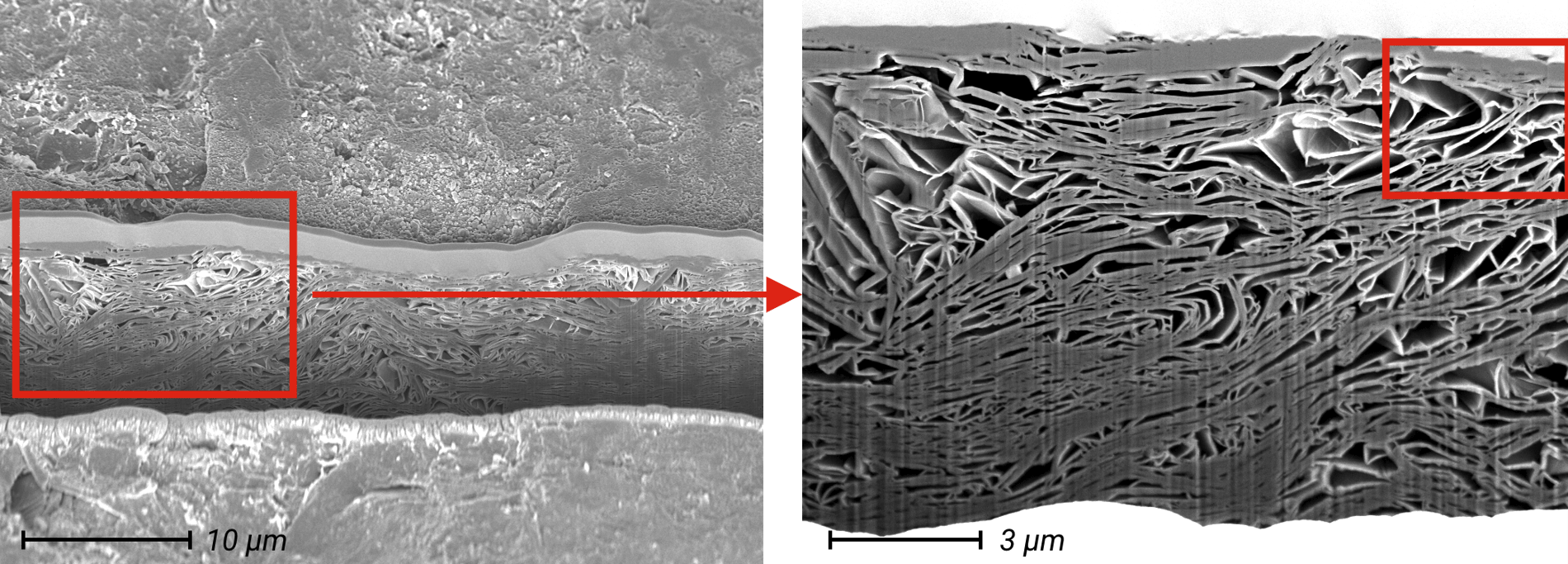}
         \caption{}
        \label{fig:SEMbeamspot}
    \end{subfigure}
    \caption{Representative SEM images at FIB cross sections: (a)~reference zone not in the primary beam-sweep area; (b)~beam-spot zone at energy density peak~\cite{P.Stefan2022Post-mortemDump}.}
    \label{fig:sigraflex-sem}
\end{figure}

Figure~\ref{fig:sigraflex-sem} presents representative SEM images of the cross sections, captured using secondary electron detectors (SESI and InLens). At low magnification, both cross sections have a similar appearance, characterized by tightly packed graphite layers with certain disordered regions containing larger pores.

At higher magnifications, a distinct contrast difference is observed near the surface, origination from the presence of a top region with a thickness between 65--\SI{125}{\nano\meter} that has a lower secondary electron emissivity, thus appearing darker~\cite{P.Stefan2022Post-mortemDump}. Energy dispersive spectroscopy (EDS) analysis revealed that this layer has a slightly higher oxygen content than the underlying graphite sheet layers. This finding aligns well with the high-sensitivity EDS results obtained from the same sample in a top-down view. To further characterize the chemical changes in the near-surface region, X-ray photoelectron spectroscopy was utilized to measure the surface chemical composition of the expanded graphite within an information depth of <\SI{10}{\nano\meter} from the surface (further experimental detail can be found in \cite{MorenoFernandez2021}). In the irradiated regions with dark optical appearance, an enhanced surface oxygen (up to 9 at.\%) and nitrogen (up to 5 at.\%) content was detected, while the unaffected regions have only an surface O content of 1 at.\% with no detectable N signal~\cite{H.MarcelX-RayXPS}. Although the beam dump vessel underwent an evacuation by rough vacuum pumps and were subsequently filled with N2 gas, sufficient oxygen-containing species were present in the residual gas for surface oxidation. It is likely that the impacting high-energy particles directly activated the gas molecules in the beam dump environment allowing them to bond to the graphite surface. Alternatively, activation of surface carbon atoms by the deposited beam energy and the resulting local material temperature increment may have triggered the dissociation of gas phase molecules, leading to a higher surface reactivity.

%It is likely that high-energy particles activated the surrounding atoms, allowing them to bond to the graphite surface within the beam dump environment~\cite{H.MarcelX-RayXPS}. Alternatively, activation of surface carbon atoms may have triggered the dissociation of gas molecules, leading to a surface reaction. Although the beam dumps underwent vacuum treatment and were subsequently filled with $N_2$ gas prior to installation, the presence of sufficient oxygen-containing species in the residual gas could potentially initiate oxidation.

Despite the evident beam sweep on the expanded graphite sheets, no signs of material degradation were detected. Surface oxidation does not appear to have any structural or density implications for the material. The difference in behavior between the expanded graphite in the beam dumps and that in the HiRadMat-43 experiment is primarily attributed to the constraints on the surfaces of the graphite sheets. In the beam dump, the stack of sheets undergoes slight compression, and this helps to prevent the delamination of the sheets upon beam impact.

\subsection{Isostatic graphite}
\begin{figure}[tbp]
\centering
\includegraphics[width=\columnwidth]{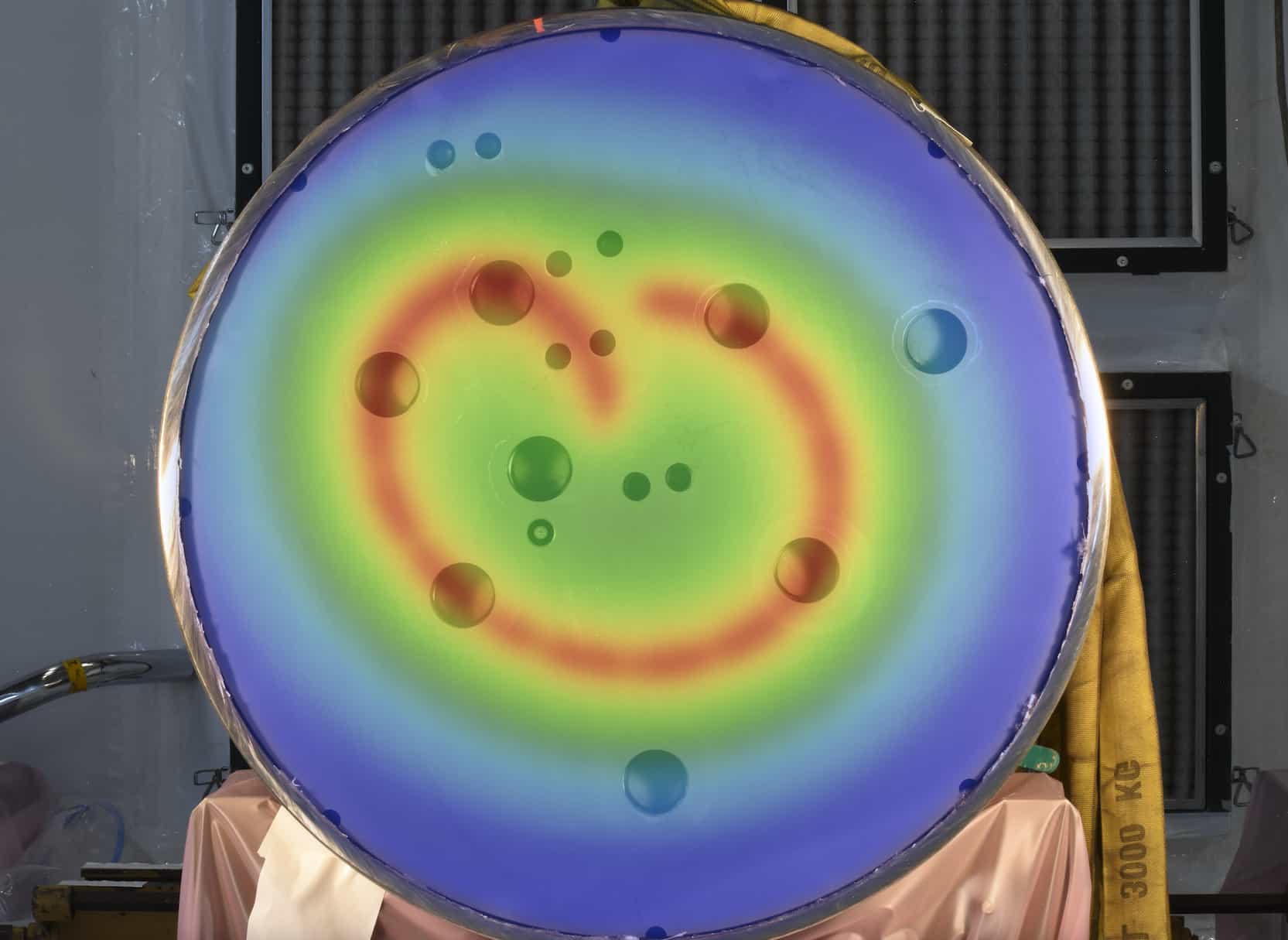}
\caption{View of the downstream face of the upstream HD-block after specimen extraction, overlaid with corresponding beam sweep to indicate corings from different energy densities.}
\label{fig:high-density1}
\end{figure}

\begin{figure}[tbp]
    \centering
    \begin{subfigure}{0.23\textwidth}
        \centering
        \includegraphics[width=\textwidth]{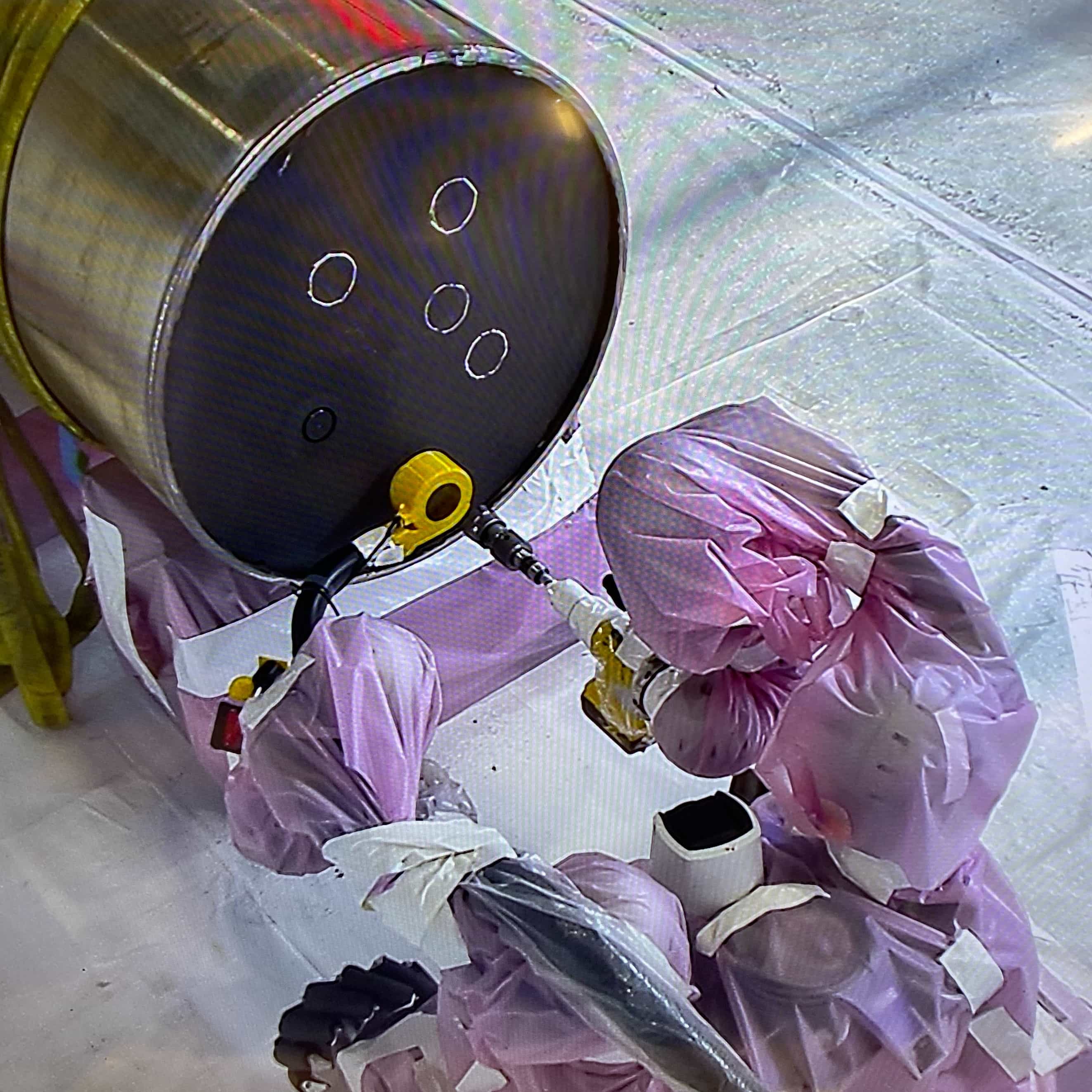}
        \caption{}
        \label{fig:coring:a}
    \end{subfigure}
    \begin{subfigure}{0.23\textwidth}
        \centering
        \includegraphics[width=\textwidth]{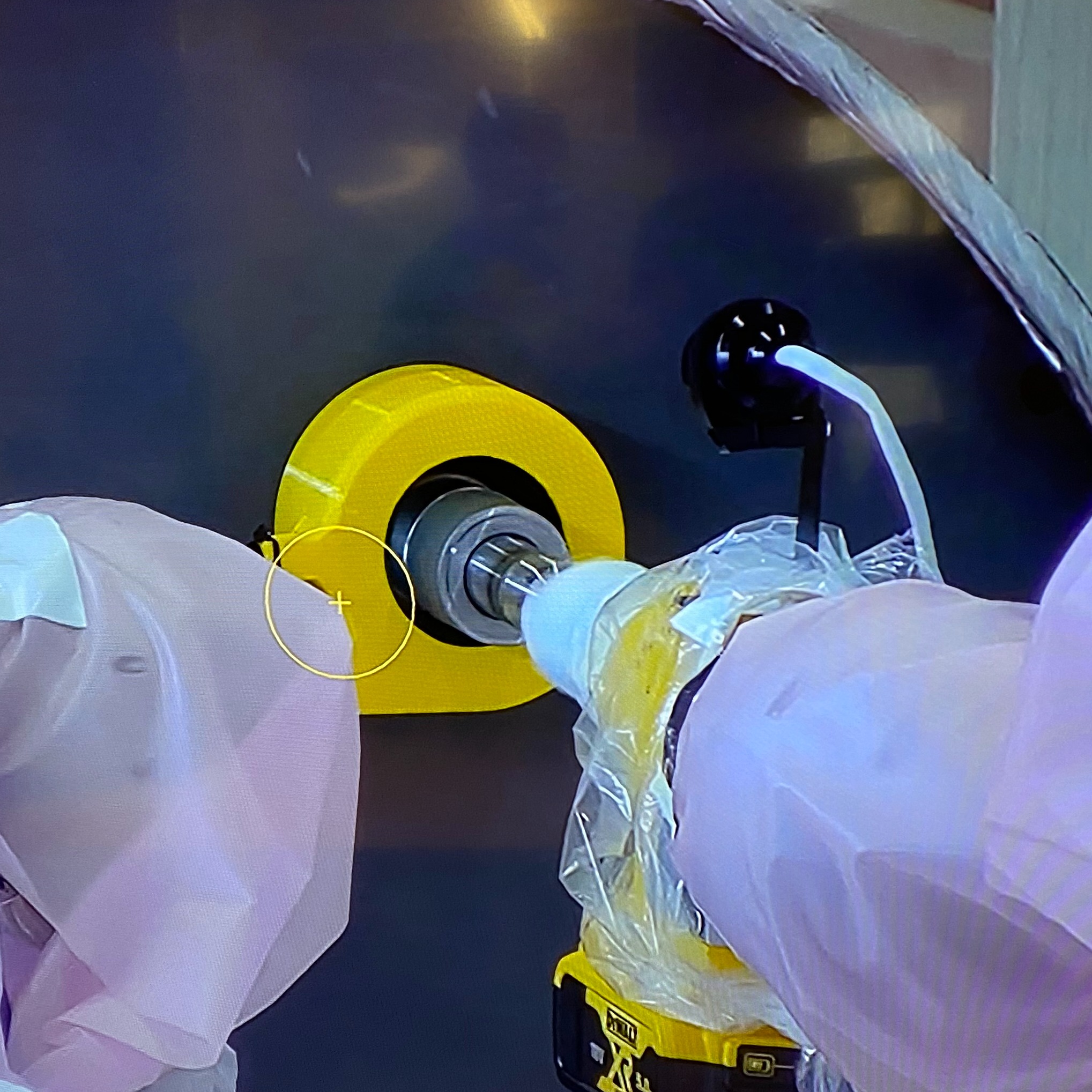}
        \caption{}
        \label{fig:coring:b}
    \end{subfigure}
    \caption{View of the CERNbot performing the coring operation for specimen extraction from the HD blocks. (a)~Overview of the coring setup, with specimen extraction locations marked by white circles. (b)~Ad-hoc drilling tool and drill dust collector surrounding the coring tool, used to minimize the spread of particles during the operation.}
    \label{fig:coring}
\end{figure}

Similar to the LD expanded graphite sheets, the surface of the isostatic graphite blocks displayed a distinct pattern of small black dots, indicating the path of the beam. This pattern eased the determination and marking of extraction positions across different energy densities, enabling a comparative assessment. Control samples were also taken outside the sweep pattern to evaluate the sensitivity of the material to beam impact in less-irradiated areas. The final coring positions, overlaid with the corresponding energy density map, are shown in Fig.~\ref{fig:high-density1}. Due to the HD blocks being shrink-fitted, the presence of the activated vessel was unavoidable, requiring the extraction to be performed by a robot with a coring tool, as shown in Fig.~\ref{fig:coring}. A vacuum cleaner was installed around the coring tool to minimize the spread of particles and potential contamination during the operation. 

\begin{figure}[tbp]
\centering
\includegraphics[width=\columnwidth]{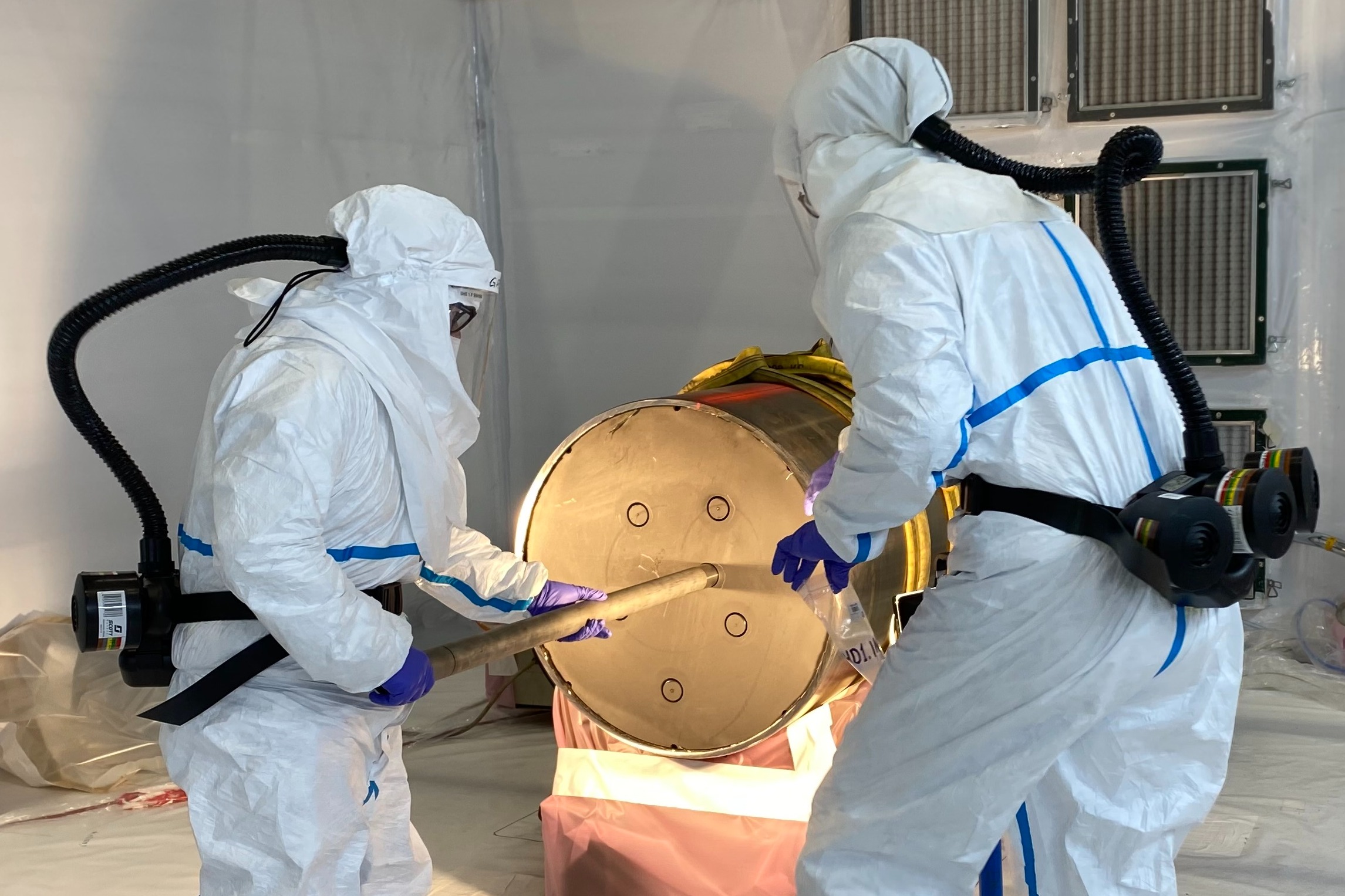}
\caption{Extraction process of specimens from the HD blocks. A dedicated pipe was used to break off the corings.}
\label{fig:coring-extraction}
\end{figure}

Upon completion of all coring operations, the specimens were extracted using a pipe to detach the cores, as illustrated in Fig.~\ref{fig:coring-extraction}. The specimens were collected from the HD blocks located on either side of the LD sector, from the surfaces that were oriented toward the LD sector. As shown in Fig.~\ref{fig:energy-density-along-beamline}, these were the surfaces where the isostatic graphite had been exposed to the highest energy densities.

\begin{figure}[tbp]
    \centering
    \includegraphics[width=\columnwidth]{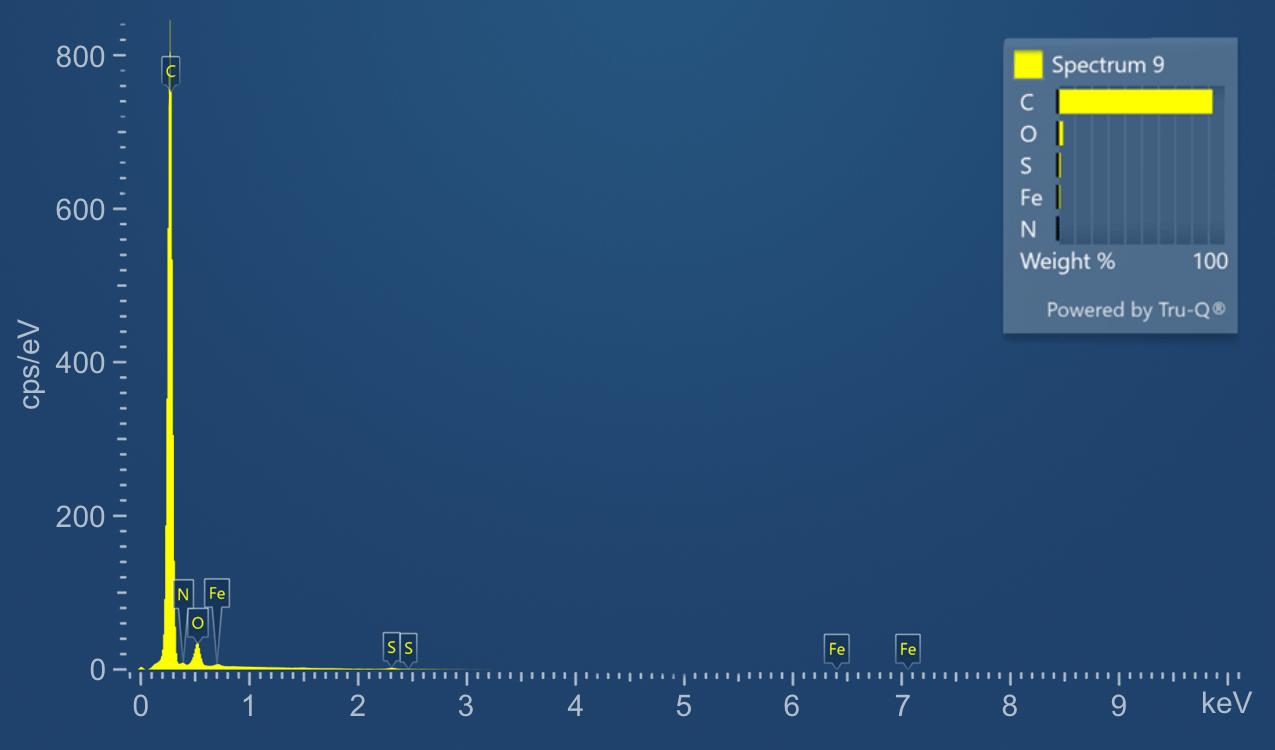}
    \caption{EDS spectrum obtained from a core extracted from the downstream face of the upstream isostatic graphite block at the transverse location with the peak energy deposition.}
    \label{fig:spectrum}
\end{figure}

SEM images were taken to assess any surface changes between the corings; however, no significant differences were observed. Figure~\ref{fig:spectrum} shows the results of a chemical analysis performed using EDS~\cite{Daks2025Sigraflex}. The darkened areas showed the presence of carbon (C), oxygen (O), and nitrogen (N), in line with the observations for the expanded graphite sheets. The hardness of all core samples ranged from 72--85 according to the Shore D standard. This standard was used as the material was nonmetallic and too porous for conventional micro- or macro-hardness testing. The hardness tests showed no clear dependence on extraction location. Furthermore, no significant correlation was observed between hardness and the energy density each sample experienced along the beam sweep.

\subsection{Extruded graphite plate} \label{extruded graphite plate}
As described in Sec.~\ref{motivation-behind-dump-autopsy}, an endoscopy conducted before the autopsy revealed a significant fracture in one of the extruded graphite plates. During the cutting operations, similar fractures were observed in all extruded graphite plates.

\begin{figure}[tbp]
    \centering
    \includegraphics[width=\columnwidth]{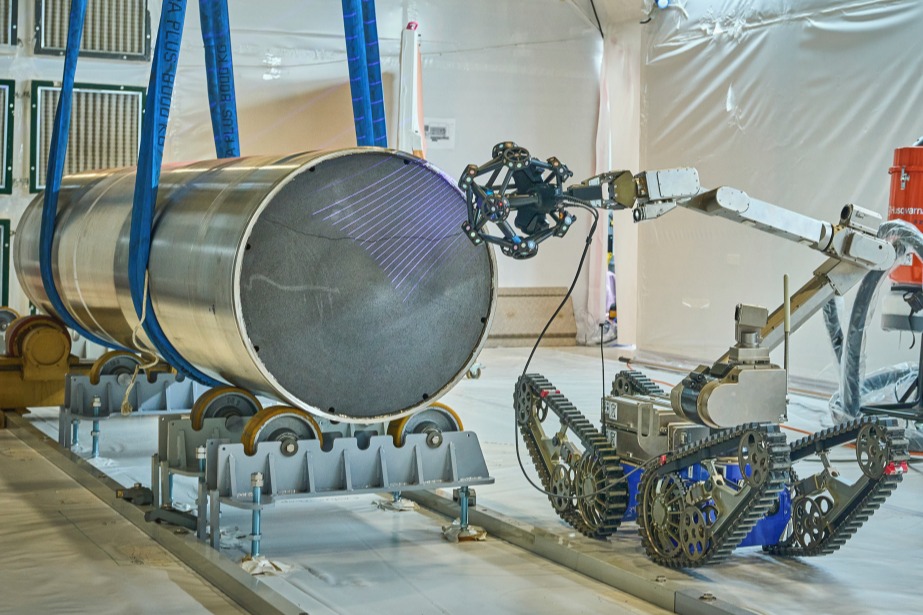}
    \caption{Photograph showing the remote optical CMM scanning of the upstream extruded graphite plate after isolation of the LD sector. A crack of \SI{50}{\centi\meter} can be observed, with its nucleation point in one of the outgassing holes~\cite{Brice2022LHCAutopsy}.}
    \label{fig:metrascan}
\end{figure}

Figure~\ref{fig:metrascan} shows the optical CMM scanning of the upstream extruded graphite plate following the separation of the LD and HD sectors. To minimise the exposure of personnel to radiation during this operation, particularly considering the presence of the highly irradiated vessel, the MetraSCAN was mounted on a mobile robot equipped with an adjustable arm, enabling full movement in all spatial directions.

\begin{figure}[tbp]
    \centering
    \includegraphics[width=\columnwidth]{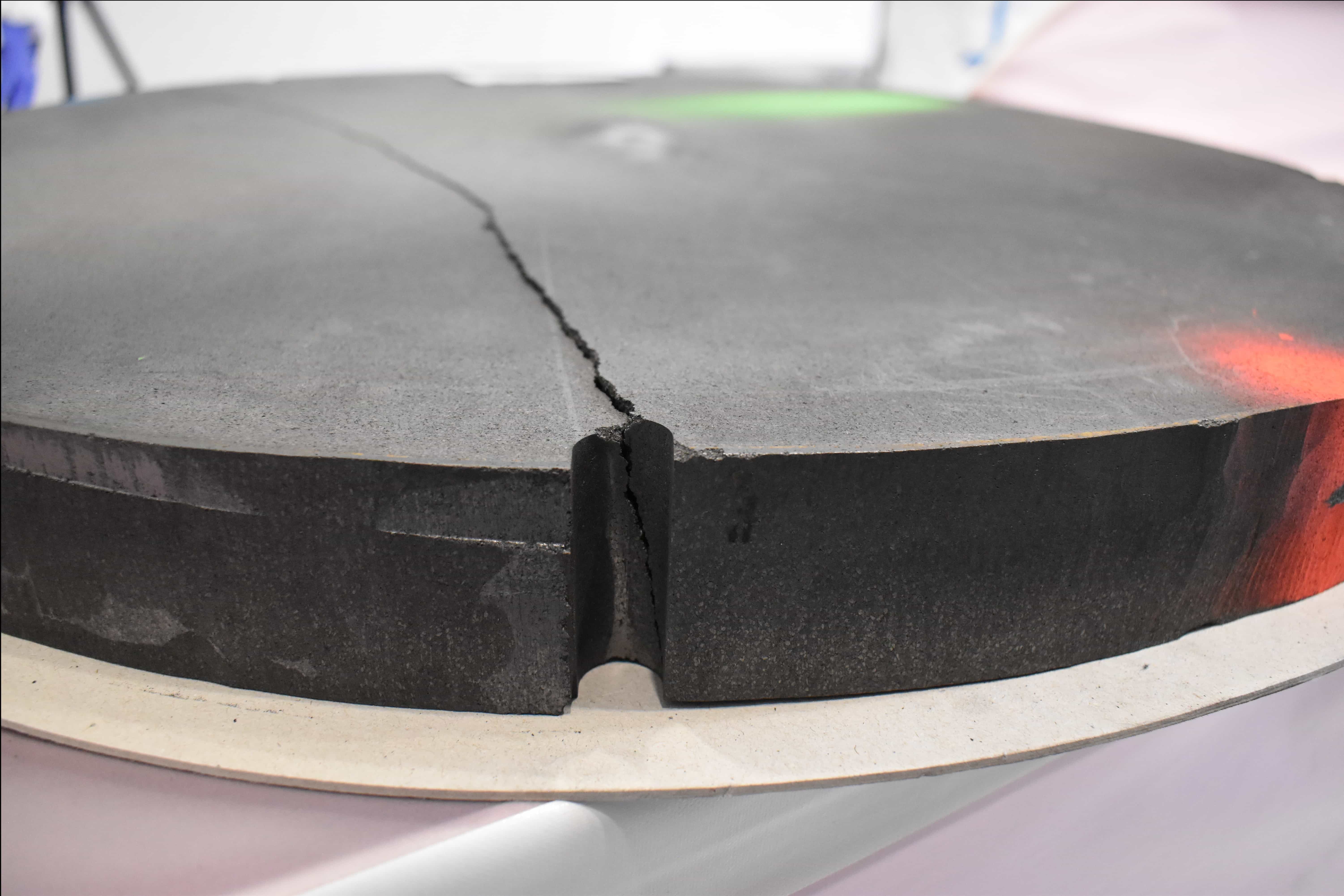}
    \caption{Detailed view of the extruded graphite after extraction. It can be seen that the crack aligns with the outgassing hole.}
    \label{fig:extrudedgraphitecrackextracted}
\end{figure}

Upon the successful longitudinal cutting of the vessel of the LD sector, the extruded graphite plates were extracted, and a second optical CMM scan was performed in the setting as shown in Fig.~\ref{fig:extrudedgraphitecrackextracted}. This optical CMM scan provided a more detailed quantification of the severity of the fracture, allowing a more comprehensive evaluation of the crack and potential failure modes. The fractures were found to align with the locations of the outgassing holes on all of the extruded graphite plates.

The mechanical properties of extruded graphite exhibit anisotropy as a result of the extrusion process, which aligns the graphite particles along the direction of extrusion. Despite being performed under high compression, the extrusion process typically does not achieve the same degree of particle compression as seen in isostatic graphite, resulting in lower density and strength. The failure mode of both isostatic and extruded graphite is well documented in the literature to be predominantly brittle, primarily due to their porosity and limited capacity for plastic deformation~\cite{Berto2013Graphite}.

The outgassing holes present in the extruded graphite plates, as seen in Fig.~\ref{fig:extrudedgraphitecrackextracted}, are thought to have acted as significant stress concentrators upon the dynamic response of the beam dumps. Due to the longitudinal vibration mode of the vessel, the outer \SI{1}{\centi\meter} edge of each extruded graphite plate would have been loaded with high forces through its contact with the retaining rings; consequently, it is believed that the cracks originated and propagated from these outgassing holes. As a result of the cracking of the plates, the expanded graphite sheets in the LD sector were subjected to a lower compression and small gaps formed between them.

A visual inspection of the plates in the beam-sweep area revealed no evidence of deterioration due to energy deposition.

\section{Implications for Run~3 operation and the development of the Run~3 spare beam dumps}
The findings from the decommissioning and post-irradiation examination of the Run~2 beam dumps have had significant implications for the continued operation of the beam dumps that are installed during Run 3 and the development of spare beam dumps for Run~3 in 2023.

The successful features of the Run~2 beam dumps have been taken into account for ongoing development, while the components that have shown signs of damage have prompted further evaluations and redesign efforts.

\subsection{Run~3 operational beam dumps}
As noted in Sec.~\ref{history}, the Run~2 beam dumps were replaced with upgraded spares during LS2 to address nitrogen leaks and to mitigate the effects of the dynamic response of the beam dump block~\cite{Maestre2021DesignMj/Dump}. The upgrades mainly focused on decoupling the beam dump from the extraction line and introducing a new support frame, which allows the beam dump to vibrate more freely~\cite{Maestre2021DesignMj/Dump}; however, the material layout in the core remained unchanged from the Run~2 beam dumps.

The results from the post-irradiation examination of the Run~2 beam dumps provided increased confidence in the performance of the carbon-based materials. In Run~3, the stored energy is up to 68\% higher than in Run~2 (320 vs \SI{539}{\mega\joule}; see Table~\ref{table:BeamParameters}), complicating direct comparisons. Nevertheless, none of the materials in the Run~2 beam dumps showed signs of beam-induced damage after a decade of operation. The concerning occurrence of delamination observed in the expanded graphite sheets during the HiRadMat-43 experiment~\cite{Maestre2021SigraflexPerspective} was absent in the Run~2 beam dumps. This was most likely due to the differing surface constraints imposed on the sheets in each case and the difference in the tested material grade.

The operational beam dumps for Run~3 continue to rely on the same design principles for the retaining rings and extruded graphite discs, which apply the necessary compression to the expanded graphite sheets, thereby preventing delamination. Nonetheless, given the increased energy deposition in Run~3, the extruded graphite discs experience higher dynamic loads, leading to the expectation that similar cracking may occur as Run~3 progresses. It has been confirmed that the graphite powder discovered within the vessel during the post-irradiation examination is attributable to these cracks rather than to a loss of density of the main beam-absorbing materials.

\subsection{Design of Run~3 spare beam dumps}
In response to the observed cracking of the extruded graphite plates, the plates for the Run~3 operational spares have been replaced by carbon-fiber-reinforced carbon (CFRC) plates. These materials, which have already been used in beam-intercepting devices exposed to high levels of energy deposition, are attractive due to their significantly greater strength and enhanced resistance to crack initiation and propagation~\cite{gruber2021carbon}. The retaining plates now feature larger outgassing holes, which are intended to reduce local stress concentrations and decrease the likelihood of crack initiation. This modification aims to ensure proper containment and compression of the LD expanded graphite sheets. In addition, a thorough assessment of the retaining rings has been conducted, resulting in a complete redesign to improve their stability and the implementation of effective measures for securing them firmly within the dumps. Furthermore, the design has been optimized for disassembly by eliminating the tack welds between the vessel and the retaining ring, instead employing a pin-lock configuration that secures the retaining ring within the vessel’s groove.

The HD isostatic graphite and LD expanded graphite both demonstrated strong performance during Run~2, with no evidence of material degradation aside from minor, nondetrimental surface oxidation. As a result, a similar core-material configuration has been retained for the Run~3 operational spares.

The layout of the Run~3 operational spare dumps has undergone slight alterations.
Specifically, an increase in the axial spacing between the HD graphite blocks has been implemented to facilitate easier dismantling of the devices.

\section{Conclusions}
As a consequence of the dismantling and replacement of the two Run~2 beam dumps in 2020, a solution was needed to decommission and assess the condition of the carbon-based cores after 9~years of operation and to safely prepare the dumps for waste disposal.

%A preliminary trial, alongside an unsuccessful market survey conducted with a view to outsourcing these activities, underscored the necessity to develop a framework to allow the internal execution of the autopsy and decommissioning of the devices.

By combining the extensive archive of historically dumped beam intensities with the advanced capabilities of FLUKA Monte Carlo simulations, comprehensive insights were gained into the RP challenges associated with these activities. These insights served as the basis for establishing the needed working environment, informing the optimal cutting sequence of the duplex stainless-steel vessel to minimize the exposure of personnel to radiation, and developing procedures to ensure the careful handling of the dumps.

A remotely controlled cutting method with high reliability, repeatability, and efficiency was developed to minimize potential hazards and radiation exposure throughout the procedure, eliminating the need for human intervention during the cutting process.

During the dismantling, it was discovered that the extruded graphite plates exhibited a consistent pattern of cracking, and the retaining rings were displaced. These structural issues were identified as the root cause of the inability to adequately contain and compress the LD expanded graphite sheet sector. The mechanical and geometrical shortcomings of the plates and rings meant that they were unable to withstand the highly dynamic response of the beam dumps when subjected to the impact of the beam. 

The post-irradiation examination of the carbon-based core materials revealed no evidence of substantial material degradation, even under exposure to energy densities of up to \SI{1.5}{\kilo\joule\per\gram}. The only noticeable change was minor surface oxidation, which had no detrimental effects. This result, contradicting the results of the previous HiRadMat-43 experiment, highlighted the sensitivity of the materials to surface constraints. As a result, the same core material configuration was retained for the Run~3 operational spares, with the exception of the extruded graphite discs used to retain the LD-sector, which were replaced by carbon fibre-reinforced carbon discs.

This work effectively demonstrated the ability to dismantle and decommission large and highly radioactive objects at CERN; it also enhances the level of confidence in the execution of such operations for future endeavors.

\section*{Acknowledgments}
The authors would like to thank the CERN Radiation Protection Group for their support and expertise throughout the decommissioning and post-irradiation examination processes. Special thanks are extended to the teams involved in the development and implementation of the cutting methods, as well as the operational support staff who ensured the safe handling and processing of the beam dumps. This work has benefited from the collaborative efforts of numerous sections across CERN, whose contributions were vital to the success of this project.

%\clearpage
\bibliography{references, RP_references}

\end{document}